\documentclass[]{elsarticle}
\usepackage{geometry}
\usepackage[titletoc,page]{appendix}
\usepackage{amsmath,amssymb,amsfonts,bm}
\usepackage{graphicx}
\usepackage[dvipsnames]{xcolor}
\usepackage[english]{babel}
\usepackage{tensor}
\usepackage{hyperref}
\usepackage[capitalise]{cleveref}
\usepackage{natbib}
\usepackage{braket}
\usepackage{mathabx}
\usepackage{tikz}
\usepackage{tensor}
\usetikzlibrary{calc}
\usepackage{comment}

\usepackage{multicol}
\usepackage[font=small,labelfont=bf]{caption}

\newcommand{\fdir}{figures/}
\graphicspath{{\fdir}}

\DeclareMathOperator{\sgn}{sgn}

\newcommand{\follows}{\ensuremath{\Rightarrow\quad}}
\newcommand{\inv}[1]{\ensuremath{\frac{1}{#1}}}

\newcommand{\intd}{\ensuremath{\mathrm{d}}}

\newcommand{\iu}{\ensuremath{\mathrm{i}}}

\newcommand{\vx}{\ensuremath{{\bm x}}}

\newcommand{\mP}{\ensuremath{\mathcal{P}}}
\newcommand{\Lag}{\ensuremath{\mathcal L}}

\newcommand{\Ddiff}{\ensuremath{D_{\text{diff}}}}

\newcommand{\Rho}{\mathrm{P}}

\newcommand{\mA}{\mathrm{A}}
\newcommand{\mB}{\mathrm{B}}

\renewcommand{\vec}[1]{\ensuremath{\bm{#1}}}

\graphicspath{{figures/classstat/}}

\biboptions{sort&compress}
\journal{Nuclear Physics B}

\begin{document}
\title{Critical dynamics of relativistic diffusion}

\address[gie]{Institut f\"ur Theoretische Physik, Justus-Liebig-Universit\"at, Heinrich-Buff-Ring 16, 35392 Gie{\ss}en, Germany}
\address[hfhf]{Helmholtz Research Academy Hessen for FAIR (HFHF), Campus Gie{\ss}en, 35392 Gie{\ss}en, Germany}
\address[bie]{Fakult\"at f\"ur Physik, Universit\"at Bielefeld, D-33615 Bielefeld, Germany}

\author[gie]{Dominik Schweitzer}
\ead{dominik.schweitzer@theo.physik.uni-giessen.de}
\author[bie]{S\"{o}ren Schlichting}
\ead{sschlichting@physik.uni-bielefeld.de}
\author[gie,hfhf]{Lorenz von Smekal}
\ead{lorenz.smekal@physik.uni-giessen.de}

\begin{abstract}
    We study the dynamics of self-interacting scalar fields with $Z_2$ symmetry governed by a relativistic Israel-Stuart type diffusion equation in the vicinity of a critical point.
    We calculate spectral functions of the order parameter in mean-field approximation as well as using first-principles classical-statistical lattice simulations in real-time.
    We observe that the spectral functions are well-described by single Breit-Wigner shapes.    Away from criticality, the dispersion matches the expectations from the mean-field approach.
    At the critical point, the spectral functions largely keep their Breit-Wigner shape, albeit with non-trivial power-law dispersion relations.
    We extract the characteristic time-scales as well as the dynamic critical exponent $z$, verifying the existence of a dynamic scaling regime.
    In addition, we derive the universal scaling functions implied by the Breit-Wigner shape with critical power-law dispersion and show that they match the data.
    Considering equations of motion for a system coupled to a heat bath as well as an isolated system, we perform this study for two different dynamic universality classes, both in two and three spatial dimensions.
\end{abstract}


\maketitle
\tableofcontents
\clearpage

\section{Introduction}
Thermodynamic systems at a critical point display characteristic features caused by large, long-range fluctuations.
As the typical length scale of these fluctuations diverges, singular contributions in various observables arise, whose functional form is dictated by scale-invariance.
Analogously, the characteristic time scale of the system's dynamics diverges, leading to singular contributions in multi-time correlation functions.
Since microscopic details of the respective systems average out at sufficiently large length and time scales, these critical phenomena show universality:
Different physical systems exhibit quantitatively equivalent critical behaviour as long as their effective macroscopic degrees of freedom have similar properties.
Thereby, studying the critical behaviour of well-behaved and relatively easy-to-control condensed-matter systems like e.g.~thin films \cite{dunlavy_critical_2005,honerkamp-smith_experimental_2012}, one can gain valuable insight into respective properties of hard-to-access systems, such as e.g.~strong-interaction matter described by the theory of quantum chromodynamics (QCD).

The phase diagram of QCD has been subject of many theoretical studies and experimental campaigns over the last decades \cite{stephanov_signatures_1998,rajagopal_condensed_2001,odyniec_rhic_2013,bzdak_mapping_2020}.
One particular challenge that remains is a precise determination of the location of the presumed QCD critical endpoint (CEP) at the end of the first-order chiral transition line in the plane of temperature and baryon chemical potential.
It is in the main focus of a number of current and future heavy-ion collision experiments at RHIC, GSI-FAIR and NICA, which are designed specifically to probe the relevant part of the QCD phase diagram.
The CEP, and with it the onset of the first-order chiral transition are expected to manifest themselves by causing a non-monotonic behaviour in event-by-event fluctuations of conserved charges when varying the beam energy or the centrality class of the collision \cite{stephanov_signatures_1998}.
While data from the STAR collaboration does show some indications of such a behaviour in fluctuations of the net-proton number \cite{star_collaboration_energy_2014,thader_higher_2016}, solid theoretical predictions are required for the interpretation of these results.
However, first-principles lattice QCD calculations currently cannot address the physics of the high-density region at large baryon chemical potential.
Thus, on has to resort to effective descriptions and theories, while using all available prior knowledge about the critical behaviour of QCD \cite{bluhm_dynamics_2020}.

Close to the QCD CEP, as the baryon diffusion rate vanishes, the relevant slow mode is a linear combination of the net-baryon density and the chiral condensate \cite{son_dynamic_2004}.
The resulting fluctuations are governed by diffusive dynamics.
Thus, an apt description of the dynamic critical behaviour of such a diffusive quantity is crucial for the interpretation of future experimental data.
In the classification scheme by Hohenberg and Halperin \cite{hohenberg_theory_1977}, the dynamic universality class of a system with a conserved order parameter without any further dynamic couplings is that of Model B.
Introducing a dynamic coupling to a non-critical conserved scalar density, e.g.~the energy density, changes this to Model D.
In the case of full QCD dynamics, one additionally has to take into account the energy-momentum tensor and concludes that it is then equivalent to Model H \cite{son_dynamic_2004} which otherwise describes the dynamics of liquid-gas transitions.

Continuing a previous study \cite{schweitzer_spectral_2020}, we investigate a relativistic scalar field theory with a second-order phase transition in the same static $Z_2$ Ising universality class as the QCD critical endpoint.
Ultimately, the goal is to devise a microscopical model which shares with QCD also the dynamic universality class of Model H.
Since modeling the complete set of relevant dynamic degrees of freedom for Model H proves more challenging, here we first focus on the slightly simpler diffusive Model B, without the additional conserved shear modes characteristic of Model H. In addition to the inherent value of a systematic characterization of critical real-time dynamics, we consider this an important further step from the previously studied relaxational Models A and C towards the full dynamics of Model H for QCD.

A powerful tool for quantifying dynamic critical phenomena is the study of real-time correlation functions of the order parameter field and in particular its spectral function.
The abundance of information encoded in spectral functions is of great interest for a wide area of research ranging from condensed matter systems to nuclear and particle physics.
Specifically, spectral functions in equilibrium contain the spectrum of possible excitations of the system, enabling one to deduce and identify the degrees of freedom most relevant for an effective description.
As the system comes close to a critical point, spectral functions strongly reflect the emergent dynamic critical phenomena dominated by slow modes.
Since the characteristic time scale diverges with a power of the spatial correlation length determined by the dynamic critical exponent $z$, one observes scale invariance in the low-energy behaviour of spectral functions which can be exactly described by universal dynamic scaling functions \cite{schweitzer_spectral_2020}.

We employ the framework of classical-statistical lattice simulations in real time \cite{aarts_spectral_2001,berges_dynamic_2010,schlichting_spectral_2019} to non-perturbatively calculate the spectral function of a single-component scalar field theory with dynamically conserved order parameter.
We thereby extend our earlier study \cite{schweitzer_spectral_2020} where we investigated the critical dynamics of a similar model, but with purely relaxational rather than diffusive dynamics.
In order to calculate the spectral function, we use the fluctuation-dissipation relation or Kubo-Martin-Schwinger (KMS)  condition \cite{kubo_statistical-mechanical_1957,martin_theory_1959} which allows us to obtain the spectral function from the statistical two-point function.
The latter can be calculated from an elementary unequal-time correlation function of classical fields.
As in Ref.~\cite{schweitzer_spectral_2020}, we study the spectral function of the model at finite spatial momenta in 2+1 and 3+1 dimensional space-time, in parallel, both with and without coupling to a heat bath included.
This allows us to extract the characteristic time scale $\xi_t$ and its divergence at the critical point, the corresponding dynamic critical exponent $z$, as well as the universal dynamic scaling functions governing the low-energy regime of the critical spectral functions for each case.

This paper is organized as follows:
After recapitulating the dynamic equations used in the re\-nor-malization-group analysis of dynamic critical phenomena \cite{hohenberg_theory_1977,folk_critical_2006}, we introduce a generalization of the dynamical models of \cite{schweitzer_spectral_2020} that allows for order parameter conservation in \cref{sec:models}.
Some subtleties of its lattice regularization are then discussed in \cref{sec:lattice}.
Beginning with \cref{sec:specfunc}, we show our numerical results for the thermal spectral functions of the order parameter at different points in the phase diagram, after which \cref{sec:op_crit} is dedicated specifically to the dynamic scaling behavior.
Finally, \cref{sec:conclusion} provides our conclusions and an outlook for further possible applications and studies.
Some lengthy algebra and derivations are given in the appendices.

\section{Dynamical Models}\label{sec:models}
We start with the scalar Landau-Ginzburg-Wilson (LGW) model in $d=2,3$ spatial dimensions.
The partition function of the order-parameter field $\phi(x) \equiv \phi(\vec x,t)$ is given by
\begin{align}
    Z &= \int \mathcal D[\phi] \exp{\left( -\beta \mathfrak H[\phi] \right)} \label{eq:LGW_equilibrium_phi}, \\
    \mathfrak H[\phi] &= \int \intd^dx \left\{ \frac{1}{2} \phi(x) \left( -\nabla^2 + m^2 \right) \phi(x) + \frac{\lambda}{4!} \phi^4(x) -J \phi(x) \right\}, \label{eq:z2_continuum}
\end{align}
where $\mathfrak H[\phi]$ is the LGW Hamiltonian, and $\beta = 1/T$ refers to the inverse temperature.
In case of vanishing explicit symmetry breaking ($J=0$), this model is invariant under the $Z_2$ transformation of the order parameter field $\phi(x) \to -\phi(x)$.
If one chooses a negative square-mass $m^2 < 0$, the symmetry is spontaneously broken for temperatures $0\leq T<T_c$ below the critical temperature $T_c$, but restored above $T_c$ after undergoing a second-order phase transition in the $Z_2$ (Ising) universality class.

Here, we consider an extension of the LGW model with an additional non-critical field $\rho(x)$, such that the effective Hamiltonian is given by
\begin{equation}
    \mathfrak H'[\phi,\rho] = \mathfrak H[\phi] + \frac{1}{2}\int \intd^d x \left[ \rho^2(x) + g \rho(x) \phi^2(x)  \right] \label{eq:modeld_lit_H}.
\end{equation}
One can integrate out $\rho(x)$ by completing the square and absorbing the shift in the anharmonicity by a redefinition of $\lambda\to \lambda' = \lambda - 3g^2$.
Therefore, the introduction of $\rho(x)$ does not change the static critical behaviour of the theory, in particular as $\lambda \to \lambda^*$ approaches a universal fixed point value anyway.

Interpreting the fields $\phi(x)$ and $\rho(x)$ as coarse-grained hydrodynamic degrees of freedom, one can write linearized equations of motion (see e.g.~\cite{rajagopal_static_1993}).
The dynamic equations are then given by
\begin{align}
    \dot \phi(\vec x, t) = F^{\phi}_{\text{rev}}[\phi,\rho] - \hat \gamma_{\phi\phi} \frac{\delta \mathfrak H'[\phi, \rho]}{\delta\phi(\vec x, t)} - \hat \gamma_{\phi\rho} \frac{\delta \mathfrak H'[\phi, \rho]}{\delta\rho(\vec x, t)} + \xi(\vec x, t), \\
    \dot \rho(\vec x, t) = F^{\rho}_{\text{rev}}[\phi,\rho] - \hat \gamma_{\rho\rho} \frac{\delta \mathfrak H'[\phi, \rho]}{\delta\rho(\vec x, t)} - \hat \gamma_{\rho\phi} \frac{\delta \mathfrak H'[\phi, \rho]}{\delta\phi(\vec x, t)} + \zeta(\vec x, t),
\end{align}
with reversible forces $F_{\text{rev}}$, dissipative forces driving the system towards a minimum of $\mathfrak H'$, and stochastic forces $\xi,\, \zeta$.
In slight abuse of the notation, the functional derivative is defined here as a $d$-dimensional one at a fixed time $t$.
Dots are used for partial time derivatives of the fields $\dot \phi(\vec x, t) \equiv 
\partial_t  \phi(\vec x, t)$.

Reversible forces are present whenever there are non-linear couplings between hydrodynamic modes. They can be expanded in terms of the non-vanishing equal-time generalized Poisson brackets $\left\{\cdot, \cdot\right\}$. At vanishing temperature, for example, this would here lead to the following form,\footnote{In addition to the zero-temperature contributions given explicitly here, which correspond to $\dot \phi(\vec x,t) = \left\{ \mathfrak H'[\phi, \rho], \phi(x) \right\}$, additional terms $\propto\, T $  might in general be needed for the reversible forces to be compatible with the equilibrium distribution $\mP[\phi, \rho] = e^{-\beta \mathfrak H'[\phi, \rho]}$  \cite{folk_critical_2006,tauber_critical_2014}.
    These can be obtained explicitly following the Mori-Zwanzig projector formalism \cite{zwanzig_ensemble_1960,zwanzig_memory_1961,mori_transport_1965,kawasaki_simple_1973}.
}
\begin{align}
    F^\phi_{\text{rev}}[\phi, \rho] = -\int \intd^d x' \frac{\delta \mathfrak H'[\phi,\rho]}{\delta \rho(\vec x', t)} \left\{ \phi(\vec x, t), \rho(\vec x', t) \right\},
\end{align}
and an analogous expression for $F^{\rho}_{\text{rev}}[\phi,\rho]$ with $\phi$ and $\rho$ interchanged.

The noise correlators of the stochastic forces are given by
\begin{align}
    \braket{\xi(\vec x,t)\xi(\vec x',t')} = 2T \hat \gamma_{\phi\phi} \delta(\vec x-\vec x') \delta(t-t'), \label{eq:noise_ff}\\
    \braket{\zeta(\vec x,t)\zeta(\vec x',t')} = 2T \hat \gamma_{\rho\rho} \delta(\vec x-\vec x')\delta(t-t'), \label{eq:noise_rr}\\
    \braket{\xi(\vec x,t)\zeta(\vec x',t')} = 2T \hat \gamma_{\phi\rho} \delta(\vec x-\vec x')\delta(t-t'). \label{eq:noise_fr}
\end{align}
Due to Onsager's principle, the (operator valued) kinetic coefficient matrix $\hat\gamma $ must be symmetric, i.e.~$\hat \gamma_{\phi\rho} = \hat \gamma_{\rho\phi}$.
If the off-diagonal terms $\hat \gamma_{\phi\rho}>0$ are non-zero, one calls $\rho$ and $\phi$ \emph{dynamically coupled}.

Further constraints on the remaining kinetic coefficients depend on whether the corresponding field is (locally) conserved or not.
If it is not conserved, the derivative expansion of the kinetic coefficient starts with an order-zero constant term  $\hat \gamma_{\phi\phi} \sim \Gamma_\phi$ which then corresponds to the Onsager relaxation coefficient.
In the case of a locally conserved density, on the other hand, the derivative expansion starts at second order. I.e.~in momentum space one then has $\hat \gamma_{\rho\rho}(\vec q) \sim \lambda_\rho \vec q^2$ in the static long-wavelength limit to describe the relaxational dynamics of the conserved density with diffusion  constant $\lambda_\rho$.\footnote{Without the requirement of locality in the conservation law, one can also have $\hat\gamma_{\psi\psi}\sim {\vec q}^{\sigma}$ with a more general exponent $\sigma>0$ leading to different dynamic critical behaviour \cite{sen_is_2002}.
}

For studies of dynamic critical phenomena in Models A-D one can furthermore assume dynamically decoupled degrees of freedom which amounts to vanishing kinetic cross-coefficients, $\hat \gamma_{\phi\rho}=0$.
As we have studied the dynamics of Models A and C with non-conserved order parameters in Sections 5 and 6 of our previous  Ref.~\cite{schweitzer_spectral_2020} already, we now move on to discuss Models B and D here.

When analyzing dynamics of Model D, one has a locally conserved order-parameter $\phi(x)$ statically coupled to a conserved non-critical secondary density $\rho(x)$.
Since the mesoscopic fields $\phi$ and $\rho$ neither have non-commuting microscopical equivalents, nor does either of them contain a generator of a symmetry group of the system, the generalized Poisson brackets $\left\{ \phi, \rho \right\}=0$ in this case vanish identically \cite{dzyaloshinskii_poisson_1980}.
Therefore, the reversible forces themselves vanish as well which thus leads to first-order equations of motion of the form
\begin{align}
    \dot \phi(\vec x, t) = \lambda_\phi \nabla^2 \frac{\delta \mathfrak H'[\phi,\rho]}{\delta \phi(\vec x, t)} + \xi(\vec x,t), \label{eq:modeld_lit_phi} \\
    \dot \rho(\vec x, t) = \lambda_\rho \nabla^2 \frac{\delta \mathfrak H'[\phi,\rho]}{\delta \rho(\vec x, t)} + \zeta(\vec x,t), \label{eq:modeld_lit_rho}
\end{align}
with the LGW Hamiltonian from \eqref{eq:z2_continuum}, and diffusion constants $\lambda_\phi $ and $\lambda_\rho $.
In order to be consistent with \cref{eq:noise_ff,eq:noise_rr,eq:noise_fr} the noise terms are also conserved with vanishing expectation value and correlators given by
\begin{align}
    &\braket{\xi(\vec x,t)\xi(\vec x',t')} = -2 T \lambda_\phi \nabla^2 \delta(\vec x-\vec x') \delta(t-t'), \\
    &\braket{\zeta(\vec x,t)\zeta(\vec x',t')} = -2 T \lambda_\rho \nabla^2 \delta(\vec x-\vec x')\delta(t-t'), \\
    &\braket{\xi(\vec x,t)\zeta(\vec x',t')} = 0.
\end{align}

Setting $g\to 0$ and thus decoupling the non-critical density, one recovers Model-B dynamics for $\phi$.
A dynamic renormalization group analysis shows that, due to the order parameter being conserved, there are no $\varepsilon$-dependent contributions to the response propagator.
Therefore conventional theory holds \cite{halperin_renormalization-group_1974,hohenberg_theory_1977,folk_critical_2006}, and one has for the dynamic critical exponent of the order parameter $z_B = 4-\eta$.
This is unchanged if one introduces the secondary, non-critical conserved quantity $\rho$, which can model e.g.~the energy density and is coupled with $g>0$ to the square of the order parameter field $\phi $.
For the secondary density $\rho$, on the other hand, the dynamic scaling hypothesis requires a different dynamic critical exponent $z_C = 2+\alpha/\nu$ \cite{tauber_critical_2014} which is the same as that for the order parameter field in Model C (in the Gaussian approximation one has $z_B  = z_C^2 = 4$).

\subsection{Microscopic field theory realizations}

In order to probe the dynamic critical behaviour using classical statistical simulations, we set out to define relativistic continuum theories whose effective dynamic degrees of freedom at the critical point match the prescriptions of the effective Models of Halperin and Hohenberg.
In \cite{schweitzer_spectral_2020}, we have studied a relativistic scalar field theory evolving under Langevin and Hamiltonian dynamics.
Specifically, we used as the stochastic evolution equation of the order parameter field $\phi(\vec x, t)$ a second-order equation of motion with uncorrelated white noise of the form
\begin{align}
    \ddot\phi(\vec x,t) &= - \frac{\delta \mathfrak H[\phi]}{\delta \phi(\vec x,t)} - \gamma \dot \phi(\vec x,t) + \sqrt{2 \gamma T} \, \eta(\vec x, t), \label{eq:ModelA_eom} \\
    \Braket{\eta(\vec x,t)} &= 0,\ \Braket{\eta(\vec x',t')\eta(\vec x,t)} = \delta(\vec x'- \vec x)\delta(t'-t). \label{eq:ModelA_noise}
\end{align}
The real parameter $\gamma$ represents the Langevin coupling to a heat bath via the Gaussian random noise $\eta(\vec x,t)$.
The conjugate momentum field in this case is identical to the time derivative of $\phi$, and we define $\pi(\vec x, t) \equiv \dot\phi(\vec x,t)$ to be used as the kinetic momentum field throughout in the following.
As we have demonstrated explicitly in \cite{schweitzer_spectral_2020}, this system shows the expected dynamic critical behaviour of Model A (C) for finite (vanishing) Langevin coupling $\gamma$.

In the case of the diffusive dynamics of Model B, when the order parameter $Q = \int\intd^dx \, \phi(\vec x, t)$ is conserved, i.e.~$\dot Q = 0$, we consider equations of motion of the form
\begin{align}
    \ddot\phi(\vec x,t) &=  \mu \nabla^2 \frac{\delta \mathfrak H[\phi]}{\delta \phi(\vec x,t)} - \gamma \dot \phi(\vec x,t) + \sqrt{2 \gamma T} \, \eta(\vec x, t), \label{eq:ModelB_eom} \\
    \Braket{\eta(\vec x,t)} &= 0,\ \Braket{\eta(\vec x',t')\eta(\vec x,t)} = - \mu \nabla^2 \delta(\vec x'-\vec x)\delta(t'-t), \label{eq:ModelB_noise}
\end{align}
where $\mu$ is the mobility coefficient (for low frequency excitations with $\omega \ll \gamma $ it reduces to $\mu = \gamma\lambda_\phi $ in the linearized equations of the previous subsection).

We note that for both Models A and B, decoupling the system from the heat bath by setting the Langevin coupling $\gamma = 0$, leads to another conserved scalar quantity in the system, which can be identified with the total energy. Due to the presence of this additional conserved quantity, this conservative limit of Model A in Eq.~(\ref{eq:ModelA_eom}) corresponds to the dynamic universality class of Model C as discussed explicitly in~\cite{hohenberg_theory_1977}. While in the limit $\gamma \to0$ the equation of motion (\ref{eq:ModelB_eom}) for Model B also features an additional conserved quantity in Eq.~(\ref{eq:K_energy_B}), this situation is clearly more subtle. Even though in this limit the model in Eq.~(\ref{eq:ModelB_eom}) features the same set of conserved quantities as Model D, the structure of excitations is completely different as for $\gamma \to 0$, Eq.~(\ref{eq:ModelB_eom}) becomes a non-linear wave-equation, which conserves the order parameter but no longer features ordinary diffusive behavior at tree level. Since the classification of the non-dissipative limit of our Model B dynamics is not obvious, we will refer to it as ``Model BC'' in the following to highlight that this dynamics emerges as the conservative (C) limit of an Israel-Stuart type diffusive dynamics (Model B).

As a brief recap, the equilibrium distribution of the order-parameter field for the standard Langevin evolution in Eq.~\eqref{eq:ModelA_eom} of course corresponds to the Boltzmann distribution, 
\begin{equation}
    \mP_\mA\left[\phi,\pi \right] = Z^{-1} \exp\bigg\{ -\beta \mathfrak H[\phi]  -\beta \int \intd^d x \,  \frac{\pi^2(x)}{2} \bigg\} \equiv  Z^{-1} \exp\big\{ -\beta H_\mA [\phi,\pi] \big \} \, . \label{eq:stationary-solution-A} 
\end{equation}
It is the stationary solution to the Fokker-Planck equation for the It\^o-Langevin process described by Eqs.~\eqref{eq:ModelA_eom} and \eqref{eq:ModelA_noise} with Model A dynamics whose drift term vanishes. One hence has Liouville's theorem
\begin{equation}
    \frac{\intd \mP_\mA}{\intd t} \, =\,  
    \frac{\partial \mP_\mA}{\partial t } - \int \intd^d x\, \bigg(\frac{\delta \mathfrak H[\phi]}{\delta \phi_{\vec x}} \frac{\delta }{\delta \pi_{\vec x}} - \pi_{\vec x}  \frac{\delta }{\delta \phi_{\vec x}}  \bigg) \, \mP_\mA[\phi,\pi] \, = \, 0\,,
\end{equation}
where the implicit time dependence is given by the equal-time Poisson bracket between $H_\mA $ and $\mP_\mA$ as usual, with subscripts $\vec x$ as shorthand notations for the spatial functional derivatives w.r.t.\ the fields at fixed times.
In general, the right hand side of the Fokker-Planck equation is given by the collision term.
For our Model A dynamics it reads,
\begin{equation}
    \frac{\intd \mP_\mA}{\intd t} \, =\, \gamma \int \intd^dx \; C_\mA(\vec x,\vec x,t)\, , \;\; \mbox{with} \;\;
    C_\mA(\vec x, \vec y , t) \, =\,  \frac{\delta }{\delta \pi_{\vec x}} \bigg[ \pi_{\vec y} \mP_\mA + T  \frac{\delta }{\delta \pi_{\vec y}} \mP_\mA \bigg]  \, ,
\end{equation}
and separately also vanishes in equilibrium, simply because
$\displaystyle
T  \frac{\delta \mP_\mA }{\delta \pi_{\vec x}} = - \pi_{\vec x} \, \mP_\mA 
$.

\medskip

By the same line of arguments, the equilibrium distribution for our diffusive Model B dynamics is in turn given by
\begin{equation}
    \mP_\mB\left[\phi,\pi \right] = Z^{-1} \exp\bigg\{ -\beta \mathfrak H[\phi]  + \beta \int \intd^d x \,  \frac{1}{2\mu} \, \pi(x) \nabla^{-2} \pi(x)  \bigg\} \, . \label{eq:stationary-solution-B} 
\end{equation}
This is the stationary solution to 
\begin{equation}
    \frac{\intd \mP_\mB}{\intd t} \, =\,  
    \frac{\partial \mP_\mB}{\partial t } + \int \intd^d x\, \bigg(\Big(\mu\nabla^2 \frac{\delta \mathfrak H[\phi]}{\delta \phi_{\vec x}}  \Big)\frac{\delta }{\delta \pi_{\vec x}} + \pi_{\vec x}  \frac{\delta }{\delta \phi_{\vec x}}  \bigg) \, \mP_\mB[\phi,\pi] \, = \, 0\,,
\end{equation}
and it also nullifies the collision integral, where the kernel now gets modified according to 
\begin{equation}
    C_\mB(\vec x, \vec y , t) \, =\,  \frac{\delta }{\delta \pi_{\vec x}} \bigg[ \pi_{\vec y} \mP_\mB - T\mu \nabla^2  \frac{\delta }{\delta \pi_{\vec y}} \mP_\mB \bigg]  \, .
\end{equation}
And finally, for completeness, away from equilibrium the full Fokker-Planck equation for our Model B process reads as follows:
\begin{equation}
    \frac{\partial \mP_\mB}{\partial t } \, = \, - \int \intd^d x\, \bigg[ \bigg(\Big(\mu\nabla^2 \frac{\delta \mathfrak H[\phi]}{\delta \phi_{\vec x}}  \Big)\frac{\delta }{\delta \pi_{\vec x}} + \pi_{\vec x}  \frac{\delta }{\delta \phi_{\vec x}}  \bigg) \, \mP_\mB[\phi,\pi] - \gamma \, C_\mB(\vec x, \vec x , t) \bigg]\, .
\end{equation}

Note that the Model A version of the equilibrium distribution in \eqref{eq:stationary-solution-A} is given by the usual Hamiltonian $H_\mA[\phi,\pi] $ of the corresponding scalar field theory with a single real field variable $\phi(x)$ and its conjugate momentum field $\pi(x) \equiv \dot \phi(x)$.
It is therefore tempting to also identify the equilibrium distribution in \cref{eq:stationary-solution-B} with the Boltzmann distribution  $\mP_\mB \,\propto \, \exp{(-\beta H_\mB)}$ of an effective total energy $H_\mB$, i.e.,
\begin{equation}
    H_\mB  = \int \intd^d x \, \bigg\{ - \frac{1}{2\mu }\,  \pi(x) \nabla^{-2} \pi(x) + \frac{1}{2} \phi(x) \left( -\nabla^2 + m^2 \right) \phi(x) + \frac{\lambda}{4!} \phi^4(x) -J \phi(x) \bigg\} \; .
    \label{eq:energy_B}
\end{equation}
However, the kinetic momentum field $\pi(x) =\dot\phi(x) $ is then no longer equal to the canonically conjugate momentum variable of the field $\phi(x)$.
Introducing a canonical momentum field $K(x)$ as the solution to $\pi(x) = -\mu \nabla^{2} K(x) $ for the scalar field conjugate to $\phi(x)$, on the other hand, it is straightforward to show that the Hamiltonian
\begin{equation}
    H_\mB \left[ \phi, K \right]  = -\int \intd^dx\, \frac{\mu}{2} \, K(x) \nabla^2 K(x)  + \mathfrak H[\phi] = \int \intd^dx \; \frac{\mu}{2} \,  \big(\nabla K(x)\big)^2 + \mathfrak H[\phi] 
    \label{eq:K_energy_B}
\end{equation}
generates the conservative part of the equation of motion \eqref{eq:ModelB_eom}  with $\dot \phi(\vx , t) = -\mu \nabla^2 K(\vx , t)$.

An intuitive interpretation of $K(\vx,t)$ is obtained from recalling that the diffusive dynamics of Model~B results from the conservation of the total magnetization $Q$, i.e.~the order parameter field obeys a continuity equation
\begin{equation}
    \dot \phi(\vx,t) + \nabla \cdot \vec J(\vx,t) = 0 ,
    \label{eq:continuity_B}
\end{equation}
where the magnetization current $\vec J(\vx,t) = \mu\nabla K(\vx,t)$, in the conservative case, is proportional to 
the gradient of the conjugate momentum field $K(x)$, related by the mobility coefficient.

The coupling of the magnetization current to the heat bath must then be consistent with \cref{eq:ModelB_eom}, such that we have for its evolution
\begin{align}
    \dot{\vec J}(\vx,t) &= \mu \nabla \dot K(\vx,t) - \gamma \vec J(\vx,t) - \sqrt{2\gamma\mu T} \, \vec\zeta(\vx,t) \label{eq:current_eom}\\
    &= -\mu \nabla \frac{\delta H_\mB}{\delta \phi(\vec x,t)} - \gamma \vec J(\vx,t) - \sqrt{2\gamma\mu T} \, \vec\zeta(\vx,t), \nonumber
\end{align}
with a $d$-component vectorial noise $\vec\zeta(\vx,t)$, related to the noise in  \cref{eq:ModelB_noise} by  $\sqrt{\mu} \, \nabla\!\cdot\!\vec\zeta = \eta$, and hence with zero mean and covariance here, i.e.~$\Braket{\zeta_i(\vx',t') \zeta_j(\vx,t)} = \delta_{ij}\delta(\vx'-\vx)\delta(t'-t)$.

\noindent In particular, this confirms that $\displaystyle  \dot K_{\vec x}  = - \frac{\delta}{\delta \phi_{\vec x}} H_\mB [ \phi, K ]$.

\subsection{Causal diffusion}
\label{sec:causal}

Without derivative terms, a standard Landau-Ginzburg Hamiltonian would yield a hyperbolic and causal field equation for the conservative forces  in (\ref{eq:ModelB_eom}) because the order of time derivatives then matches that of the spatial ones. This is different when we use the LGW Hamiltonian of \cref{eq:z2_continuum} which already includes the second-order derivative term which is of order $\vec p^2$ in momentum space, so that the right hand side of our diffusive field equation in (\ref{eq:ModelB_eom}) is of order $\vec p^4$ while the left hand side is only of order $\omega^2$. This mismatch of orders in time versus spatial derivatives is a source of acausal ultraviolet modes and limits the range of applicability of the model to an effective-low energy theory.  While this is sufficient for our main focus on the critical infrared dynamics in this paper, a causal ultraviolet extension is possible with minor modifications.
In order to avoid acausal diffusion in the relativistic limit we have to replace our Hamiltonian equations of motion for the reversible forces from $H_\mB $ in (\ref{eq:K_energy_B}) by relaxation-type equations. Introducing  a small relaxation time $\tau_r $ which will not affect the critical dynamics studied in this paper as long as $\tau_r \ll \tau_R\equiv 1/\gamma$, we then write  
\begin{align}
     \tau_r\, \ddot\phi(\vx,t) &+ \dot \phi(\vx,t) - \frac{\delta H_\mB}{\delta K_{\vec x}}
          = 0 \, , \label{eq:relax_phi}\\
       \tau_r\, \ddot K(\vx,t) &+ \dot K(\vx,t) + \frac{\delta H_\mB}{\delta \phi_{\vec x}} = 0\, .
    \label{eq:relax_K}
\end{align}
As a result, however, already the conservative equations of motion, for $\gamma = 0$ here, can then no-longer be described by Hamilton dynamics as we did for the purely diffusive dynamics in the previous subsection, and there is no strict energy conservation anymore because  $\frac{\intd }{\intd t} H_\mB = \mathcal O(\tau_r) $.
The first relaxation equation for $\phi$ then reads
\begin{equation}
    \tau_r\, \ddot\phi(\vx,t) + \dot \phi(\vx,t) + \mu \nabla^2  K(\vx,t) = 0 \, , \label{eq:relax_phi_old}
\end{equation}
which  for $K \sim \phi$ would represent a hyperbolic heat equation, and replaces $\dot\phi(\vx,t) = - \mu \nabla^2 K(\vx,t)$ above. With the continuity equation in~(\ref{eq:continuity_B}) this becomes 
\begin{equation}
 \dot{\vec J}(\vx,t) \, = -\frac{1}{\tau_r} \big( \vec J(\vx,t)  -  \mu\nabla K(\vx,t)\big)\, ,
  \label{eq:relax}
\end{equation}
thus also representing a relaxation equation for the magnetization current to replace the acausal constitutive relation  $\vec J(\vx,t) = \mu\nabla K(\vx,t)$ used implicitly above.  With damping and noise, Eqs.~(\ref{eq:relax_phi}) and (\ref{eq:relax_K}) together then lead to 
\begin{equation}
  \big(\tau_r \partial_t + 1  \big)^2 \, \partial_t^2 \phi(\vec x,t)    = \mu \nabla^2 \frac{\delta \mathfrak H[\phi]}{\delta \phi_{\vec x}} - \gamma \partial_t \phi(\vec x,t) + \sqrt{2 \gamma T} \, \eta(\vec x, t)\, ,
  \label{eq:ModelB_eom_causal}
\end{equation}
as the regularized causal version of Eq.~(\ref{eq:ModelB_eom}).
Using the continuity equation again, the regularized replacement for the evolution of the magnetization current in Eq.~(\ref{eq:current_eom}) simply becomes
\begin{align}
   \big(\tau_r \partial_t + 1  \big)^2  \,
      \partial_t \vec J(\vx,t)   &= -\mu \nabla \frac{\delta \mathfrak H[\phi]}{\delta \phi_{\vec x}} - \gamma \vec J(\vx,t) - \sqrt{2\gamma\mu T} \, \vec\zeta(\vx,t)\, , \label{eq:current_eom_causal}
\end{align}
which is also consistent with Eq.~(\ref{eq:relax}) for $\gamma=0$.

With the main focus on critical dynamics, from the critical low-frequency excitations, we will neglect the problem with possible acausal diffusion of high-frequency ultraviolet excitations and therefore consider the microscopic theory as an effective low-energy theory for the critical dynamics. 
We can then safely set $\tau_r \to 0$ in the following, so that the causal versions of all the equations for relativistic diffusion in this subsection reduce to those from the previous subsections again, where we have reformulated the diffusion process in terms of Hamiltonian dynamics with the (approximately) conserved energy given by the effective Hamiltonian $H_\mB $ in \cref{eq:K_energy_B}.

\subsection{Covariant formulation}

Both dynamical models can be written in a Lorentz-covariant manner.
The model without conserved order parameter is described by the usual Lagrangian density of a self-interacting relativistic scalar field, 
\begin{align}
    \Lag_\mA = \dot \phi \, \frac{\delta H_\mA}{\delta \pi} - \frac{1}{2} \pi^2 - \frac{1}{2}(\nabla \phi)^2  - V(\phi) 
    &= \frac{1}{2} (\partial_{\mu}\phi) \partial^{\mu}\phi - V(\phi)\, , \;\; \mbox{with} \;\; 
    V(\phi) = \frac{m^2}{2}\phi^2 + \frac{\lambda}{4!}\phi^{4} \, ,\label{eq:cov_Lag_c}
\end{align}
and metric with signature $(+,-,-,-)$.
The Euler-Lagrange equation yields the equation of motion 
$ \partial_{\mu}\partial^{\mu}\phi +V'(\phi)=0 $ for the non-dissipative system for $\gamma = 0$ with Model C dynamics.
Adding the coupling to the heat bath, we have to specify its local rest frame.
Denoting the four-velocity of the bath by $u^{\mu}$, with $u_{\mu}u^{\mu}=1$, we can then write,
\begin{align}
    0\,  =\,  \partial_{\mu}\partial^{\mu}\phi +    V'(\phi) + \gamma\, u_{\mu}\partial^{\mu}\phi - \sqrt{2\gamma T}\, \eta \label{eq:cov_eom_a}
\end{align}
as the covariant version of the equation of motion for our realization of Model A dynamics, where \cref{eq:ModelA_eom} is recovered with $u^{\mu} = (1,0,0,0)$ in the rest frame of the heat bath.

In order to translate the Hamiltonian \eqref{eq:K_energy_B} and the equation of motion \cref{eq:ModelB_eom} with Model B dynamics to covariant form, we introduce some notation from relativistic hydrodynamics:
Along with the local rest-frame velocity $u_{\mu}$, we denote the corresponding timelike derivative in the local rest frame by $D_\tau \equiv u_{\mu}\partial^{\mu}$.
For the spacelike gradient one first introduces the 4-dimensionally transverse projector, $\Delta^{\mu\nu} \equiv g^{\mu\nu}-u^{\mu}u^{\nu}$ and with this, $\nabla^{\mu} \equiv \Delta^{\mu\nu}\partial_{\nu}$, so that $\partial^\mu = u^\mu D_\tau + \nabla^\mu $.
The corresponding spatial Laplacian is analogously written as  $ \Delta = -  \nabla_\mu \Delta^{\mu\nu} \nabla_\nu = - \nabla_\mu \nabla^\mu $.
Moreover, we introduce the spacelike 4-vector
\begin{equation}
    \nu^\mu \equiv \Delta^{\mu\nu} J_\nu = -\mu \nabla^\mu K \, , \;\; \mbox{such that} \;\;
    u_\mu \nu^\mu   = 0 \,. \label{eq:spacelike_J}
\end{equation}
This relation again holds as it stands first without dissipation.
Including the coupling to the heat bath, we can then write our equation of motion for Model B, analogous to 
\cref{eq:current_eom}, in the following form,
\begin{equation}
    D_\tau \nu^\mu \, =\, - \mu D_\tau (\nabla^\mu K ) -\gamma \nu^\mu  - \sqrt{2\gamma\mu T}\, \zeta^\mu_\perp \, , \label{eq:covcurrent_eom}
\end{equation}
where the spacelike noise vectors, with $u_\mu \zeta^\mu_\perp = 0 $, now obey
\begin{equation}
    \Braket{\zeta^\mu_\perp (x) \zeta^\nu_\perp(x')} = \Delta^{\mu\nu} \delta(x-x')\, ,     
\end{equation}
with $d+1$ dimensional $\delta$-function.
They are related to the scalar noise $ \eta $ by $ \sqrt{\mu} \, \nabla_\mu \zeta^\mu_\perp = \eta $, whose variance is now given by the covariant form of the spatial Laplacian $\Delta =  -  \nabla_\mu \nabla^\mu $, 
\begin{equation}
    \Braket{\eta (x) \eta(x')} = - \mu \Delta  \, \delta (x-x') \, .
\end{equation}
In the spacelike projection of \cref{eq:covcurrent_eom} we can now use $\Delta^{\mu\nu} D_\tau \nabla_\nu = \nabla^\mu D_\tau $ to commute timelike and spacelike derivatives of the momentum field $K$ on the right hand side.
For the timelike derivative of $K$ we furthermore use
\begin{equation}
    D_\tau K = \Delta \phi - V'(\phi) 
        \, ,
\end{equation}
where we now have  $\phi \equiv u_\mu J^\mu$.
 \cref{eq:covcurrent_eom} thus now becomes
\begin{equation}
    \Delta^{\mu\nu}    D_\tau \nu_\nu \, =\, -\gamma \Big( \nu^\mu - \frac{\mu}{\gamma} \, \nabla^\mu \big( V'(\phi)  - \Delta \phi \big) \Big)   - \sqrt{2\gamma\mu T}\, \zeta_\perp^\mu \,.        \label{eq:covcurrent_eom_v2}
\end{equation}
In this hydrodynamic form, the conserved current  $J^\mu $ in the continuity equation  \eqref{eq:continuity_B},  $\partial_\mu J^\mu = 0 $, is thus decomposed as $J^\mu = \phi u^\mu + \nu^\mu $, and \cref{eq:covcurrent_eom_v2} assumes the role of an Israel-Stewart type relaxation equation \cite{israel_transient_1979,israel_thermodynamics_1981} with relaxation time $1/\gamma$ and vector force $I^\mu = \nabla^\mu\big(V'(\phi) - \Delta\phi \big)$.
In the non-interacting scalar field theory, for example, the corresponding diffusion rate is thus given by $ \Ddiff(\vec k) = (\mu/\gamma) \,(m^2+ \vec k^2)  =  (\mu/\gamma)\,\chi^{-1}(\vec k) $, i.e.~inversely proportional to the respective static susceptibility $\chi(\vec k)$ as expected.

The analogous procedure as used for our Model A or Model C (without dissipation for $\gamma = 0$) Lagrangian $\Lag_\mA $ in \cref{eq:cov_Lag_c} above, now first leads to a Lagrangian for the non-dissipative ($\gamma=0$) part of our theory with conserved order parameter which is of the form,
\begin{align}
    \Lag_\mB \, &= \, - \frac{1}{2\mu }\, \dot\phi \nabla^{-2}\dot \phi    - \frac{1}{2}(\nabla \phi)^2  - V(\phi) \nonumber \\
    &= \frac{\mu}{2}\, K \nabla^2 K  + K \dot \phi  - \frac{1}{2}(\nabla \phi)^2  - V(\phi) \, , \label{eq:L_B}
\end{align}
where we have reintroduced $K$ as a Gaussian auxiliary field whose equation of motion implements the constraint 
\begin{equation}
    \frac{\partial \Lag_\mB}{\partial K} \, =\,     \dot\phi + \mu\nabla^2 K = 0\,. \label{eq:phi_acausal}
\end{equation}
Together with the equation of motion for the $\phi$ field,
\begin{equation}
    \frac{\partial \Lag_\mB}{\partial\phi} \, =\,     \nabla^2 \phi - V'(\phi)  \, = \, \dot K \, , \label{eq:K_acausal}
\end{equation}
we thus recover the non-dissipative part of \cref{eq:ModelB_eom},
\begin{equation}
    \ddot\phi\,=\, -\mu \partial_t \nabla^2 K \,=\, -\mu \nabla^2 \big( \nabla^2 \phi - V'(\phi)\big) \, . 
    \label{eq:conseom_causal}
\end{equation}
To avoid acausal diffusion at high frequencies, as discussed in Subsection~\ref{sec:causal}, we need to replace 
$\dot\phi \to \dot \phi + \tau_r \, \ddot\phi $ and $\dot K \to \dot K + \tau_r\, \ddot K $ in the equations of motion (\ref{eq:phi_acausal}) and (\ref{eq:K_acausal}) which then agree with
(\ref{eq:relax_phi}) and (\ref{eq:relax_K}), and lead to the causal equation of motion (\ref{eq:ModelB_eom_causal}) instead of its low-frequency approximation in Eq.~(\ref{eq:ModelB_eom}) for the order parameter in Model B. 

Moreover, note that we had to commute the time derivative with the spatial Laplacian in  (\ref{eq:conseom_causal}), in order to get from (\ref{eq:phi_acausal}) and (\ref{eq:K_acausal}) to (\ref{eq:ModelB_eom}). 
This becomes a bit subtle as well in the covariant formulation when the local rest-frame  velocity is spacetime dependent, for the same reason that we needed the transverse projection in the Israel-Stewart equation \eqref{eq:covcurrent_eom_v2} in order to be able to commute the timelike and spacelike derivatives on $K$.
In the covariant version of the equation of motion for the scalar field $\phi $, the necessary commutator is readily worked out to be
\begin{equation}
    D_\tau \Delta \, = \, \Delta D_\tau +  (\partial_\mu u^\mu) a^\nu \nabla_\nu  \, ,
\end{equation}
where the spacelike vector $a^\mu = D_\tau u^\mu $ describes the acceleration of the local fluid element, and $\partial_\mu u^\mu$ its expansion.
Hence, the spatial Laplacian $\Delta = - \nabla_\mu \nabla^\mu $ commutes with the timelike derivative $D_\tau = u^\mu \partial_\mu $ for incompressible fluids with $\partial_\mu u^\mu =0 $, and we observe that the non-dissipative part of \cref{eq:ModelB_eom} describes the diffusive dynamics of an incompressible fluid.

For causal diffusion we again use the decomposition of the four-vector  $J^\mu = \phi u^\mu +\nu^\mu $, which is inverted by $\phi = u_\mu J^\mu $ and $\nu^\mu = \Delta^{\mu\nu} J_\nu$, but now together with the covariant version of the relaxation-type equation, \cref{eq:relax} of Subsection~\ref{sec:causal},
\begin{equation}
   \Delta^{\mu\nu}  D_\tau \nu_\nu \, = -\frac{1}{\tau_r} \big( \nu^\mu + \mu \nabla^\mu K \big) \, ,
\end{equation}
which by itself resembles Israel-Stewart hydrodynamcis with \cref{eq:spacelike_J} as the corresponding Navier-Stokes limit, see App.~\ref{sec:israel-stewart}. 
One then readily verifies that the current conservation law in covariant form reads,
\begin{align}
    \partial_\mu J^\mu \, 
    &=\, D_\tau\phi + (\partial_\mu u^\mu) \phi+ \partial_\mu  \nu^\mu  \nonumber\\
    &=\,\tau_r\,  D_\tau^2 \phi +  D_\tau \phi -  \mu \, \partial_\mu \nabla^\mu K\,  =\, 0 \, ,  \label{eq:cov_currcons}
\end{align}
where we have again assumed incompressibility ($\partial_\mu u^\mu = 0$), and vanishing acceleration ($a^\mu = D_\tau u^\mu = 0$) in the second line. The covariant version of the Lagrangian (\ref{eq:L_B}) for this conservative Model BC dynamics,
\begin{equation}
    \Lag_\mB \, 
    = \frac{\mu}{2}\, (\nabla_\mu K) \nabla^\mu K + K D_\tau\phi 
        + \frac{1}{2}\, (\nabla_{\mu}\phi) \nabla^{\mu}\phi - V(\phi) \, , \label{eq:constraint_eom_B}
\end{equation}
can only generate the reversible part of the equations of motion, valid for $\tau_r \to 0$. Causality then requires replacing $D_\tau \to \tau_r D_\tau^2 + D_\tau$ analogous to the procedure explained in Subsection~\ref{sec:causal} above, which yields 
  \begin{align}
     \tau_r\, D_\tau^2 K + D_\tau  K  +\partial_\mu\nabla^\mu \phi  + V'(\phi)  &= \, 0 \, , \label{eq:cov_ModelD_eom_causal} \\
   \tau_r\, D_\tau^2 \phi + D_\tau\phi - \mu\, \partial_\mu\nabla^\mu  K &= \, 0 \, , \label{eq:cov_ModelD_aux_causal}
\end{align}
as a coupled set of hyperbolic heat equations for $\phi$ and $K$ with the conservative force $-V'(\phi)$ acting as a source.
Finally note that, in general, $\partial_\mu\nabla^\mu \not= \nabla^\mu \partial_\mu$ and neither of the two is equal to the (negative) spatial Laplacian $\nabla_\mu\nabla^\mu = -\Delta $. Only for an incompressible fluid without acceleration they are all the same and these distinctions are luckily unnecessary.\footnote{With acceleration  $a^\mu = D_\tau u^\mu $ and expansion $\partial_\mu u^\mu $ one has
$    \partial_\mu\nabla^\mu \, = \, \nabla_\mu \nabla^\mu - a^\mu\partial_\mu $ and 
  $ \nabla^\mu \partial_\mu \, = \, \nabla_\mu \nabla^\mu + (\partial_\mu u^\mu) D_\tau $.} 
From these equations of motion, adding damping and noise again, one can furthermore  derive the covariant version of \cref{eq:current_eom_causal},
\begin{equation}
 \big( \tau_r \Delta^{\mu\rho }    D_\tau + g^{\mu\rho } \big) \big( \tau_r \Delta_{\rho\sigma }    D_\tau + g_{\rho \sigma} \big) \Delta^{\sigma\nu}    
 D_\tau \nu_\nu \, =\, -\gamma \Big( \nu^\mu - \frac{\mu}{\gamma} \, \nabla^\mu \big( V'(\phi)  - \Delta \phi \big) \Big)   - \sqrt{2\gamma\mu T}\, \zeta_\perp^\mu \,.        \label{eq:covcurrent_eom_v2_causal}
\end{equation}
which yields the analogous causal extension of our Israel-Stewart type equation in (\ref{eq:covcurrent_eom_v2}).

\subsection{Lattice Regularization}\label{sec:lattice}

We employ a lattice regularization to supply a UV cutoff to possible spatial variations of the order parameter field $\phi(\vec x, t)$.
The LGW Hamiltonian then becomes a sum over the field $\phi(\vec x, t) \to \phi_x(t)$ at discrete lattice sites
\begin{equation}
    \mathfrak H[\phi_x] = \sum\limits_x a^d \left\{ - \frac{1}{2 a^2} \sum\limits_{y\sim x} \phi_x \phi_{y} + \left( \frac{m^2}{2} + \frac{d}{a^2} \right)\phi_x^2 + \frac{\lambda}{4!}\phi_x^4 + J\phi_x \right\},
    \label{eq:lattice_LGW}
\end{equation}
where the sum $\sum_{y\sim x}$ runs over all nearest neighbour sites $y$ attached to the site $x$, and $\sum_x a^d$ denotes the sum over the spatial volume with lattice spacing $a$.
For the lattice Laplacian we use $\nabla^2 \phi_x \equiv \nabla_b \cdot \nabla_f \phi_x = a^{-2}\left(\sum_{y\sim x}\phi_{y} - 2d \phi_x\right)$ which corresponds to applying one forward and one backward derivative defined as $\partial^i_f \phi_x \equiv (\phi_{x+\hat i} - \phi_x)/a \equiv \partial^i_b \phi_{x+\hat i}$, where $\hat i$ is a lattice-unit vector in the direction $i$. We use periodic boundary conditions so that the rules of partial integration with $\partial_b^i = - (\partial_f^i)^\dagger $ on the spatial lattice apply.
To simplify the notation, we from here on let the lattice spacing $a=1$ be unity.
If not stated otherwise, the model parameters in these lattice units are set to $m^2 = -1$, $\lambda=1$, as well as $J=0$.

For Model C, where the order parameter is not conserved, the discretization of the Hamiltonian is straightforward, and we obtain for the lattice Hamiltonian and the equations of motion
\begin{align}
    H_\mA &= \sum\limits_x \frac{\pi_x^2}{2} - \frac{1}{2} \sum\limits_{y\sim x} \phi_x \phi_{y} + \left( \frac{m^2}{2} + d \right)\phi_x^2 + \frac{\lambda}{4!}\phi_x^4 + J\phi_x, \label{eq:lattice_hamiltonian_A}\\
    \dot \phi_x &= \frac{\partial H_\mA}{\partial \pi_x} = \pi_x,\\
    \dot \pi_x &= -\frac{\partial H_\mA}{\partial \phi_x} -\gamma \pi_x + \sqrt{2\gamma T}\eta_x \label{eq:pi_eom_A_lattice}\\
    -\frac{\partial H_\mA}{\partial \phi_x} &= -\sum_{x\sim y} \left( \phi_y - \phi_x \right) + \left( m^2 + \frac{\lambda}{6}\phi_x^2 \right)\phi_x + J, \label{eq:partialH_A}
\end{align}
where $\eta_x$ is a zero-mean Gaussian white noise at every lattice site with $\Braket{\eta_{x'}(t') \eta_x(t)} = \delta_{x'x} \delta\left( t'-t \right)$.

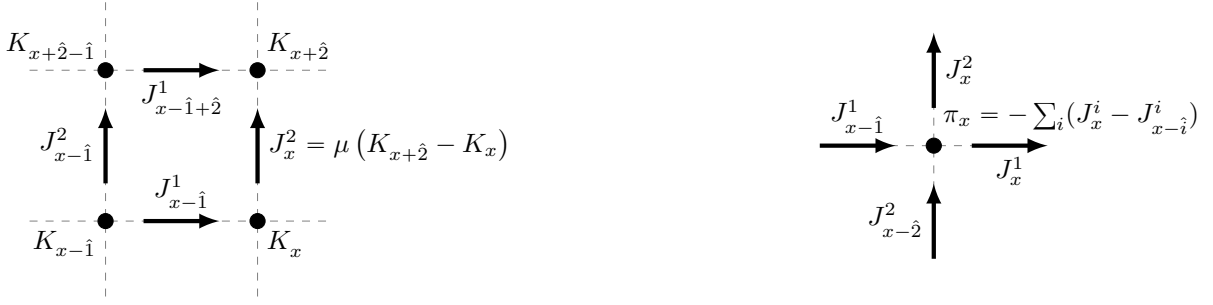
\begin{figure}
    \begin{minipage}{.67\textwidth}
        \begin{tikzpicture}
            \def\Xmin{-1}
            \def\Xmax{0}
            \def\Ymin{0}
            \def\Ymax{1}
            \coordinate (Origin)   at (0,0);

            \draw[style=help lines,dashed] (2*\Xmin - 1, 2*\Ymin-1) grid[step=2cm] (2*\Xmax + 1, 2*\Ymax+1);
            \foreach \x in {\Xmin,...,\Xmax}{
                \foreach \y in {\Ymin,...,\Ymax}{
                    \node[draw,circle,inner sep=2pt,fill] at (2*\x,2*\y) {};
                }
            }

            \node at (-2,0) [below left] {$K_{x - \hat 1}$};
            \node at (-2,2) [above left] {$K_{x + \hat 2 - \hat 1}$};
            \node at (0,2) [above right] {$K_{x + \hat 2}$};
            \node at (0,0) [below right] {$K_{x \vphantom{\hat k}}$};

            \draw [ultra thick,-latex] (-2,.5) -- (-2,1.5) node [left,midway] {$J^2_{x - \hat 1}$};
            \draw [ultra thick,-latex] (-1.5,0) -- (-.5,0) node [midway,above] {$J^1_{x - \hat 1}$};
            \draw [ultra thick,-latex] (-1.5,2) -- (-.5,2) node [midway,below] {$J^1_{x - \hat 1 + \hat 2}$};
            \draw [ultra thick,-latex] (0,.5) -- (0,1.5) node [right,midway] {$J^2_x = \mu \left( K_{x+\hat 2} - K_x \right)$};
        \end{tikzpicture}
    \end{minipage}
    \begin{minipage}{.33\textwidth}
        \begin{tikzpicture}
            \def\Xmin{0}
            \def\Xmax{0}
            \def\Ymin{0}
            \def\Ymax{0}
            \coordinate (Origin)   at (0,0);

            \draw[style=help lines,dashed] (2*\Xmin - 1, 2*\Ymin-1) grid[step=2cm] (2*\Xmax + 1, 2*\Ymax+1);
            \foreach \x in {\Xmin,...,\Xmax}{
                \foreach \y in {\Ymin,...,\Ymax}{
                    \node[draw,circle,inner sep=2pt,fill] at (2*\x,2*\y) {};
                }
            }

            \node at (0,0) [above right] {$\pi_x = -\sum_i (J^i_x - J^i_{x - \hat i})$};

            \draw [ultra thick,-latex] (0,-1.5) -- (0,-0.5) node [left,midway] {$J^2_{x - \hat 2}$};
            \draw [ultra thick,-latex] (-1.5,0) -- (-.5,0) node [midway,above] {$J^1_{x - \hat 1}$};
            \draw [ultra thick,-latex] (.5,0) -- (1.5,0) node [midway,below] {$J^1_x$};
            \draw [ultra thick,-latex] (0,.5) -- (0,1.5) node [right,midway] {$J^2_x$};

        \end{tikzpicture}
    \end{minipage}
    \caption{Lattice fields in the model with diffusive dynamics (Model B/D).
        The left panel shows how the components of the order parameter current $\vec J$ are obtained via forward derivative of the conjugate momentum field $K_x$, cf.~\cref{eq:current_lattice}.
        Applying a backward derivative for the lattice divergence, one obtains the time derivative $\dot\phi_x=\pi_x$ of the order parameter field from the continuity equation, \cref{eq:continuity_lattice}.
    }
    \label{fig:lat_d}
\end{figure}

For Models B and BC, we begin the discretization by considering again the pair of conjugate variables $\phi(\vec x,t), K(\vec x,t) \to \phi_x(t), K_x(t)$, which are defined on the sites of the lattice.
For the magnetization current, $\vec J  = \mu \nabla K $ without dissipation, the forward derivative is used as the discrete version of the gradient which  defines the components of the discretized magnetization current $J^i(\vec x, t)\to J^i_x(t)$ on the forward links from site $x$ to $x + \hat i $, i.e. 
\begin{equation}
    \vec J_x = \mu \nabla_f K_x
    \label{eq:current_lattice}
\end{equation}
is given by the \emph{forward} derivative, so that $ \vec J_x $ defines an exact lattice one-form as the discrete exterior  derivative of the zero-form site variable $K$ on the lattice.
For the continuity equation we need the lattice divergence of the discrete current, i.e.~the lattice version of the exterior co-derivative of the lattice one-form given by the link variable $\vec J_x$, to the time derivative $\pi_x$ on the sites. This is achieved by the  \emph{backward} derivative so that the discretized version of the continuity equation \eqref{eq:continuity_B} becomes
\begin{align}
    \pi_x + \nabla_b \cdot \vec J_x &= 0. \label{eq:continuity_lattice}
\end{align}
These basic discrete exterior (co-)derivative operations are illustrated in \cref{fig:lat_d}.
Finally, we then have the equations of motion with heat bath, for the discretized current and time derivative,
\begin{align}
    \dot{\vec J}_x &=  \mu \nabla_f \dot K_x -\gamma \vec J_x + \sqrt{2\mu \gamma T}\vec\zeta_x = -\nabla_f \frac{\partial H_\mB}{\partial \phi_x} -\gamma \vec J_x + \sqrt{2\mu \gamma T}\vec\zeta_x, \label{eq:current_eom_lattice}\\
    \follows \dot \pi_x &= - \nabla_b \dot{\vec J}_x = \mu \nabla_b\nabla_f \frac{\partial H_\mB}{\partial \phi_x} + \gamma \nabla_b \vec J_x - \sqrt{2\mu\gamma T}\nabla_b\vec\zeta_x \nonumber \\
    &= \mu \nabla^2 \frac{\partial H_\mB}{\partial \phi_x} - \gamma \pi_x - \sqrt{2\mu \gamma T}\nabla_b\vec\zeta_x.
    \label{eq:pi_eom_lattice}
\end{align}
In summary, we thus have the lattice Hamiltonian and equations of motion for Model B:
\begin{align}
    H_\mB &= \sum\limits_x \left\{ - \frac{\mu}{2} \sum\limits_{y\sim x} K_x \left(K_{y} - K_x\right) - \frac{1}{2} \sum\limits_{y\sim x} \phi_x \phi_{y} + \left( \frac{m^2}{2} + d \right)\phi_x^2 + \frac{\lambda}{4!}\phi_x^4 + J\phi_x \right\}, \label{eq:lattice_hamiltonian_B}\\
    \dot \phi_x &= \frac{\partial H_\mB}{\partial K_x} =  -\mu \sum_{y\sim x} (K_y - K_x) = - \mu \nabla^2 K_x \equiv \pi_x,\\
    \dot \pi_x &= \mu \sum_{y\sim x} \left( \frac{\partial H_\mB}{\partial \phi_y}  - \frac{\partial H_\mB}{\partial\phi_x}\right) -\gamma \pi_x + \sqrt{2\mu \gamma T}\nabla_b\vec\zeta_x \label{eq:pi_eom_B_lattice},
\end{align}
where $\partial H_\mB/ \partial\phi_x = \partial H_\mA/\partial\phi_x$ is given in \cref{eq:partialH_A}.
For practical reasons, we work with the variables $\phi_x$ and $\pi_x$ in both cases.
We also set the mobility coefficient in our numerical calculations to $\mu = 1$ in lattice units, i.e.~in other units $\mu = a^2 $ according to its canonical dimension $-2$.

Concerning the noise term in \cref{eq:pi_eom_lattice}, we remark that by the affine transformation of a $d$-component vector of Gaussian random numbers $\vec\zeta_x(t) $ with unit covariance,
we generate random variables with a distribution approaching the correct continuum limit of \cref{eq:ModelB_noise},
\begin{align}
    \eta_x(t) &\equiv \nabla_b \cdot \vec\zeta_x(t) \, , \;\; \text{with}\;\;  \Braket{\zeta^i_x(t)\zeta^j_y(t')} = \delta_{ij}\delta_{xy}\delta(t-t'), \\
\Rightarrow \,     \braket{\eta_x(t)\eta_y(t')} &= \Braket{\left(\nabla_b \cdot \vec{\zeta}_x(t)\right) \nabla_b \cdot \vec{\zeta}_y(t') } = (\underbrace{\nabla_b^\dagger}_{=-\nabla_f}  \nabla_b) \delta_{xy}\delta(t-t') = -\nabla^2 \delta_{xy}\delta(t-t').
    \label{eq:diff_noise_lattice}
\end{align}

In order to generate a thermal distribution of the time derivative field $\pi$ matching the stationary solution for Model A/C, the lattice variables are drawn from a Gaussian multivariate distribution with the diagonal covariance matrix $T \delta_{xy}$.
In case of Model B/D, however, the covariance matrix is no longer diagonal, but of the form $-T \mu \nabla^2 \delta_{xy}$
Similar to the generation of the conserving noise, this is realized by taking the backward derivative of a vector noise with the Gaussian distribution,
\begin{align}
        \pi_x(t=0) = \sqrt{T\mu} \, \nabla_b \cdot \vec \zeta_x = \sqrt{T\mu} \sum_i^d \left(\zeta^i_x - \zeta^i_{x-\hat i}\right)\\
    \Rightarrow\; \Braket{\pi_x(t=0) \pi_y(t=0)} = -T\mu \nabla^2 \delta_{xy}. \label{eq:pi_initial}
\end{align}

\subsection{Static critical behavior}

The static critical behavior of the scalar field theory in the $Z_2$ Ising universality class was analyzed within our framework for classical-statistical lattice simulations in detail in \cite{schweitzer_spectral_2020}, where we investigated expectation values, equal-time correlation functions, and  spectral functions of the order parameter for the dynamics without a conserved order parameter, i.e.~for Models A and C.
Since the static critical behavior of the model is the same for all dynamical models considered here, the required static results are identical to those reported in \cite{schweitzer_spectral_2020}.
The ones relevant for our present study are compactly summarized in \cref{tab:statics} for convenience here again.

\begin{table}
    \centering
    \begin{tabular}{l r r}
        & \multicolumn{1}{c}{2D} & \multicolumn{1}{c}{3D} \\
        \hline\hline
        $T_c$ & 4.4629(10) & 9.3707(3) \\
        $f_{\xi}^+$ & 0.918(8) & 0.92(3) \\
        $\nu$ & 1 & 0.629971 \\
        $\eta$ & 0.25 & 0.036298
    \end{tabular}
    \caption{Critical temperatures and scaling amplitudes of the correlation length obtained in \cite{schweitzer_spectral_2020}, alongside relevant critical exponents.
        Non-universal amplitudes are obtained by fits of the correlation length as obtained from plane-correlation functions.
        For the model function, we used an ansatz containing some corrections to scaling as $\xi(\tau) = f_{\xi}^{+} \tau^{-\nu}\left( 1+f^+_1 \tau^{\omega\nu} \right)$ for $\tau > 0$; see \cite{schweitzer_spectral_2020} for details.
        Critical exponents in 2D are known analytically from Onsager's solution of the Ising model \cite{onsager_crystal_1944}.
        High-precision results for the 3D Ising exponents were obtained by the conformal bootstrap approach \cite{kos_precision_2016,komargodski_random-bond_2017}.
    }
    \label{tab:statics}
\end{table}

\section{Spectral functions}\label{sec:specfunc}
We now turn to the investigation of real-time spectral functions of the order parameter field $\phi$.
In the following, we outline our approach to obtain numerical data on the spectral functions.
Using the classical-statistical lattice formulation discussed in \cref{sec:lattice}, the calculation of the spectral function can be performed in a straightforward way, as described in detail in Sec.~3 of \cite{schweitzer_spectral_2020}.
Based on the fluctuation-dissipation relation for a (classical) equilibrium system, the spectral function can be determined from the un-equal time correlation function 
\begin{equation}
    \rho(t-t', \vec x-\vec x', T) = -\frac{1}{T}\Braket{\pi(\vec x, t) \phi(\vec x', t')}_T,  \label{eq:def_sf}
\end{equation}
where $\langle . \rangle_T$ denotes an average over a thermal ensemble.

While \cref{eq:def_sf} enables a direct calculation of the spectral function $\rho(t, \vec x, T)$ in the time domain, it is in most cases more natural to study the spectral function in the Fourier domain, i.e.
\begin{equation}
    \rho(\omega, \vec p, T) = -\iu \int \intd t \intd^d x \, e^{\iu \left( \omega t - \vec{p x} \right)} \rho(t, \vec x, T), \label{eq:def_FT_rho}
\end{equation}
where, following the notation in \cite{schweitzer_spectral_2020}, we introduce an additional factor of $-\iu$ to ensure that the spectral function is real in both the time and frequency domain.
We restrict our analysis to positive frequencies, since any results can be trivially extended to negative frequencies by the symmetry property of the spectral function $\rho(-\omega) = -\rho(\omega)$.
In practice, we evaluate the thermal expectation value by preparing $\sim 30$ independent configurations of the field $\phi_x$ using a Hybrid Monte-Carlo method \cite{duane_hybrid_1987}.
Subsequently, these thermal initial conditions are then evolved using an Euler-Maruyama scheme for times on the order of $10^4a$ to $10^5a$, where $a$ denotes the spatial lattice spacing.
We note that in order to avoid discretization errors accumulating on large time scales, in particular near the critical point, the time step $\Delta t$ in the integrator has to be chosen sufficiently small.
If not stated otherwise, we employ $\Delta t = .00625a$, for which we checked that time-discretization errors are negligible for the results presented in the following.
By recording the time histories of spatial Fourier modes of the order parameter field for each classical trajectory, we then compute the spectral function as the multi-time correlator of the classical fields in \cref{eq:def_sf}.
Statistical errors are estimated by taking the point-wise average over $\rho(t,\vec p)$ resp.~$\rho(\omega, \vec p)$ over different configurations.
Since results for the dynamics of a non-conserved order parameter (Models A/C) have already been reported in \cite{schweitzer_spectral_2020}, we will focus on the dynamics for a conserved order parameter, both in the presence $(\gamma>0)$ and absence $(\gamma=0)$ of the coupling to an external heat bath (Models B/D).

Before we present our numerical results, it is useful to consider the mean-field limit of the spectral function. 
We note that for negative square mass parameter $m^2<0$, one has non-trivial minima of the Hamiltonians where $\phi^2 = {\bar{\phi}}\,^2 = -6m^2/\lambda$.
Expanding the fields around $\phi = \bar \phi$ yields a zero-temperature mean-field squared mass of $\bar m^2 = -2m^2$.

With this mean-field mass we can write down the spectral function, which has a Breit-Wigner shape,
\newcommand{\rhomf}{\rho_{\text{mf}}}
\begin{equation}
    \rhomf(\omega,p) = \frac{\gamma \omega \mu \vec p^2}{(\omega^2-\mu \vec p^2 (\bar{m}^2+\vec p^2))^2+\gamma^2\omega^2},
    \label{eq:rho}
\end{equation}
with dispersion $\omega^2_{p} = \mu \vec p^{2}\left(\bar m^2 + \vec p^2\right)$.
This form of the spectral function is obtained from the imaginary part of the corresponding retarded propagator, which we derive from Israel-Stewart hydrodynamics in Appendix \labelcref{sec:israel-stewart} with the result,
\begin{align}
    G(z, \vec p) = \frac{(1-\iu z \tau_R)\chi(\vec p)}{\Ddiff(\vec p)\, \vec p^2 - \tau_R z^2 - \iu z},
    \label{eq:IS_propagator_0}
\end{align}
where we have introduced the relaxation time $\tau_R\equiv 1/\gamma $, and used the static mean-field susceptibility $\chi(\vec p) = (\bar m^2 + \vec p^2)^{-1}$  as well as the corresponding  momentum-dependent diffusion rate $\Ddiff(\vec p)$,
\begin{align}
          \Ddiff(\vec p) \equiv \frac{\mu}{\gamma}(\bar m^2 + \vec p^2) = \frac{\mu}{\gamma \chi(\vec p)}.
\end{align}
The poles of the propagator are therefore located at
\begin{equation}
    z=\frac{-\iu\gamma}{2} \pm \frac{\iu\gamma}{2} \sqrt{1-4 \frac{\mu}{\gamma^2}(\bar m^2 + \vec p^2){\vec p}^2}\;.
    \label{eq:imagpoles}
\end{equation}
If the spatial momentum is small compared to the Langevin damping, at fixed mobility and mass for 
$\mu \vec p^2 \ll \gamma^2/\bar m^2$ , the pole with smaller imaginary part corresponds to the hydrodynamic Navier-Stokes mode at $z_{\text{hydro}} \approx -\iu \Ddiff(0) \,\vec p^2$, with another short-lived non-hydrodynamic mode at $z_{\text{non-hydro}} \approx - \iu/\tau_R$.
However, for large spatial momentum $\mu \vec p^2 \gg \gamma^2/\bar m^2$, one finds the pair of poles located at 
\begin{align}
    z = \frac{-\iu\gamma}{2} \mp \sqrt{ \mu \vec p^2(\bar m^2 + \vec p^2)}, \label{eq:realpoles}
\end{align}
corresponding to damped propagating waves.
We therefore conclude that at any finite $\gamma > 0$, the infrared limit always contains the classical Navier-Stokes diffusion dynamics.
However, the non-dissipative limit $\gamma \to 0$ fundamentally changes the structure of the infrared  dynamics.
We are then in the limit of infinite relaxation time $\tau_R=1/\gamma\to\infty$, with propagating modes and no diffusion. The analogous short-lived non-hydrodynamic mode 
will be determined by the small relaxation time $\tau_r$ needed for causality in this case, as discussed in Subsection \ref{sec:causal}, while the low-momentum dispersion relation of the remaining modes will always be linear in momentum, with real poles at $z\approx \pm \bar m \sqrt{\mu \vec p^2}$. We conclude that -- in contrast to the usual Model D, which is realized by coupling the diffusive dynamics of Model B to an additional conserved quantity and exhibits the same dynamic critical behavior as Model B~\cite{halperin_renormalization-group_1974,hohenberg_theory_1977,folk_critical_2006} -- it is not the presence of an additional conserved quantity but rather the absence of a diffusive pole in the limit $\gamma \to 0$ that can be expected to change the dynamic critical behavior of the theory. Since we are not aware of an analogous model in the classification scheme of Halperin and Hohenberg, we will refer to this conservative limit of the relativistic Israel-Stuart type diffusion as Model BC.

\subsection{Overview of numerical results}

\begin{figure}
    \renewcommand{\fdir}{figures/sf_overview/}
    \graphicspath{{\fdir}}
    \begin{minipage}[t]{.50\linewidth} \centering{$d=2$} \end{minipage}
    \begin{minipage}[t]{.50\linewidth} \centering{$d=3$} \end{minipage}
    \begin{minipage}[t]{.5\linewidth} \includegraphics{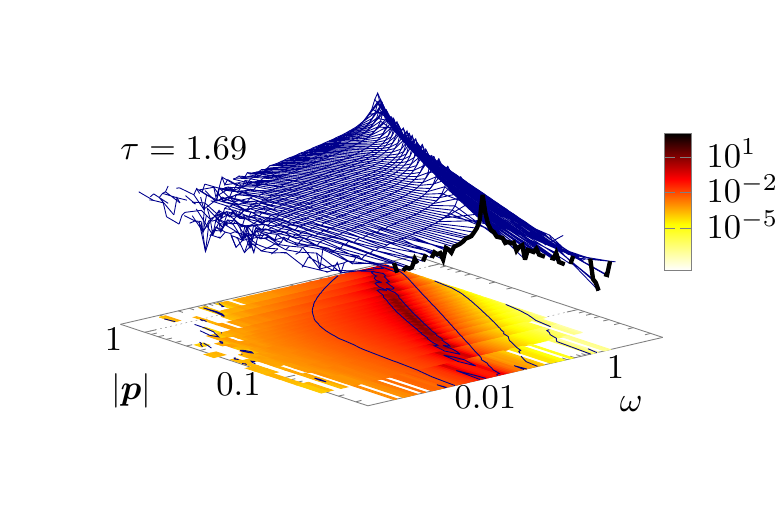} \end{minipage}
    \begin{minipage}[t]{.5\linewidth} \includegraphics{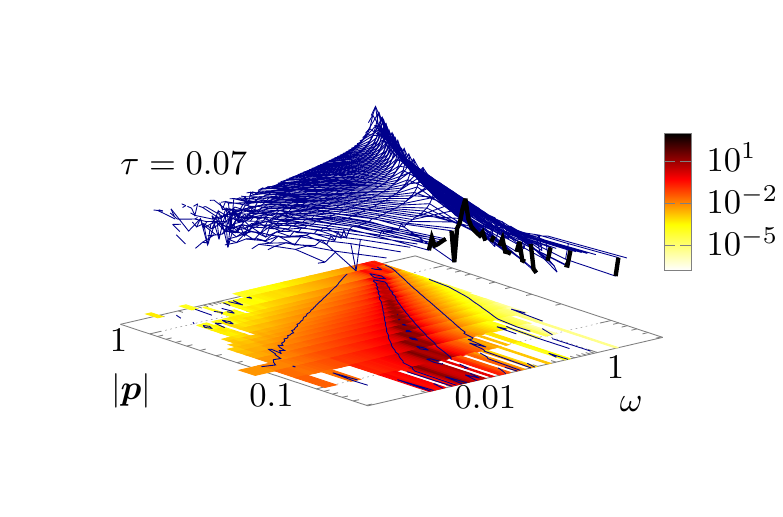} \end{minipage}\vspace{-15mm}
    \begin{minipage}[t]{.5\linewidth} \includegraphics{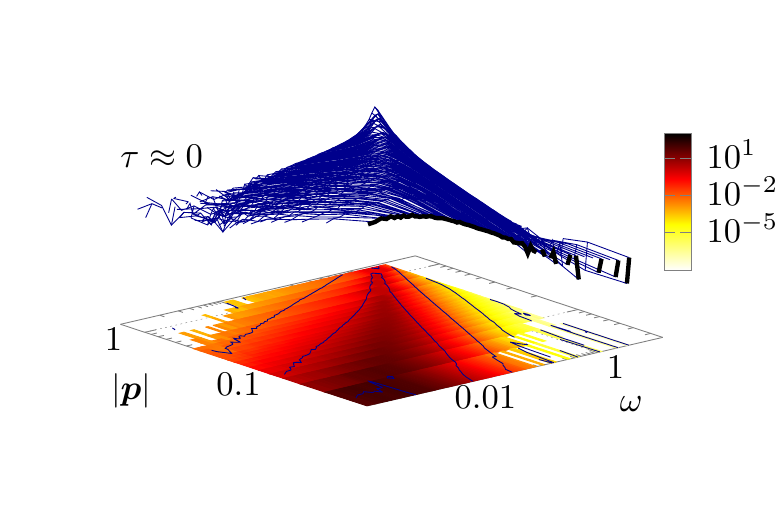} \end{minipage}
    \begin{minipage}[t]{.5\linewidth} \includegraphics{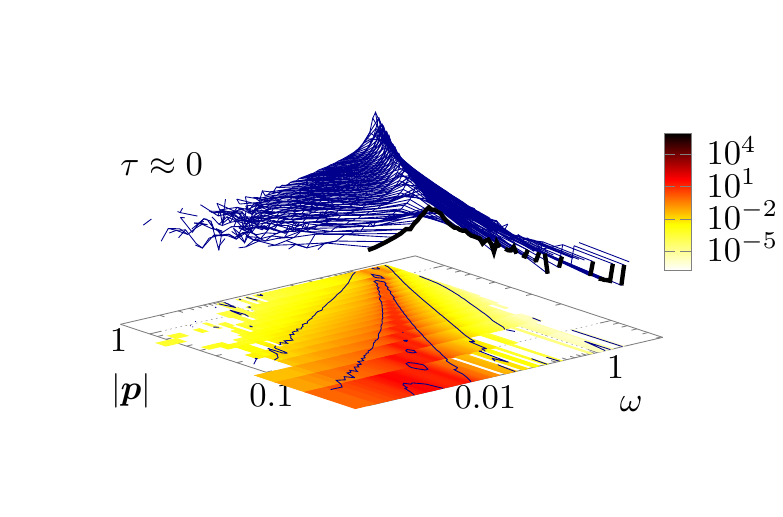} \end{minipage}\vspace{-15mm}
    \begin{minipage}[t]{.5\linewidth} \includegraphics{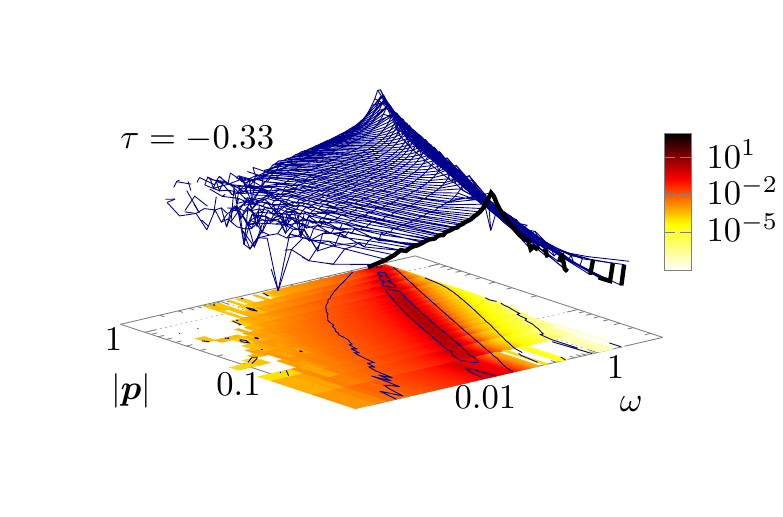} \end{minipage}
    \begin{minipage}[t]{.5\linewidth} \includegraphics{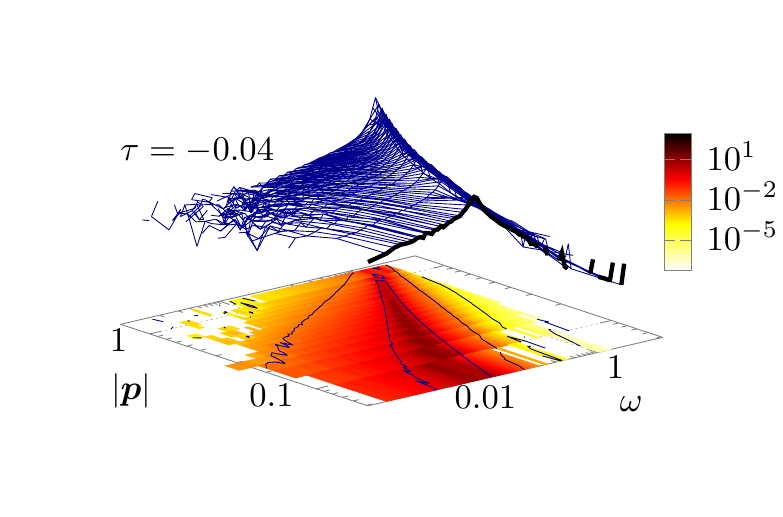} \end{minipage}
    \caption{Overview of spectral functions $\rho(\omega,\vec p)$ for diffusive dynamics without dissipation ($\gamma = 0$) at different points in the phase diagram. 
        Heat maps at the bottom of each panel visualize support and spectral strength in the $(\vec p,\omega)$ plane.
        The axes are scaled logarithmically, and the smallest non-zero momentum modes are highlighted by a black solid line on the front boundary of the surface in the 3D plots.
        Spectral functions away from the critical temperature are dominated by  Breit-Wigner structures with dispersion relation $\omega_p^2 = \mu \vec p^2 (m^2 + \vec p^2)$.
        Away from criticality the spectral functions are dominated by Breit-Wigner structures with dispersion relation $\omega_p^2 = \mu \vec p^2 (m^2 + \vec p^2)$, 
        and the decay widths decrease with a power of $|\vec p|$, leading to narrow peaks at the lower end of the spatial momentum range.
        Close to criticality, the effective thermal mass vanishes the widths appear to become regular in $|\vec p|$.
        The shift of the central frequencies towards the infrared produces broad structures at low spatial momenta, although
        in 3+1D, the peak structure in the low-momentum modes appears to survive at the critical point.
    }
    \label{fig:sf-overview-d}
\end{figure}

\begin{figure}
    \renewcommand{\fdir}{figures/sf_overview/}
    \graphicspath{{\fdir}}
    \begin{minipage}[t]{.50\linewidth} \centering{$d=2$} \end{minipage}
    \begin{minipage}[t]{.50\linewidth} \centering{$d=3$} \end{minipage}
    \begin{minipage}[t]{.5\linewidth} \includegraphics{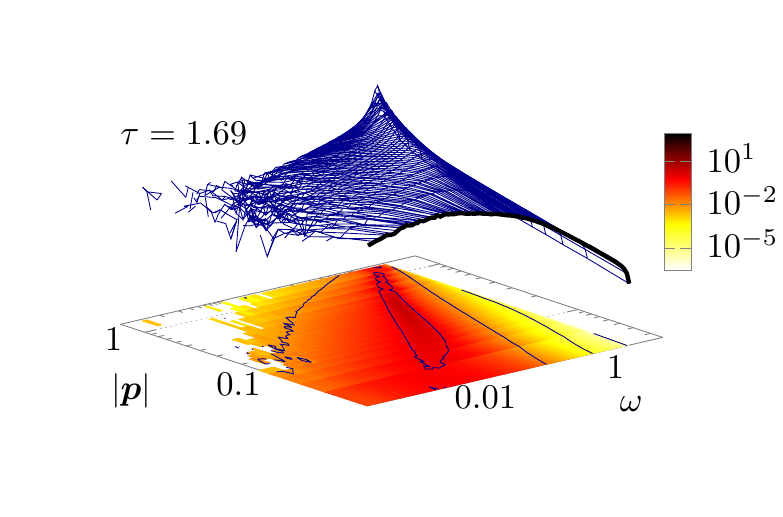} \end{minipage}
    \begin{minipage}[t]{.5\linewidth} \includegraphics{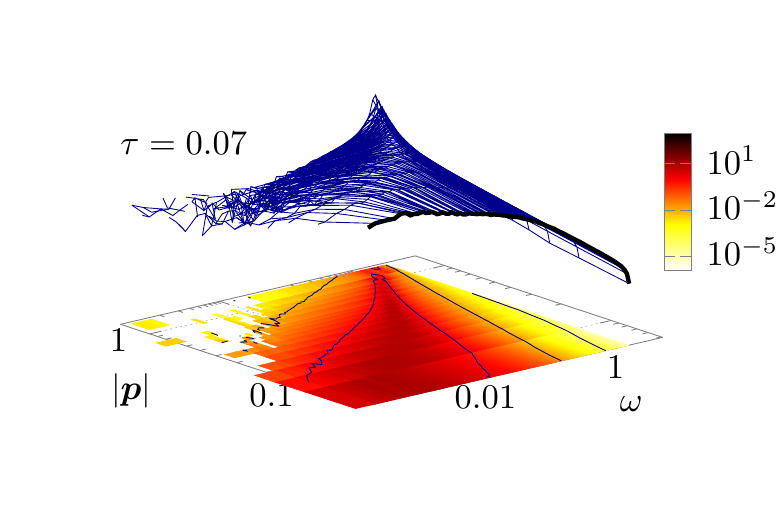} \end{minipage}\vspace{-15mm}
    \begin{minipage}[t]{.5\linewidth} \includegraphics{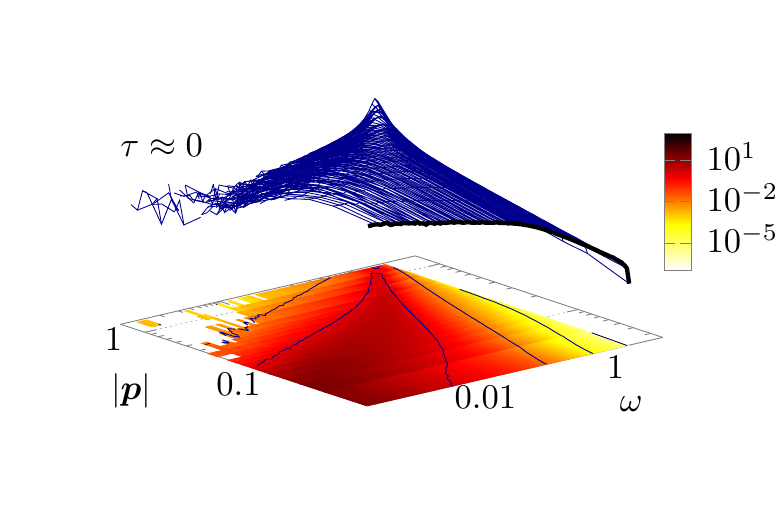} \end{minipage}
    \begin{minipage}[t]{.5\linewidth} \includegraphics{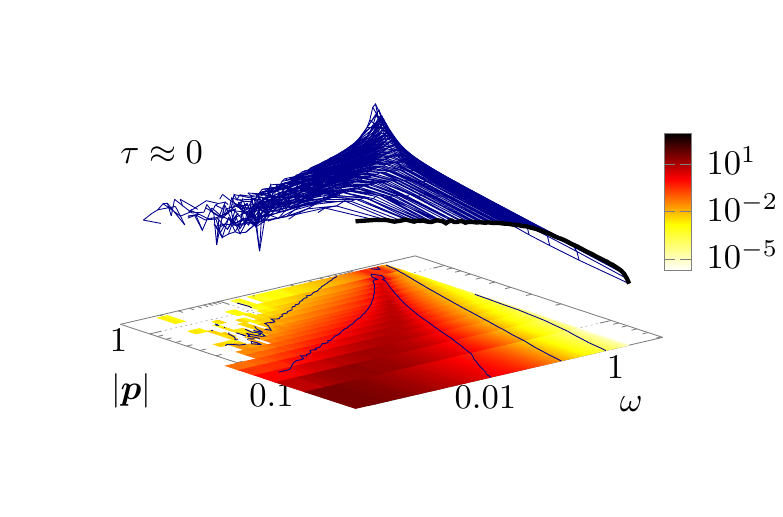} \end{minipage}\vspace{-15mm}
    \begin{minipage}[t]{.5\linewidth} \includegraphics{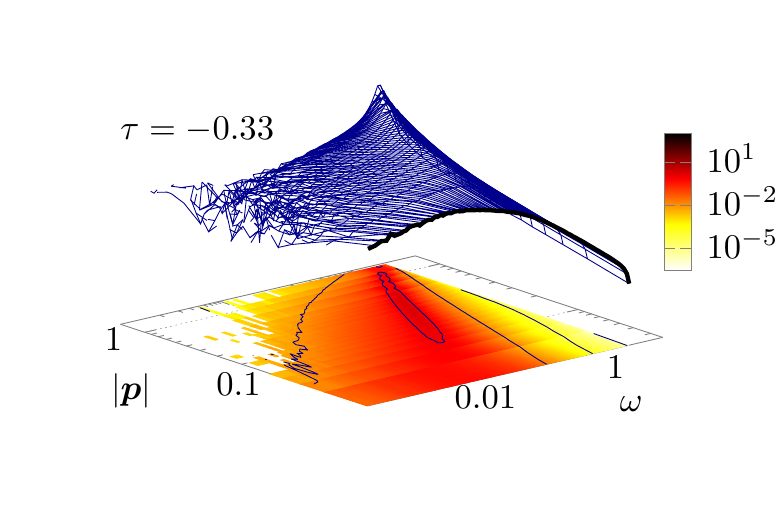} \end{minipage}
    \begin{minipage}[t]{.5\linewidth} \includegraphics{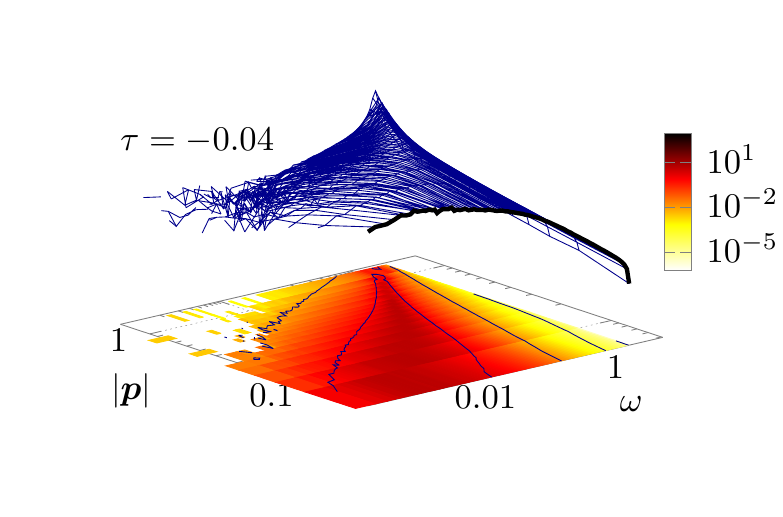} \end{minipage}
    \caption{Overview of spectral functions $\rho(\omega,\vec p)$ for diffusive dynamics now with dissipation (Model B, $\gamma = 0.1$) presented in the same style and order as in \cref{fig:sf-overview-d}. Away from the criticality they still follow
    the Breit-Wigner shapes with dispersion $\omega_p^2 = \mu \vec p^2 (m^2 + \vec p^2)$. The decay widths are now bounded from below by the Langevin damping $\gamma$, and otherwise increase with some power of $|\vec p|$.
    Since the central frequencies do not have a lower bound here
    (in contrast to the purely dissipative dynamics of Model A), while the widths are now bounded by $\gamma$, the structures become relatively broad in the infrared.
        Close to criticality the dispersion relation changes, and the central frequencies at low spatial momenta shift even further into the infrared, leading to dominating broad low-frequency structures.
    }
    \label{fig:sf-overview-b}
\end{figure}

In this subsection we give an overview of our numerical data for the spectral function of the order parameter field and discuss their general shape and structure.   
If not stated otherwise, the numerical data shown in this section was obtained on lattices of size $256^2$ and $128^3$ for $d=2$ and $3$ spatial dimensions, respectively. With these lattice sizes any remaining finite volume effects are so small that they become hard to be observable at our present accuracy.
Generally, we find that the spectral functions are well described by a single Breit-Wigner structure over a wide range of parameters.
Specifically, for the case of vanishing Langevin coupling $\gamma=0$ shown in \cref{fig:sf-overview-d}, we find that, except for the immediate vicinity of the critical point, the peaks in the spectral functions are generally very narrow, which is indicative of the presence of propagating modes related to the real poles of the retarded Greens function in \cref{eq:realpoles}.
Conversely, if the heat-bath coupling is set to a finite value of $\gamma=0.1$, the situation changes dramatically as can be seen from \cref{fig:sf-overview-b}, where in all cases the spectral function at low spatial momentum becomes much broader, while high-momentum modes stay narrow.
Based on our discussion above, the broad low-momentum structure for $\gamma=0.1$ can be associated with the presence of the hydrodynamic mode in \cref{eq:imagpoles}, indicating the diffusive character of the dynamics of the order parameter. 

Close to the critical point, the peak of the spectral function at low spatial momentum visibly shifts towards the infrared, indicating a change in the dispersion relation akin to a drop in the effective mass.
We find that for finite Langevin coupling $\gamma=0.1$, the absolute decay widths stay approximately the same.
However, since the central frequencies decrease, the low-frequency part of the spectral functions is then dominated by structures with large relative widths, which closely fit the overdamped limit of the mean-field spectral function with   
$\rhomf(\omega,\vec p)\to \mu \vec p^{2}/\gamma \omega$ for $\gamma \to \infty $ in \eqref{eq:rho}.

\begin{figure}
    \centering
    \renewcommand{\fdir}{figures/bwplots}
    \includegraphics{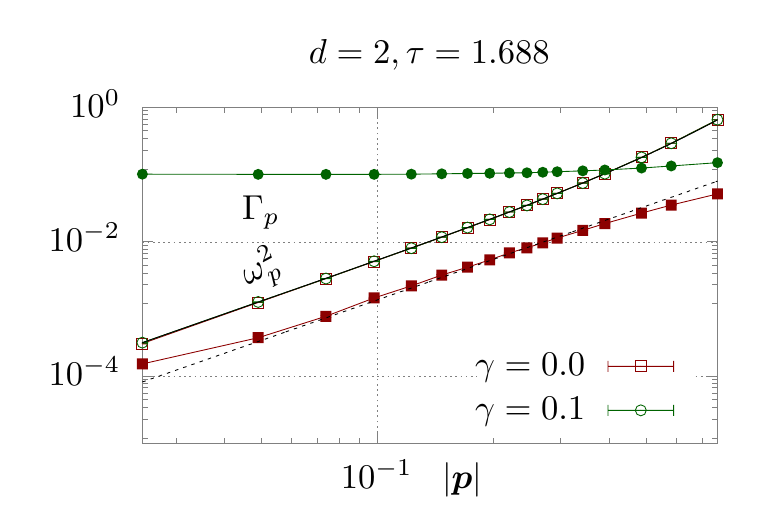}
    \includegraphics{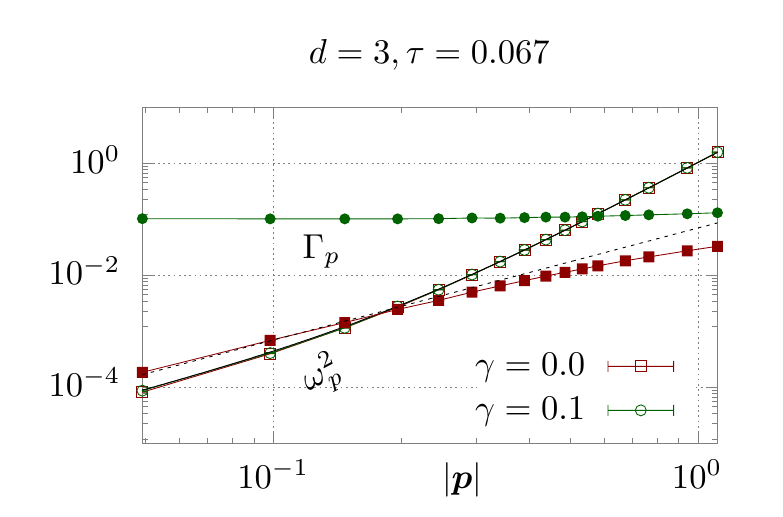}
    \includegraphics{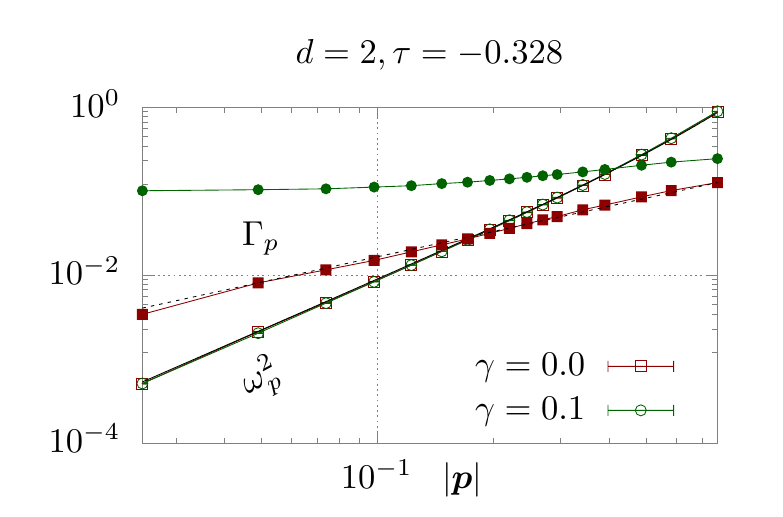}
    \includegraphics{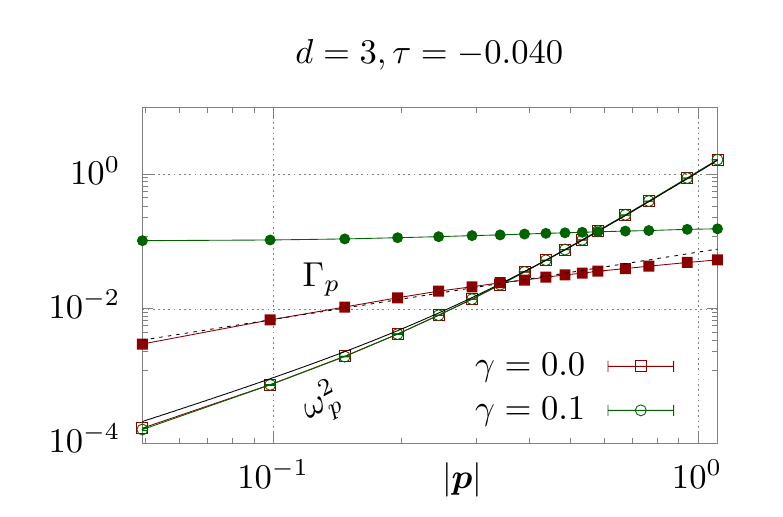}
    \caption{%
       Central frequencies $\omega_p^2$ (open symbols) and decay widths $\Gamma_p$ (filled symbols) resulting from the Breit-Wigner fits in the diffusive Model B (green circles, Langevin coupling $\gamma=0.1$) and the conservative Model BC (red squares, $\gamma = 0$) in two (left) and three (right) spatial dimensions; top row above and bottom row below  $T_c$. The results in the ordered phase are almost perfectly described by mean-field spectral functions of the form in  \cref{eq:rho}.
        Central frequencies are fitted to the same form $\omega_p^2 = \mu\vec p^2(m^2(T)+\vec p^2)$ (solid lines), with temperature-dependent mass parameters $m^2(T)$, above and below $T_c$.
        The momentum dependent widths are fitted to $\Gamma_p \,= \gamma + \bar\Gamma(T) \, \vec p^2$
        the disordered phase above, and 
        to $\Gamma_p \,= \gamma + \bar\Gamma(T) \, \,|\vec p|$ below $T_c$, respectively (dotted lines).
    }
    \label{fig:bwplots_diff}
\end{figure}
We note that, in contrast to our precursor study of the systems with non-conserved order parameter (Models A/C) \cite{schweitzer_spectral_2020}, where an additional collective excitation was observed below $T_c$, there are hardly any additional excitations visible anywhere in the phase diagram, neither 2+1D nor 3+1D.
Hence, in order to further characterize the temperature dependence of the spectral function, we can fit the spectral functions with a Breit-Wigner ansatz
\begin{equation}
    \rho_{\text{BW}}(\omega, \vec p) = \frac{\mu\vec p^2 \Gamma_p \, \omega}{\left( \omega^2 - \omega_p^2 \right)^2 + \Gamma_p^2\omega^2}
    \label{eq:def_BW}
\end{equation}
where the central frequency $\omega_p$ and decay width $\Gamma_p$ are used as the free parameters.
Results deep in the symmetric and ordered phase are shown in \cref{fig:bwplots_diff}.
We find that our results for the central frequencies $\omega_p^2$ at very low resp.~very high temperatures nearly perfectly satisfy the mean-field--like dispersion
\begin{equation}
    \omega_p^2 = \mu \vec p^2(m^2(T) + \vec p^2),
    \label{eq:mf_dispersion_BD}
\end{equation}
with no significant dependence on the Langevin coupling.
Conversely, for the decay width $\Gamma_{p}$ of the spectral function we find that the Langevin coupling $\gamma$ appears as an additional momentum-independent shift
\begin{equation}
    \Gamma_p(\gamma) = \Gamma_p(0) + \gamma,
    \label{eq:Gamma_langevin_dep}
\end{equation}
and we obtain for the momentum dependence of the decay width without the heat bath
\begin{equation}
    \Gamma_p(0) = \bar\Gamma(T) \cdot
    \begin{cases}
        |\vec p|, & T\ll T_c, \\
        \vec p^2, & T\gg T_c.
    \end{cases}
    \label{eq:Gamma_momentum_dep}
\end{equation}
which is indicated by a dashed line in \cref{fig:bwplots_diff}.
In fact, \cref{eq:mf_dispersion_BD,eq:Gamma_langevin_dep,eq:Gamma_momentum_dep} capture the momentum dependence of the spectral function so well, that one can confidently describe the spectral functions at different temperatures and damping constants $\gamma$ by just two parameters, namely the temperature dependent effective mass $m(T)$ and amplitude $\bar\Gamma(T)$ of the power-law decay of the width $\Gamma_p \to\gamma $ in the long wavelength limit.
\begin{figure}
    \centering
    \renewcommand{\fdir}{figures/bwplots}
    \includegraphics{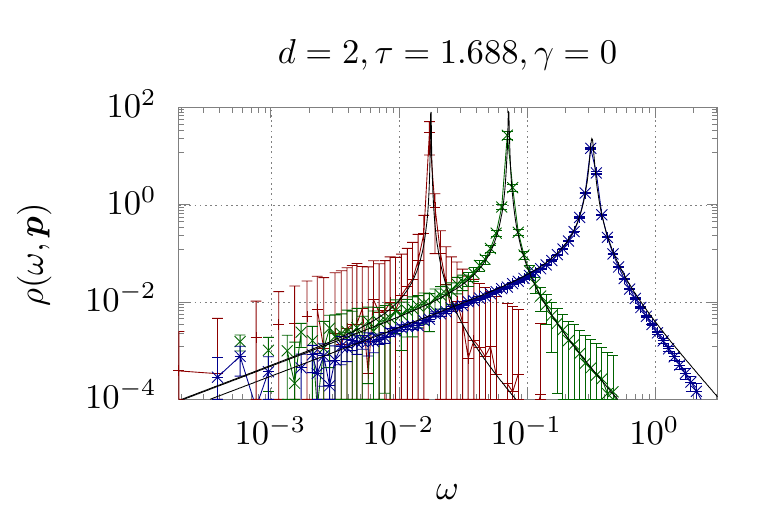}
    \includegraphics{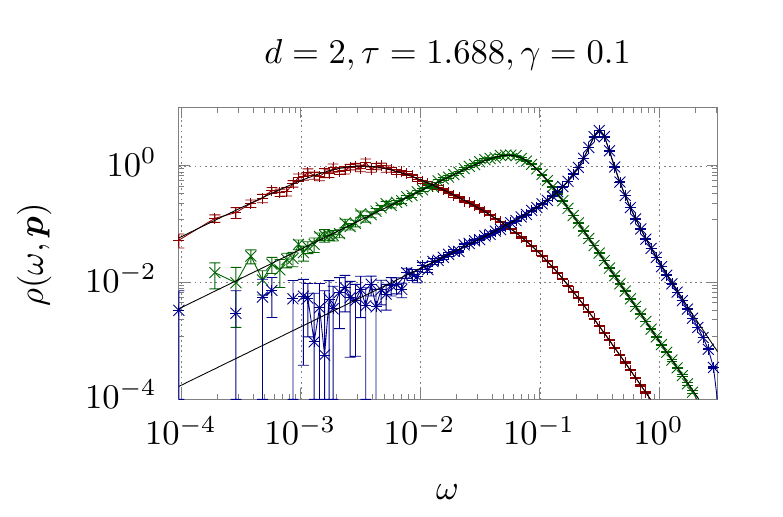}
    \includegraphics{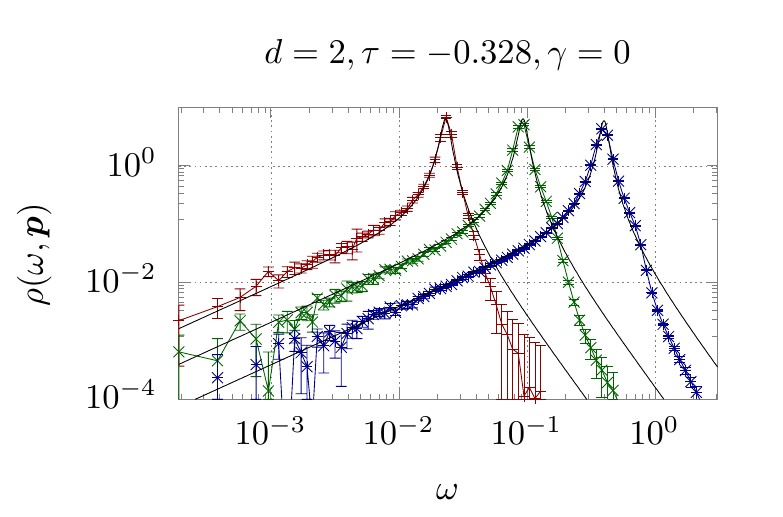}
    \includegraphics{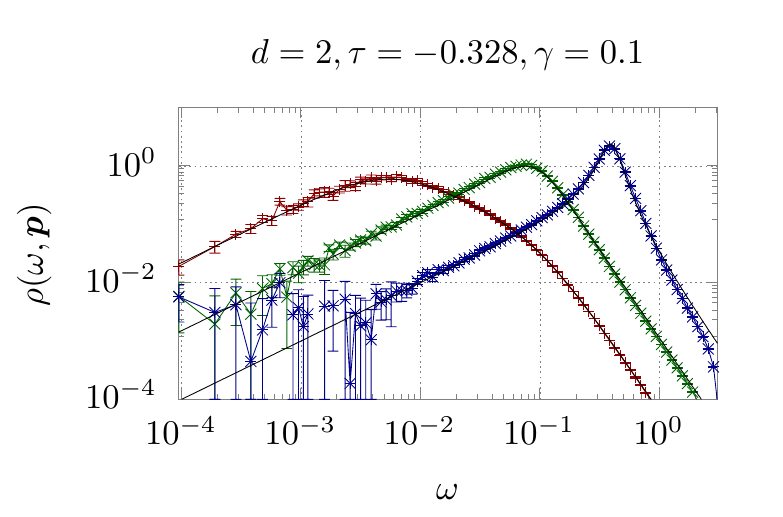}
    \caption{Frequency dependence of spectral functions $\rho(\omega,\vec p)$  in 2+1D at fixed spatial momenta $\vec p$, with $|\vec p|=.1$ \textcolor{BrickRed}{(red)}, $|\vec p|=.4$ \textcolor{OliveGreen}{(green)}, and $|\vec p|=1.4$ \textcolor{MidnightBlue}{(blue)}; deep in the symmetric (top row) and the ordered phase (bottom row), both for vanishing ($\gamma=0$, left) and finite ($\gamma=0.1$, right) heat-bath coupling $\gamma$. 
    Black lines represent fits to the Ansatz \eqref{eq:def_BW} with the thermal mass $m(T)$ and the amplitude of the decay widths $\bar\Gamma(T)$, defined in \cref{eq:mf_dispersion_BD,eq:Gamma_langevin_dep,eq:Gamma_momentum_dep},
    as the only two free parameters per row. Apart from some small deviations at large frequencies for $\gamma=0$ at low $T$ (bottom left) by and large these global fits describe all data very well.
    }
    \label{fig:fit_comparison_BD}
\end{figure}
This is demonstrated in \cref{fig:fit_comparison_BD}, where we compare the resulting fits with these two parameters to \cref{eq:def_BW,eq:mf_dispersion_BD,eq:Gamma_langevin_dep,eq:Gamma_momentum_dep} for the frequency dependence of spectral functions at three different spatial-momentum values, both deep in the ordered and the symmetric phase.
Even though there are only two free parameters per row in this figure, the fit describes the data nearly perfectly, with minor deviations at low temperatures and large frequencies only for $\gamma=0$. This is exemplified in \cref{fig:fit_comparison_BD} for $d=2$ spatial dimensions, but essentially the same quality of two-parameter mean-field fits away from criticality is obtained in 3+1D spacetime as well.

\begin{figure}
    \centering
    \renewcommand{\fdir}{figures/bwplots}
    \includegraphics{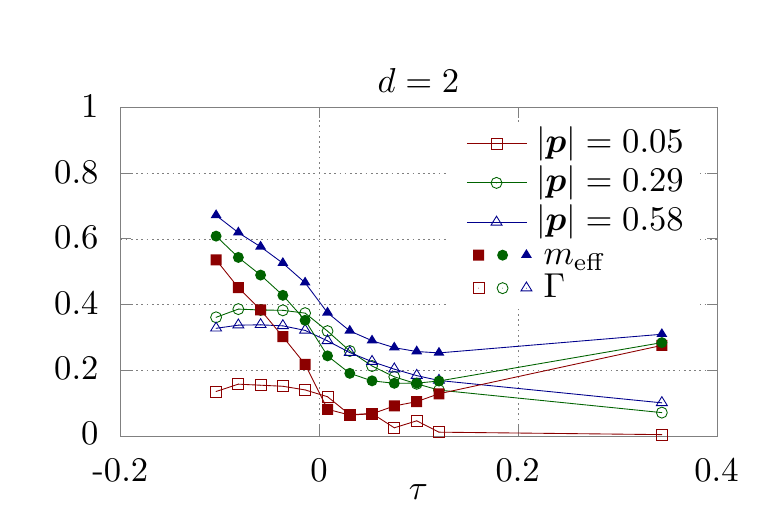}
    \includegraphics{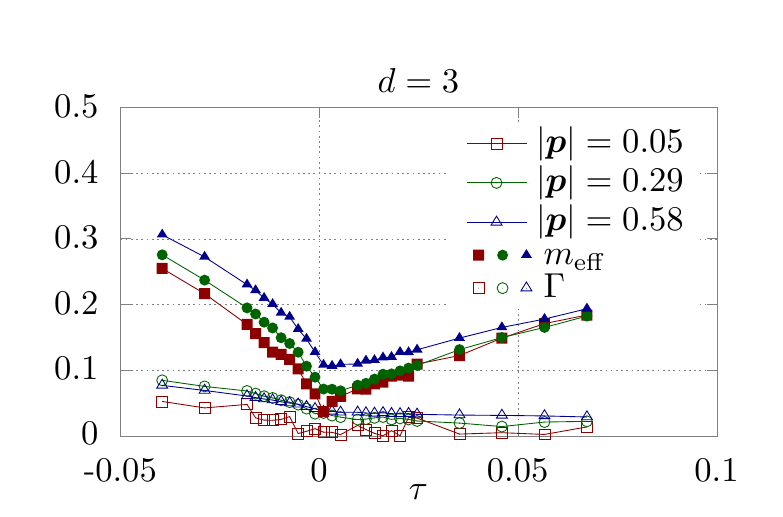}
    \caption{Temperature dependence of the effective mass parameter $m_{\text{eff}} \equiv (\omega_p^2 / \mu \vec p^2 - \vec p^2)^{1/2}$ and damping rates $\Gamma_p $ at a set of three fixed momenta as a function of reduced temperature around the critical point, here for vanishing Langevin coupling $\gamma=0$ (which would affect only the damping rates).
        At zero temperature, resp.~$\tau\to -1$, the effective mass and damping start at $m_\text{eff} =\bar m= \sqrt{2}$ and $\Gamma_p =0 $. With increasing temperature, damping rates first increase before they start to drop again when  
        approaching the critical point. The effective masses drop until they reach a minimum closer and closer to criticality as the spatial momentum (here acting as an effective infrared cutoff) is gradually reduced.    
        At high temperatures, the widths vanish again and the effective masses approach one another as the dispersion relation \cref{eq:mf_dispersion_BD} gets restored.
    }
    \label{fig:smplot}
\end{figure}

We conclude our overview discussion of the spectral functions by investigating the temperature dependence of the central frequencies $\omega_{p}$ and decay widths $\Gamma_p$ across the transition. 
Here, we conversely define a temperature dependent effective mass via $m_{\textrm{eff}}^2(T) = {\omega_p^2/\mu \vec p^2 - \vec p^2}$, such that curves for different spatial momenta $\vec p$ will coincide only if peaks in the spectral functions satisfy the mean-field dispersion relation \eqref{eq:mf_dispersion_BD}, which is the case sufficiently far away from criticality as shown in \cref{fig:bwplots_diff,fig:fit_comparison_BD}.
The resulting effective masses and widths across the transition are shown for $\gamma=0$ in \cref{fig:smplot}.
In the limit of low temperatures, $T\to0$, there are no fluctuations left in the classical field-theory system and the spectral functions will reduce to corresponding delta-function peaks, with $\Gamma_p(T=0,\gamma=0) \to 0$ and the effective mass approaching its classical mean-field value $m_{\textrm{eff}}(0) = \bar m= \sqrt{-2m^2}$.
Turning on temperature, the effective mass starts to decrease with a non-zero thermal width building up and increasing at first.

We observe from \cref{fig:smplot} that close to the critical point, the mean-field dispersion relation in \cref{eq:mf_dispersion_BD} is no-longer satisfied, although -- at least at the lowest temperatures shown in \cref{fig:smplot} -- the effective masses are still rather large.\footnote{Note that this is not in conflict with results shown in \cref{fig:fit_comparison_BD}, since we are now much closer to the critical point.}
As the system approaches the critical temperature $T_c$ from below, i.e.~for $\tau \to 0^-$, the thermal decay widths $\Gamma_p(T)$ turn around and start to decrease rather smoothly across the transition. The effective masses decrease more rapidly, and the low-momentum modes can no-longer be well described by the Breit-Wigner Ansatz \eqref{eq:def_BW} with mean-field dispersion \cref{eq:mf_dispersion_BD} across the transition.
As the temperature increases above the critical point, 
the effective masses reach their minima at a \emph{pseudo-critical} $T_{pc} > T_c $ which is closer to criticality for the lower momentum modes. In the long-wavelength limit, these minimal values tend to zero at criticality, i.e.~$m_\text{eff}(T_{pc}) \to 0 $ with $T_{pc}\to T_c $ for $|\vec p| \to 0$. The spatial momentum therefore effectively acts as an infrared cutoff similar to finite-size effects here.
For $T>T_{pc}$ the increasing effective masses  above the minima converge against one another so that the dispersion relation \eqref{eq:mf_dispersion_BD} is gradually satisfied better and better and the corresponding Breit-Wigner shape is restored again as temperature further increases. The effective mass then continues to increase monotonously with temperature for $T \gg T_c$.
Notably, the process is smoother for modes with larger spatial momenta in-line with the interpretation of $|\vec p|$ acting as an effective infrared cutoff.
While the effective masses of the higher momentum modes also reach a minimum somewhere above the critical temperature at $\tau=0$, these modes retain their mean-field Breit-Wigner shape all across the transition.

In summary, the measured spectral functions show by and large the expected behavior in the non-critical regime.
The dispersion relation obtained from the mean-field analysis is fulfilled over a wide range of temperatures, and we find that the effective masses show the expected temperature dependence on either side of the transition.
Close to the critical point, however, the dispersion relation changes as the effective mass at low momenta nearly vanishes.
We will find in the next section that the changes in the infrared dispersion are compatible with critical scaling laws.
The decay widths exhibit a non-trivial momentum dependence that changes its analytic form as the system crosses the critical point, cf.~\cref{eq:Gamma_momentum_dep}.
Similar to our findings for the non-critical spectral functions of Models A and C in \cite{schweitzer_spectral_2020}, the spectral functions at large spatial momenta retain their shape and move across the phase transition continuously, while at low momenta their spectral shape changes significantly and exhibits dynamic critical behavior as we will discuss next.

\clearpage
\section{Critical dynamics of the order parameter}\label{sec:op_crit}

We now continue to investigate the critical behavior of the spectral functions of the order parameter.
Starting with a brief summary of previous studies, we will first demonstrate the existence of a critical scaling regime, and subsequently focus on the determination of the dynamic critical exponent $z$ and, where possible, the determination of a universal scaling function for the spectral function of the order parameter.

Empirical investigations of dynamic critical phenomena in general and in Ising-like models in particular exist since the late 60s, with the first high-precision numerical studies emerging in the 90s \cite{dammann_dynamical_1993,matz_dynamic_1994,wang_study_1995,nightingale_dynamic_1996}.
Most of those studies were concerned with Glauber-like dynamics, where the order parameter is not conserved over time (Models A/C).
Experiments on thin films allow accessing the critical dynamics of 2D systems; Dunlavy and Venus found $\nu z = 2.09 \pm .06$ using ferromagnetic films \cite{dunlavy_critical_2005}.
For Ising-like systems with a conserved order parameter, on obtains the exact result $z=4-\eta$ using the dynamic renormalization group framework \cite{halperin_renormalization-group_1974}.
However, numerical and experimental measurements are scarce. 
An early numerical study by Yalabik and Gunton \cite{yalabik_monte_1982} applied the Monte-Carlo renormalization group approach on a 2D Ising model with Kawasaki dynamics, i.e.~nearest-neighbour spin flips, finding $z=3.80$, in very good agreement with the result $z=4-\eta = 3.75$ from the dynamic renormalization group.
In 2001, Zheng \cite{zheng_monte_2000} conducted a study on the critical dynamics of the two-dimensional Ising model with Kawasaki dynamics, and found that short-time correlations exhibit scaling behavior with a dynamic critical exponent $z=3.95(10)$, slightly larger than expected in 2+1D.
When changing to a different dynamic scheme, where spin exchanges happen over larger distances and the spin is no longer locally (but still globally) conserved, they found a different, much smaller exponent $z=2.325(10)$.
A study on a quasi-2D lipid bi-layer in water \cite{honerkamp-smith_experimental_2012} (Models B/H/HC) found that the exponent of the time scale of time-dependent correlation functions changed from $z_{\textrm{eff}} \sim 2$ to $z_{\textrm{eff}}\sim 3$, depending on the ratio of the correlation length of the fluctuations over a hydrodynamic length scale set by transport coefficients.
Drastic changes of the dynamic critical exponent $z$ upon seemingly slight changes of the dynamics are therefore not unheard of in systems with conserved order parameter.

In a precursor study \cite{schweitzer_spectral_2020}, we observed dynamic critical behavior of Models A and C based on the dynamic equation \eqref{eq:ModelA_eom}, where the order parameter is not conserved.
While Model A describes the dynamic critical behavior of a system where both the order parameter and energy density are fluctuating, e.g.~\cref{eq:ModelA_eom} with a finite heat-bath coupling $(\gamma>0)$, Model C applies e.g.~to Hamiltonian systems ($\gamma=0$), where the order parameter can fluctuate but the total energy is conserved.
By changing the dynamic equations to \cref{eq:ModelB_eom}, such that the order parameter is conserved, the classification scheme by Hohenberg and Halperin \cite{hohenberg_theory_1977} suggests that in the case where the theory is coupled to a heat bath ($\gamma>0$), we are dealing with the dynamics of Model B, describing a system with diffusive dynamics of the order parameter without additional conserved quantities. While the conservative limit of the relativistic Model B evolution in \cref{eq:ModelB_eom} features the same set of conserved quantities as the usual Model D, i.e.~a system with diffusive dynamics of the order parameter together with a conserved energy density, it turns out that -- as discussed in \cref{sec:specfunc} -- setting $\gamma=0$ here, changes the low-energy spectrum of the theory on Since the classification of this theory is far from obvious, we will simply refer to it as the conservative limit of our Model B or in brief Model BC. When considering the critical behavior of Model BC, we will indeed find dynamic critical exponents that are much smaller than the Model B and D value $z_B = 4-\eta$, as in the limit $\gamma = 0 $ we obtain values of the critical exponents, that are much closer to those of Models A or C here.

To simplify notation, we remark that, generally, the spectral function does not depend on the direction of the spatial momentum, and we therefore write $\rho(\omega, p, T)$, with $p\equiv |\vec p| = \sqrt{ \vec p^2}$ denoting the magnitude of spatial momentum from now on.
The data presented in this section was obtained on lattices of size $1024^2$ and $256^3$ for $\gamma \leq 0.1$, as well as $256^2$ and  $64^3$ for $\gamma=1.0$, in $d=2 $ and $3$ spatial dimensions, respectively.

\subsection{(Auto-)Correlation time} \label{sec:xi_t}
We start our study of the critical dynamics of the order parameter fluctuations by analyzing the divergence of the characteristic timescale $\xi_t$ in the vicinity of the critical point.
Specifically, we consider the behavior of the momentum-dependent auto-correlation time at criticality, defined as
\begin{align}
    \xi_t(p) = \frac{\int_{0}^{\infty}t\,  \rho(t, p, T_c) \, \intd t}{\int_{0}^{\infty} \rho(t, p,T_c) \, \intd t} \, ,
\end{align}
i.e.~as a function of the spatial momentum $p$ at the critical temperature $T=T_c $.
Since the spatial correlation length of the system $\xi$ diverges at the critical point, the relevant infrared cut-off at $\tau=0$ is again imposed by the finite spatial momentum $p$ here.
We can therefore expect the momentum dependence of the (auto-)correlation time $\xi_t $ to be given by
\begin{align}
    \xi_t(\bar p) = f_t\bar p^{-z} , \;\; \mbox{with} \;\;\; 
    \bar p \equiv f_{\xi}^{+} p \,.
    \label{eq:xit_p}
\end{align}
Here, the dimensionless momentum scale $\bar p$ is defined relative to the amplitude in the power-law  divergence (for $\tau\to 0^+$) of the correlation length listed in \cref{tab:statics} for the static critical behaviour of our scalar field theory in $d=2$ and $3$ spatial dimensions. 
\cref{eq:xit_p} furthermore defines the universal dynamic critical exponent $z$ and a non-universal amplitude $f_{t}$ characterizing the typical time scale for critical dynamics.

\begin{figure}
    \centering
    \renewcommand{\fdir}{figures/xits}
    \includegraphics[width=0.5\linewidth]{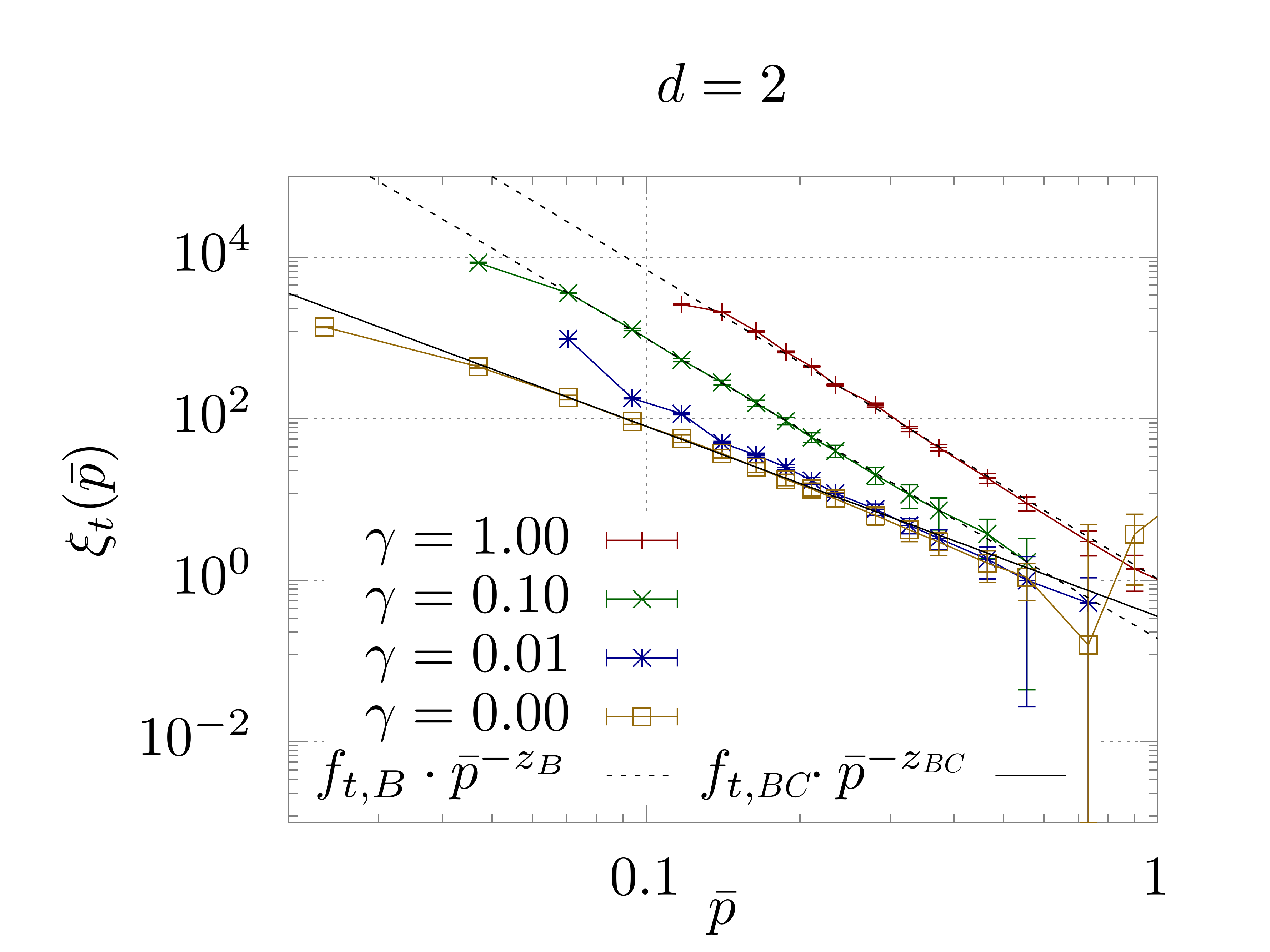}\hskip -.1cm
    \includegraphics[width=0.5\linewidth]{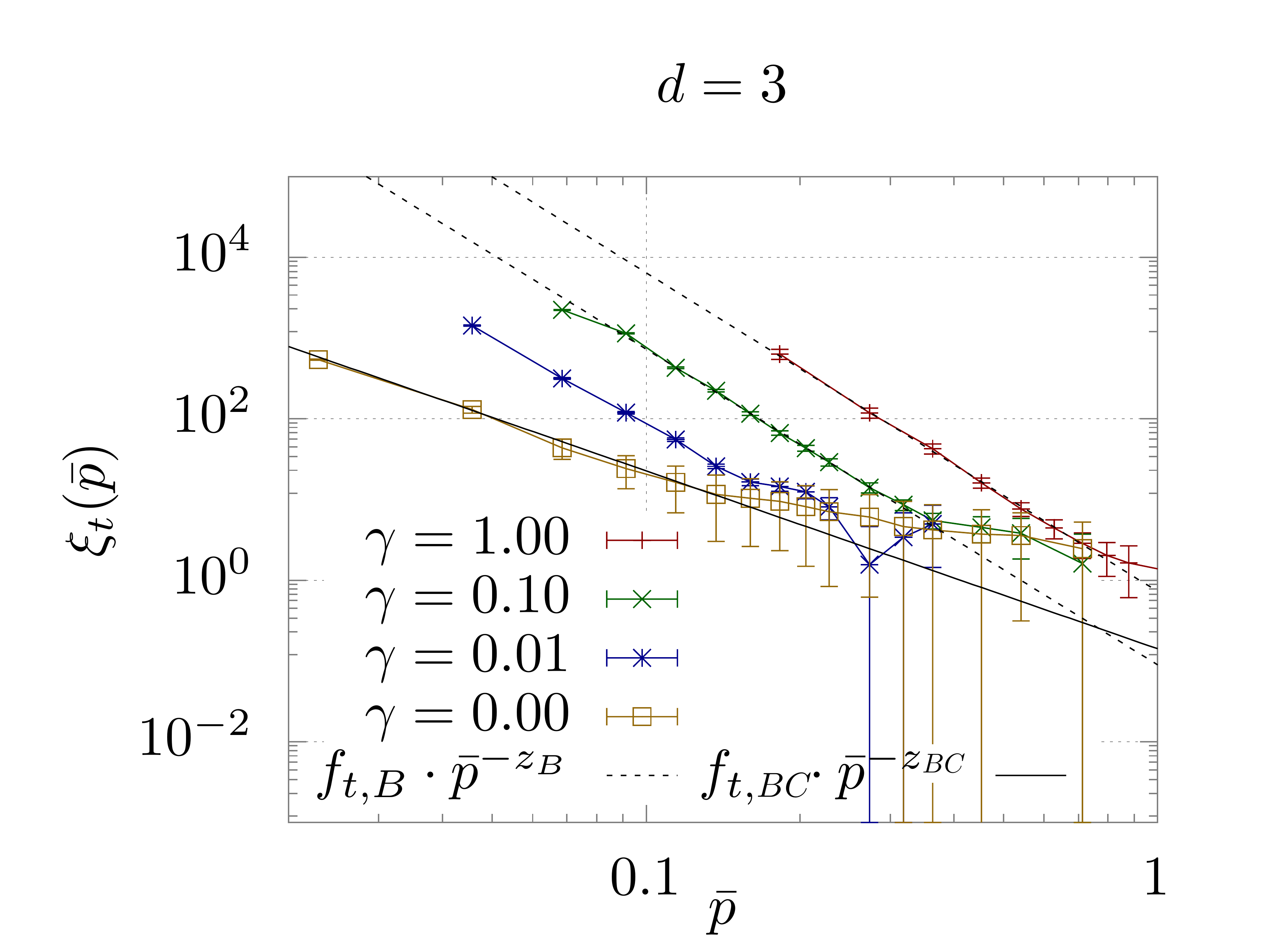}
    \caption{
        Momentum-dependent correlation time $\xi_t$ at criticality ($\tau=0$) over  the dimensionless momentum scale $\bar p$ at different values of the Langevin coupling $\gamma$ in $d=2$ (left) and 3 (right) spatial dimensions.
        Dashed lines indicate power-law fits to the low-momentum limits of the data, the resulting amplitudes and exponents are summarized in \cref{tab:intxits}.
        The exponents for finite heat-bath coupling $\gamma$ are consistent with $4-\eta$ for Model B dynamics and considerably larger than those for $\gamma=0$. The amplitudes $f_{t,B} $ scale approximately linearly with the Langevin coupling. At higher momenta, here visible especially in $d=3$, a second power law emerges with a smaller exponent (close to that for $\gamma=0$) and a prefactor which is practically independent of $\gamma$.
            }
    \label{fig:intxits_bd}
\end{figure}

\begin{table}
    \centering
    \begin{tabular}{l | r r r r r r}
        $d$ & $z(\gamma=1.0)$& $f_t(\gamma=1.0)$& $z(\gamma=0.1)$& $f_t(\gamma=0.1)$& $z(\gamma=0.0)$& $f_t(\gamma=0.0)$\\ \hline
2 & 3.83(10)& 1.04(14)& 3.716(17)& 0.190(8)& 2.354(23)& 0.358(19)\\
3 & 3.95(8)& 0.73(6)& 3.91(6)& 0.090(10)& 2.20(13)& 0.14(5)
    \end{tabular}
    \caption{
        Amplitudes and exponents in $d=2$ and $3$ spatial dimensions obtained from fits to the data in \cref{fig:intxits_bd}.
        While the exponents obtained from the data at large $\gamma$ agree well with the expected dynamic critical exponents $z=4-\eta$ of Model B, those from the data at vanishing $\gamma=0$ are much smaller, closer to Models A or C.
    }
    \label{tab:intxits}
\end{table}
Our results for the integrated auto-correlation times obtained from the measured spectral functions are illustrated in \cref{fig:intxits_bd}.
In both 2+1 and 3+1 dimensions, the correlation times $\xi_t(\bar p)$ show the expected power-law behavior of \cref{eq:xit_p}.
In particular, for finite heat-bath coupling $\gamma>0$, the results at low spatial momenta clearly exhibit a power law $\xi_t(\bar p)\sim \bar p^{-z_B}$ consistent with the dynamic critical exponent $z_B = 4-\eta$ of Model B, which smoothly merges into a second power law with a much smaller scaling exponent at higher momentum scales.
Evidently, the amplitudes $f_t$ of the critical power law at low momentum strongly depend on the value of $\gamma$, such that for smaller values of $\gamma$ the transition between the two power laws occurs at lower momenta.
When considering the case $\gamma=0$, the critical behavior of Model B ceases to exist, and therefore only the second power of the Model BC remains.

Based on our analysis of auto-correlation times in \cref{fig:intxits_bd}, we extract the dynamic critical exponent $z$ and non-unverisal amplitude $f_t$ from a $\chi^2$-fit to a power law of the form of \cref{eq:xit_p}.
Our results for the exponents and amplitudes are given in \cref{tab:intxits}.
While the exponents for finite heat-bath coupling $\gamma>0$ confirm the prediction by Model B, namely $z_B = 4-\eta$, the exponents for our conservative Model BC with $\gamma=0$, denoted by $z_{BC}$ in  \cref{fig:intxits_bd}, are much smaller.
Our analysis for the momentum-dependent correlation times is overall consistent with the existence of two competing power laws in the infrared, with leading exponent $z_B$ and subleading exponent $z_{BC} < z_B$, where the momentum scale of the transition between the two at 
$p \sim \gamma^{1/(z_B-z_{BC})}$ vanishes in the non-dissipative limit for Langevin coupling $\gamma\to 0 $.

\subsection{Critical spectral function}\label{sec:crit_sf}
\begin{figure}
    \centering
    \renewcommand{\fdir}{figures/bwplots}
    \includegraphics{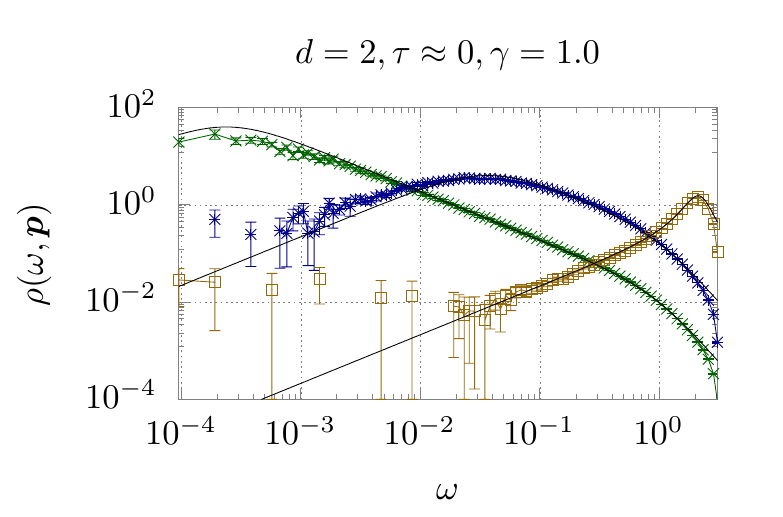}
    \includegraphics{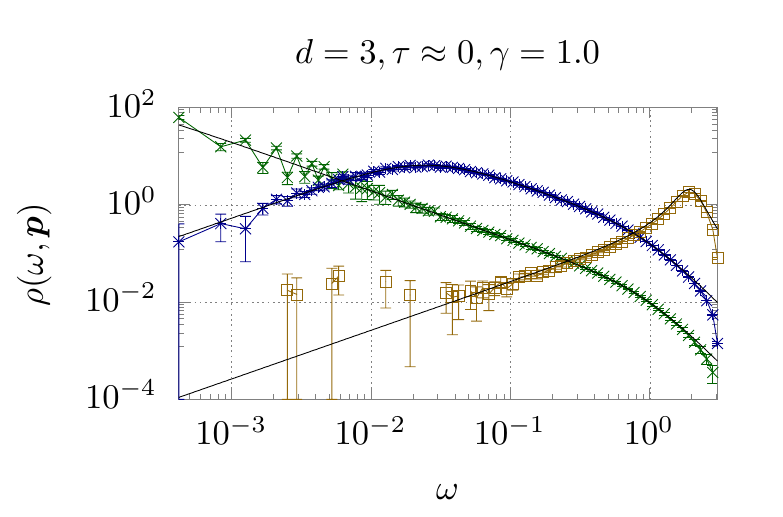}
    \includegraphics{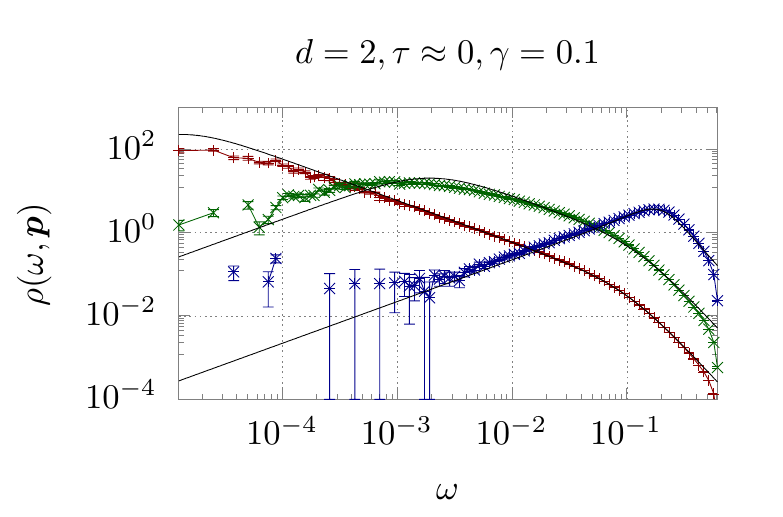}
    \includegraphics{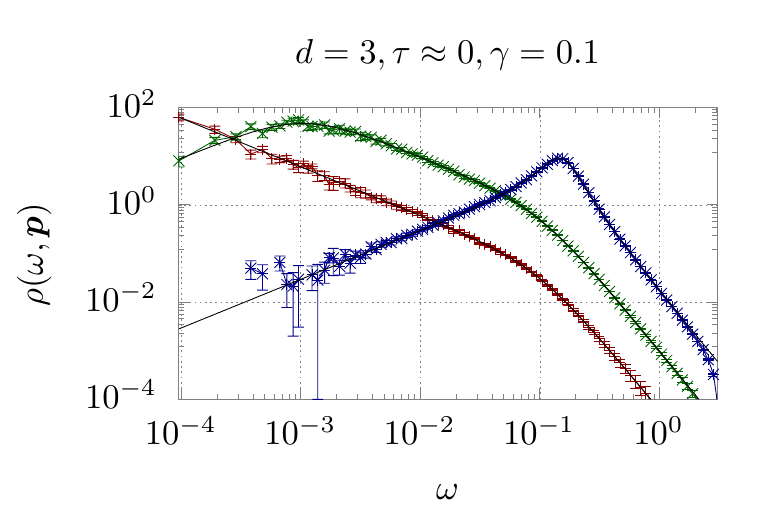}
    \includegraphics{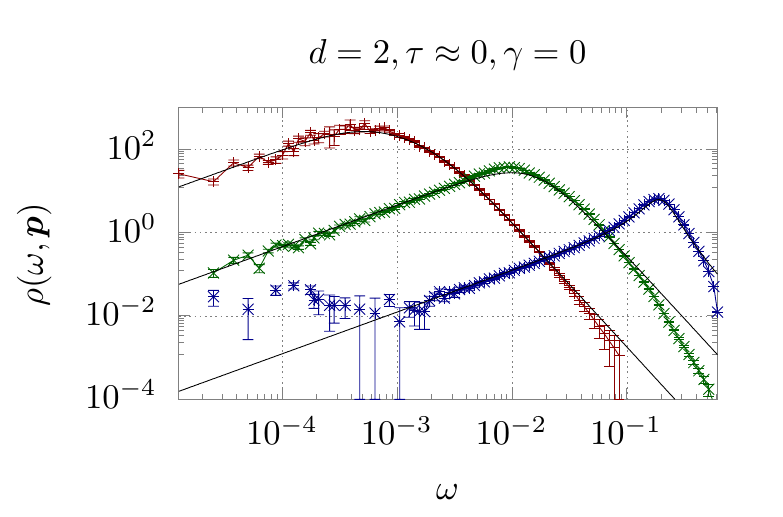}
    \includegraphics{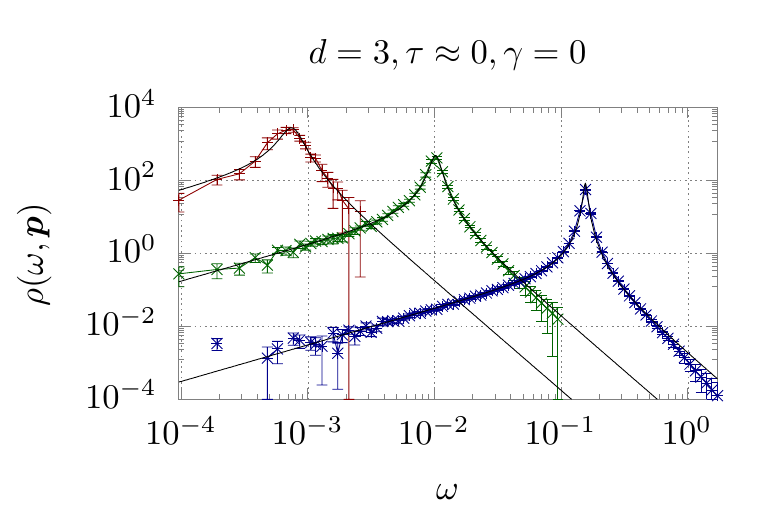}
    \caption{%
        Spectral functions of the order parameter at criticality for fixed spatial momenta $p=.025$ \textcolor{BrickRed}{(red)}, $p=.098$  \textcolor{OliveGreen}{(green)}, $p=.390$ \textcolor{MidnightBlue}{(blue)} and $p=1.414$ \textcolor{YellowOrange}{(yellow)}.
        Solid black lines represent fits to \cref{eq:def_BW}.
        The extracted fit parameters are illustrated in \cref{fig:bwplots_diff_tc} below.
    }
    \label{fig:fit_comparison_BD_tc}
\end{figure}
\begin{figure}
    \centering
    \renewcommand{\fdir}{figures/bwplots}
    \includegraphics{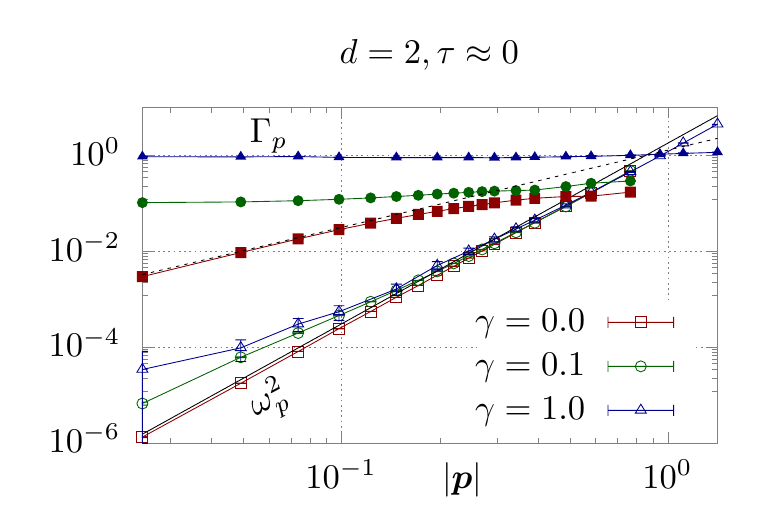}
    \includegraphics{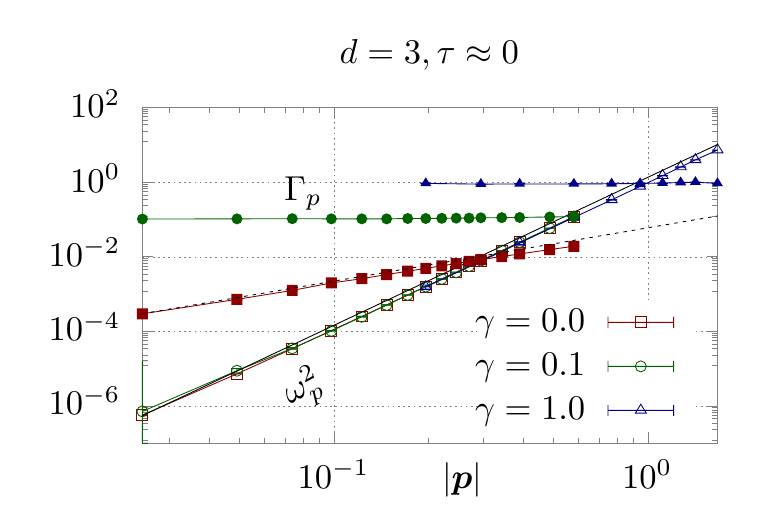}
    \caption{%
        Central frequencies and widths extracted from Breit-Wigner 
        fits to critical spectral functions for Langevin coupling $\gamma=1.0$ (blue triangles), $\gamma = 0.1$ (green circles), and without heat bath with $\gamma = 0$ (red squares). Open symbols denote the central frequencies $\omega_p^2$, and filled symbols the corresponding decay widths $\Gamma(p)$.
        The central frequencies show the power-law behavior of \cref{eq:omega_scaling} (for $\gamma>0$ in 2+1D the Breit-Wigner ansatz does not fit the data at low momenta perfectly well, compare e.g.~the center-left panel of \cref{fig:fit_comparison_BD_tc}, leading to systematic errors in determining the central frequencies). 
        The widths for $\gamma>0$ approach $\Gamma_p\to \gamma$ in the long-wavelength limit. At vanishing heat-bath coupling $\gamma=0$ they follow the power law in \cref{eq:gamma_scaling}. Together with \cref{eq:omega_scaling} these power laws predict for        the conservative Model BC the dynamic critical exponent $z_{BC} =  z_\omega - z_\Gamma$, see text. 
        The resulting exponents and amplitudes for $\gamma=0$ are summarized in \cref{tab:bwfit}, leading to the estimates $ z_{BC} \approx 2.2 $ in 2+1D and $ z_{BC} \approx 2.5 $ in 3+1D.
    }
    \label{fig:bwplots_diff_tc}
\end{figure}

\begin{table}
    \centering
    \begin{tabular}{l | r r r r}
$d$  & $z_{\omega}$& $\omega^2_0$& $z_{\Gamma}$& $\Gamma_0$\\ \hline
2 & 3.769(13)& 1.81(6)& 1.612(21)& 1.29(7)\\
3 & 3.9660(20)& 1.3430(25)& 1.43(6)& 0.059(9)
    \end{tabular}
    \caption{Extracted parameters from the power laws \cref{eq:omega_scaling,eq:gamma_scaling} controlling the momentum dependence of $\omega_p^2$ and $\Gamma_p$ at vanishing Langevin coupling $\gamma=0$.
        We find that the scaling exponent of $\omega_p^2$ matches $z_\omega = 4-\eta$ exceptionally well.
        Using the Breit-Wigner prediction for low momenta, $\xi_t(p) = \Gamma_p/\omega_p^2$ (as shown in Appendix~\labelcref{app:xit_BW}), and comparing with \cref{eq:xit_p}, one can read off the resulting dynamic critical exponent $z(\gamma =0) = z_\omega-z_\Gamma$.
    }
    \label{tab:bwfit}
\end{table}

We continue to analyze the frequency and momentum dependence of the spectral function at criticality.
Exemplary data for the spectral functions at different momenta $p$ and different values of the Langevin coupling $\gamma$ in 2+1D and 3+1D are shown in \cref{fig:fit_comparison_BD_tc}.
We find that the critical spectral functions still largely follow a Breit-Wigner shape, as illustrated by the black lines in \cref{fig:fit_comparison_BD_tc} which represent fits to the Breit-Wigner form of \cref{eq:def_BW}  but now with the fit parameters $\Gamma_p$ and $\omega_p$ (instead of the thermal mass from the mean-field dispersion \cref{eq:mf_dispersion_BD}).
Visible deviations from the Breit-Wigner shape then only emerge at low momenta and finite heat bath coupling $\gamma>0$ in 2+1D. 
We will see shortly that this part of the spectral function is controlled by an underlying universal scaling function.

Next, in order to characterize the momentum dependence of the critical spectral function, we again fit the Breit-Wigner ansatz \eqref{eq:def_BW} to the data, and show the extracted fit parameters $\omega_p^2$, $\Gamma_{p}$ as function of the spatial momentum $p$ in \cref{fig:bwplots_diff_tc}.
Similar to the results deep in the ordered or symmetric phases, the dispersion of $\omega_p$ is approximately independent of the Langevin coupling $\gamma$ within the considered range of parameters.
However, at criticality the dispersion relation changes, we no-longer find the mean-field-like relation \eqref{eq:mf_dispersion_BD}.
Instead, the momentum dependencies of the central frequencies are themselves controlled by a power law now,
\begin{equation}
    \label{eq:omega_scaling}
    \omega_{p}^2=\omega_{0}^2 \, \bar{p}^{z_{\omega}}
\end{equation}
with a scaling exponent $z_{\omega}$ matching the dynamic critical exponent  $z_B=4-\eta$ of Model B with high precision, independent of the heat-bath coupling.

However, a cruicial difference between dissipative ($\gamma>0$) and non-dissipative dynamics ($\gamma=0$) emerges when considering the momentum dependencies of the decay widths $\Gamma_{p}$, which are also shown in \cref{fig:bwplots_diff_tc}.
While in the dissipative systems, the decay widths $\Gamma_{p}$ approach the finite dissipation rate $\gamma$ in the long-wavelength limit, without dissipation the decay widths $\Gamma_{p}$ follow a power-law behaviour as well,
\begin{equation}
    \label{eq:gamma_scaling}
    \Gamma_{p}=\Gamma_{0}\, \bar{p}^{z_{\Gamma}}.\;
\end{equation}
We find that the corresponding scaling exponent is given by $z_{\Gamma}\approx 1.6$ in 2+1D and $z_{\Gamma}\approx 1.4$ in 3+1D. 
Our results for exponents $z_\omega$, $z_\Gamma$ and amplitudes $\omega_0$, $\Gamma_0 $ in the power laws \cref{eq:omega_scaling} and \cref{eq:gamma_scaling} of central frequencies and decay widths from corresponding fits to the data of \cref{fig:bwplots_diff_tc} are summarised in \cref{tab:bwfit}.

Notably, the differences in the low-momentum scaling behaviour of the decay widths $\Gamma_{p}$ can also explain the observed differences in the behaviour of the auto-correlation times $\xi_{t}(p)$ shown in \cref{fig:intxits_bd}.
By inserting the Breit-Wigner Ansatz \eqref{eq:def_BW} into the formula for the integrated correlation time \eqref{eq:xit_p}, one obtains the correlation time to be equal to the ratio 
\begin{equation}
\xi_{t}(p) = \Gamma_{p}/\omega^2_{p} \label{eq:BWscale}
\end{equation}
for low  momenta $p$ in the infrared (see Appendix \ref{app:xit_BW} for a sketch of the derivation).
We can therefore compare our findings for the auto-correlation times with those for the Breit-Wigner parameters here:
Because for the dissipative systems the decay widths $\Gamma_{p}$ approach the finite Langevin damping $\gamma$ in the long-wavelength limit, one concludes $\xi_t(\bar p) = \gamma/\omega_p^{-2} \sim \bar p^{-4+\eta}$ with an amplitude linearly dependent on the Langevin damping $\gamma$.
Conversely, for the non-dissipative dynamics of our Model BC, the momentum dependence of the decay width $\Gamma_{p}$ becomes relevant, and one has ${\xi_t(\bar p) = \Gamma_{p}/\omega_{p}^2 \sim \bar p^{-4+\eta+z_{\Gamma}}}$.
This explains the different scaling exponents in the momentum dependent correlation times to be observed for $\gamma > 0$ and $\gamma=0$ at least qualitatively, cf.~\cref{fig:intxits_bd,tab:intxits}.
Quantitatively, the Breit-Wigner prediction from \cref{eq:BWscale} with  
$z(\gamma=0) \approx z_{\omega} - z_{\Gamma}\approx 2.16 $ agrees reasonably well and within the errors with the data from the autocorrelation time analysis in 2+1D. On the other  hand, there is some tension between the same estimate $z(\gamma=0) \approx z_{\omega} - z_{\Gamma}\approx 2.54 $ and the data in 3+1D, cf.~\cref{tab:intxits}.
We suspect that this might be caused by uncertainties in the auto-correlation times $\xi_t(p)$, where too few data points effectively contribute to the fit, see the $\gamma=0$ data in right panel of \cref{fig:intxits_bd}.

\subsection{Universal scaling functions}

Clearly, the divergence of the auto-correlation time and the results of the Breit-Wigner fits of the spectral function are indicative of the emergence of critical scaling behavior in the vicinity of the critical point.
We therefore apply the dynamic scaling hypothesis to the spectral functions to extract underlying universal scaling functions.
Since the spectral function can be defined as the imaginary part of the two-point correlation function, one expects the following scaling form \cite{berges_dynamic_2010,schlichting_spectral_2019,schweitzer_spectral_2020}
\begin{equation}
    \rho \left(\omega, p, \tau\right) =  s^{2-\eta} \rho_0 \Rho \left( s^z \bar \omega, s \bar p, s^{\inv \nu} \tau \right)
    \label{eq:rho_scaling}
\end{equation}
in the limit of small $\omega$, $p$ and $\tau$, where $s$ is a dimensionless scale parameter and $\Rho$ is a universal scaling function with the model-dependent amplitude $\rho_0$.
Since we will only concern ourselves with spectral functions at non-vanishing spatial momentum $p$, finite size effects are negligible, and we may omit any residual dependencies on the finite volume.

We will give our results for the universal scaling functions in terms of dimensionless scaling variables normalized by corresponding non-universal amplitudes. As in (\ref{eq:xit_p}) before, we use $\bar p = f_\xi^+ p $ where $f_\xi^+$ denotes the $\tau>0$ amplitude of the static correlation length $\xi$  listed in \cref{tab:statics}. 
Convenient definitions for dimensionless time and frequency variables turn out to be
\begin{equation}
\bar t = t / f_t
  \;\;\mbox{and}  \;\;\;     \bar \omega = f_t \, \omega   ,
\end{equation}
where $f_t$ is the amplitude of the momentum-dependent auto-correlation time at criticality, $\xi_t(\bar p)=f_t \bar p^{-z} $, cf.~\cref{eq:xit_p}, with the numerical values given in \cref{tab:intxits}.
Note that, in \cite{schweitzer_spectral_2020} the dimensionless time and frequency variables for Models A and C were defined with a normalization $f_{t}^{+}$ determined from the divergence of the auto-correlation time $\xi_{t}(p=0,\tau)=f_{t}^{+} |\tau|^{-\nu z}$ in the high-temperature phase.
However, in the case of Models B and BC with conserved order parameters, the spectral function at zero momentum is trivial, and this amplitude is not readily accessible.
The two amplitudes are however related via a universal ratio, which we call $Q_t^+$.
This is easily seen by writing down the scaling form of the generalized auto-correlation time $\xi_t$ at non-vanishing  $\bar p$ and $\tau$,
\begin{equation}
    \xi_t(\bar p, \tau) = s^{z} \xi_{t}^0 X(s \bar p, s^{1/\nu} \tau),
\end{equation}
again with a model-dependent amplitude $\xi_t^0$ and a universal scaling function $X(x,y)$.
One can then identify the relation between the measured amplitudes and the scaling functions as
\begin{align}
    \xi_t(\bar p, 0) &\equiv \xi_t(\bar p) =  f_t \bar p^{-z} =   \bar p^{-z} \xi_t^0 X(1, 0), \\
    \xi_t(0, \tau) &= f^{\pm}_t |\tau|^{-\nu z} =  |\tau|^{-\nu z} \xi_t^{0} X(0,\pm 1), \\
    &\Rightarrow \quad \frac{f_t}{f_t^{\pm}} = \frac{X(1,0)}{X(0, \pm 1)} \equiv Q^{\pm}_t = \text{univ.}
\end{align}
Re-examining the Model-A/C data, we find that $Q_t^+$ seems to be of order $\sim 1$ there.

The scaling law in \cref{eq:rho_scaling} connotes three alternative dynamic scaling functions, which can be obtained by choosing the scale parameter $s$ to eliminate one parameter dependence at a time.
Explicitly, following~\cite{schweitzer_spectral_2020} one finds
\begin{align}
    \rho\left(\omega, p, \tau\right) &= {\bar \omega}^{-(2-\eta)/z}   \, f_\omega \left(\bar p^z/\bar \omega, \tau/\bar \omega^{1/\nu z} \right) \label{eq:rho_scale_omega},\\
    \rho\left(\omega, p, \tau\right) &= {\bar p}^{-(2-\eta)}\, f_p \left( \bar \omega/\bar p^z, \tau/\bar p^{1/\nu} \right) \label{eq:rho_scale_p},\\
    \rho\left(\omega, p, \tau\right) &= |\tau |^{-\gamma} \, f^\pm_\tau \left( \bar \omega/|\tau |^{\nu z}, \bar p^{1/\nu}/|\tau |\right) \label{eq:rho_scale_tau},
\end{align}
where the scaling functions $f_{\omega}(x_{\omega},y_{\omega})=\rho_0 \Rho(1,x^{1/z}_{\omega},y_{\omega})$, $f_p(x_{p},y_{p})=\rho_{0}\Rho(x_{p},1,y_{p})$ and $f_{\tau}^{\pm}(x_{\tau},y_{\tau})=\rho_{0}  \Rho(x_{\tau},y^{\nu}_{\tau},\pm 1)$ for $\sgn\tau = \pm 1 $ are universal up to the model-dependent amplitude $\rho_0$, and $\gamma$ is the static susceptibility exponent with $\gamma=\nu(2-\eta)$ from static scaling relations.

Evidently, mapping out the full two-dimensional structure of the scaling functions represents a formidable task, and we will therefore follow the strategy of~\cite{schweitzer_spectral_2020} and focus on the behavior along one of the coordinate axes $(x=0~\text{or}~y=0)$.
Since in case of Models B and BC, the spectral function of the order parameter at either zero spatial momentum or zero frequency is trivial,~$\rho(\omega=0,p,\tau)=\rho(\omega,p=0,\tau)=0$, the scaling function $f_{\tau}^{\pm}$ vanishes identically along each of its coordinate axes.
We will thus focus on the functions $f_{\omega}$, $f_p$, which are related to each other via
\begin{align}
    f_p(x_p,y_p) &=  x_p^{-(2-\eta)/z} \, f_\omega\big(1/x_p,y_p/x_p^{1/\nu z}\big) = x_{\omega}^{(2-\eta)/z} f_{\omega}(x_{\omega}, y_{\omega})\, ,\label{eq:fp_of_fw}
\end{align}
where we renamed the parameters as
\begin{align}
    x_{\omega}=\bar p^z / \bar\omega\;,~ x_{p}=\bar{\omega}/\bar{p}^{z}\;,~ y_{\omega}=\tau/\bar{\omega}^{1/\nu z}\;,~ y_{p}=\tau/\bar{p}^{1/\nu}.
\end{align}
Notice that the choice of normalization in combination with \cref{eq:xit_p} lets us interpret the parameters as $x_{\omega}^{-1} = x_p \equiv \omega \xi_t(\bar p)$.

Before we turn to the discussion of the numerical results, it proves insightful to consider some general properties of the scaling functions $f_{p}$ and $f_{\omega}$.
We first note that, since in Models B and BC the order parameter is conserved, the spectral function $\rho(\omega,p=0,\tau)$ vanishes trivially, indicating that $f_{\omega}(0,y_{\omega})=0$ for all values of $y_{\omega}$.
Clearly, this is in contrast to the behavior in Models A and C reported in~\cite{schweitzer_spectral_2020}, where $f_{\omega}(0,0)$ approaches a finite constant.
By considering the Breit-Wigner ansatz for the spectral function in \cref{eq:def_BW}, along with the critical scaling laws of the central frequency $\omega_{p}$ and decay rate $\Gamma_{p}$, we can further determine the expected general shape of the scaling functions.
Based on \cref{eq:omega_scaling,eq:gamma_scaling} with $z_\omega > z_\Gamma $, one finds that at sufficiently low momentum scales $\bar{p} \ll 1$, the central frequency $\omega_{p}$ is much smaller than the decay rate $\Gamma_{p}$, giving rise to two distinct scaling windows for frequencies $\omega \ll \omega_{p}$ and $\omega_{p} \ll \omega \ll \Gamma_{p}$.
We first focus on the low frequency behavior $\omega \ll \omega_{p}$, where the form in \cref{eq:def_BW} predicts for spectral function  approaches the limit
\begin{equation}
    \lim_{\omega \to 0}\rho(\omega,p,\tau=0)= \frac{\mu p^2\Gamma_{p}}{\omega_{p}^4}~\omega\;.
\end{equation}
Using this limit together with the low momentum behavior of $\omega_{p}$ and $\Gamma_{p}$ in \cref{eq:omega_scaling,eq:gamma_scaling} to evaluate the left-hand side of \cref{eq:rho_scale_omega,eq:rho_scale_p} then gives rise to the following behavior of the scaling functions at criticality ($y_{p}=y_{\omega}=0$) for small $x_{p}$ and large $x_{\omega}$
\begin{align}
    \label{eq:fomega_bw_D_xinf}
    f_{\omega}(x_{\omega} \gg 1,0) &= \bar{p}^{4-\eta-2z_{\omega}+z+z_{\Gamma}} \frac{\Gamma_0}{(f_{\xi}^{+})^2 f_{t}\omega_0^{4}}\,  x_{\omega}^{-1-(2-\eta)/z}\;, \\
    \label{eq:fp_bw_D_x0}
    f_{p}(x_{p}\ll 1,0)&=\bar{p}^{4-\eta-2z_{\omega}+z+z_{\Gamma}} \frac{\Gamma_0}{(f_{\xi}^{+})^2 f_{t}\omega_0^{4}}\,  x_{p}\;.
\end{align}
where the additional factors of $f_{\xi}^{+}$ and $f_{t}$ originate from re-expressing $p=\bar{p}/f_{\xi}^{+}$ and $\omega=\bar{\omega}/f_{t}$. Because, by definition, the scaling functions $f_{\omega}$ and  $f_{p}$ dependent on the momentum variable $\bar{p}$ only implicitly, here via the scaling variables  $x_{\omega}$ and $x_{p}$, for the relations in \cref{eq:fomega_bw_D_xinf,eq:fomega_bw_D_x0} to be compatible with this definition, the common exponent of $\bar p$ in the prefactors must vanish, i.e.
\begin{equation}
      {4-\eta-2z_{\omega}+z+z_{\Gamma}} = 0 \, . \label{eq:screl1}
\end{equation}
Conversely, to analyze the dynamic critical behavior for large $x_{p}$ or small $x_{\omega}$, one needs to consider the intermediate range of frequencies $\omega_{p} \ll \omega \ll \Gamma_{p}$, which can be formally obtained by taking the limit $\Gamma_{p} \to \infty$ of the Breit-Wigner spectral function, yielding
\begin{equation}
    \lim_{\Gamma_p \to \infty} \rho(\omega,p,\tau=0)=\frac{\mu p^2}{\omega \Gamma_p}\;.
\end{equation}
By again using \cref{eq:rho_scale_omega,eq:rho_scale_p,eq:omega_scaling,eq:gamma_scaling}, this gives now rise to the following behavior of the scaling function
\begin{align}
    \label{eq:fomega_bw_D_x0}
    f_{\omega}(x_{\omega} \ll 1,0) &= \bar{p}^{4-\eta-z_{\Gamma}-z} \frac{f_{t}}{(f_{\xi}^{+})^2\Gamma_0}\,  x_{\omega}^{1-(2-\eta)/z}\;, \\
    \label{eq:fp_bw_D_xinf}
    f_{p}(x_{p} \gg 1,0) &= \bar{p}^{4-\eta-z_{\Gamma}-z} \frac{f_{t}}{(f_{\xi}^{+})^2\Gamma_0} \, x_{p}^{-1}\;, 
\end{align}
and the exponent of $\bar p $ in the prefactors must vanish by the same argument as above again, which now entails
\begin{equation}
     {4-\eta-z_{\Gamma}-z} = 0 \, . \label{eq:screl2}
\end{equation}
The Breit-Wigner shape \cref{eq:def_BW} together with the critical scaling laws \cref{eq:omega_scaling,eq:gamma_scaling}
 from \cref{eq:screl2} therefore predicts $z=4-\eta-z_{\Gamma}$ and with this in \cref{eq:screl1}  $z_{\omega}=4-\eta $ as we observed numerically in  \cref{sec:crit_sf}. 
Specifically, for Model B, where the decay width is constant in the infrared (hence $z_{\Gamma}=0$), one obtains the standard result $z=z_{\omega}=4-\eta$, whereas for our Model BC, where the $\Gamma_p$ exhibits a non-trivial momentum dependence ($z_{\Gamma} >  0$), the dynamic critical exponent $z$ is instead determined by $z=4-\eta-z_{\Gamma}$ which is smaller than the Model B value by precisely the value of the additional exponent $z_\Gamma$ of the momentum-dependent width $\Gamma_p$.

\begin{figure}
    \centering
    \renewcommand{\fdir}{figures/rhouni}
    \includegraphics{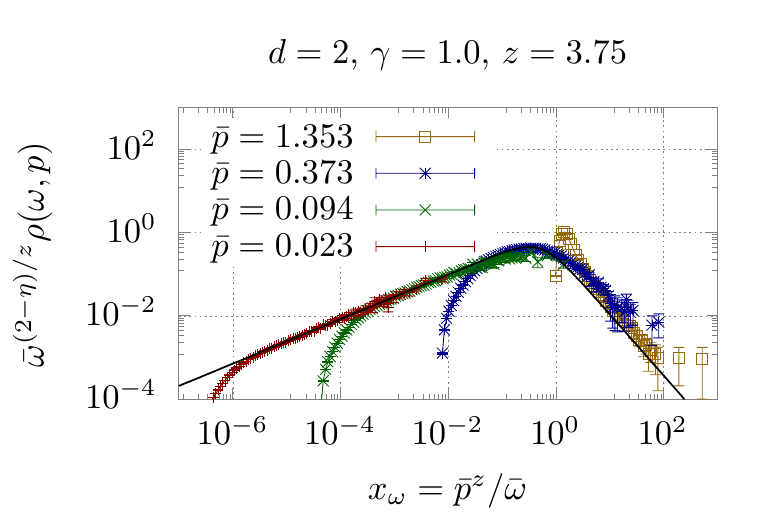}
    \includegraphics{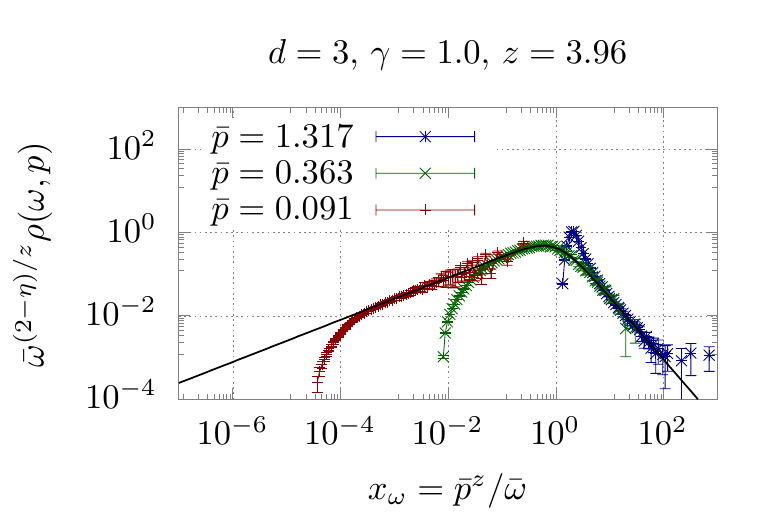}
    \includegraphics{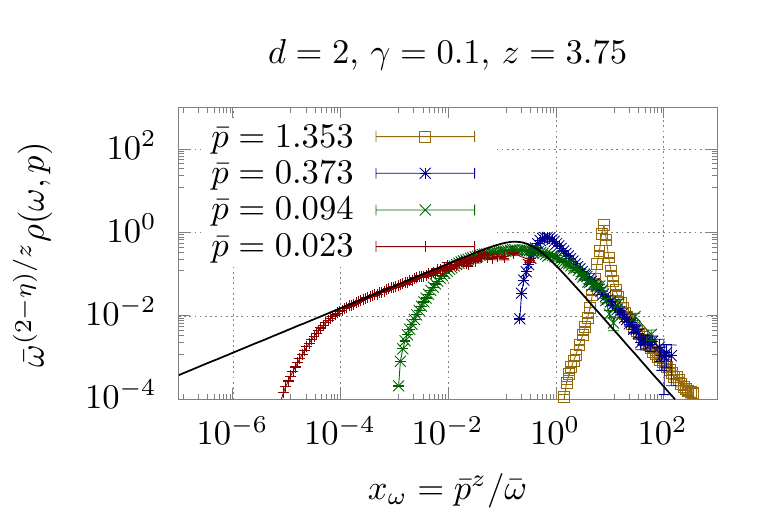}
    \includegraphics{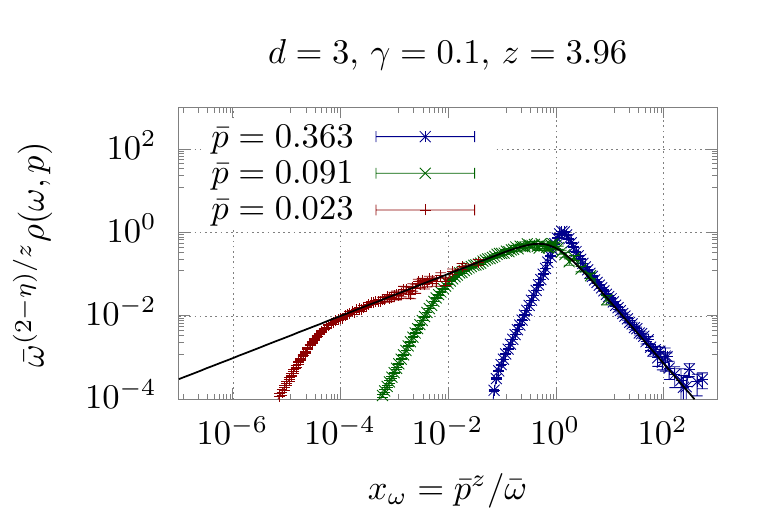}
    \includegraphics{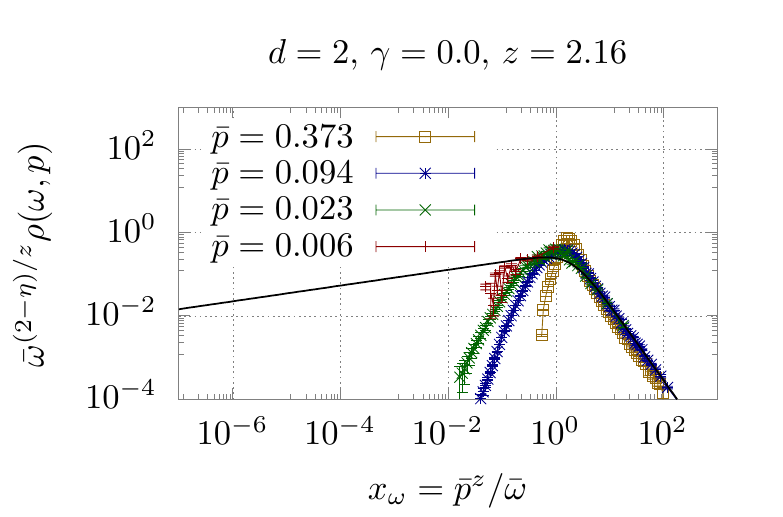}
    \includegraphics{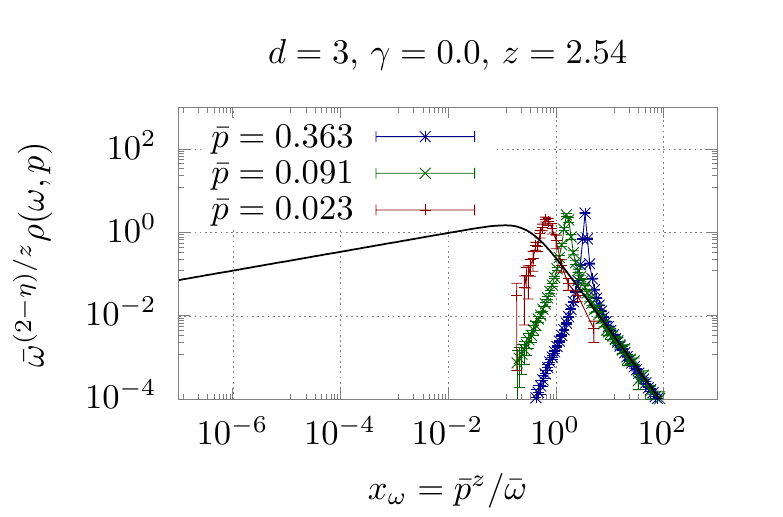}
    \caption{%
        Rescaled critical spectral functions at fixed momenta.
        Regions of overlapping data indicate the approach towards the universal scaling function $f_\omega(x_\omega, 0)$.
        Due to availability of data, we use $\tau=0.0009(2)$ ($d=2$) resp.~$\tau = 0.00008(5)$ (d=3) as proxy for the critical temperature.
        The data was obtained on lattices of size $1024^2$ respectively $256^3$.
        Shown as a solid line is the scaling function $f_{\omega}(x_\omega, 0)$ obtained by applying the relation \cref{eq:fp_of_fw} to the ansatz for $f_p(x_p, 0)$ in \cref{eq:fp}.
        Note that the parameters are taken from fits to the data in \cref{fig:intxits_bd,fig:bwplots_diff_tc}.
        For 3+1D, the agreement between data and the ansatz is excellent for $\gamma>0$.
        In 2+1D, only the limit $x_\omega \to 0$ agrees with the data, while there are discrepancies in the other limit for $\gamma>0$.
        At $\gamma=0$ in both 2+1D and 3+1D however, in the large-$x_\omega$ limit data and ansatz agree very well, but one needs to go to much lower spatial momenta to be able to observe the expected scaling behaviour for small $x_\omega$.
    }
    \label{fig:fomega_modelbd}
\end{figure}
\begin{figure}
    \centering
    \renewcommand{\fdir}{figures/rhouni}
    \includegraphics{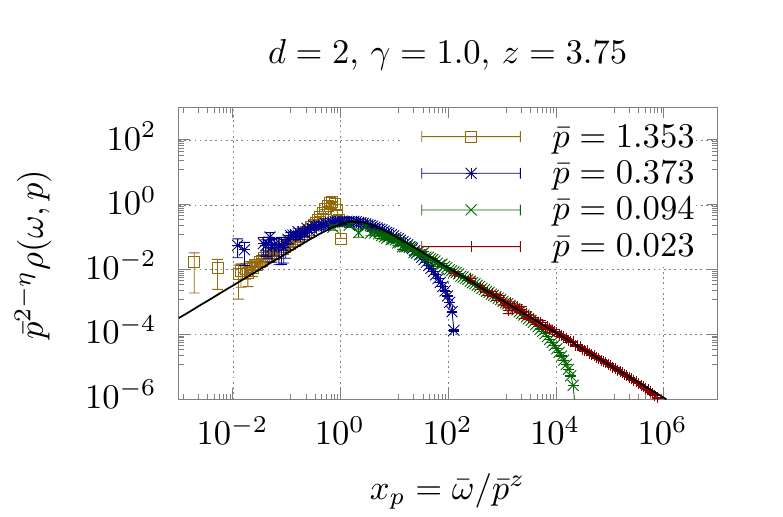}
    \includegraphics{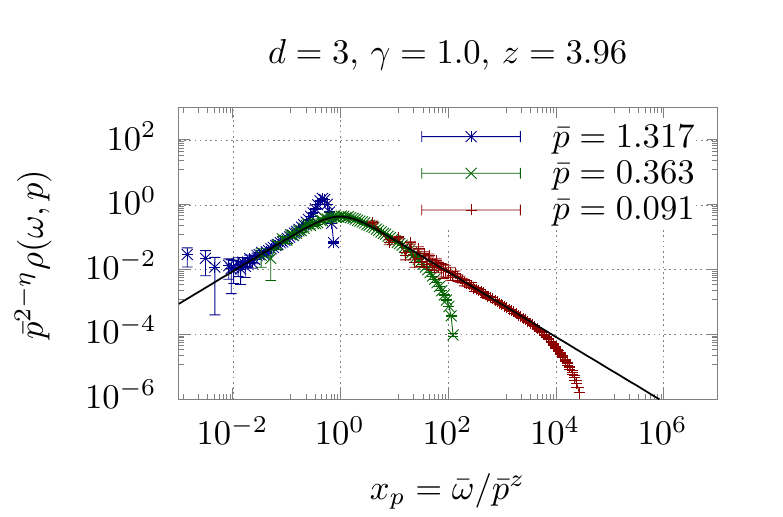}
    \includegraphics{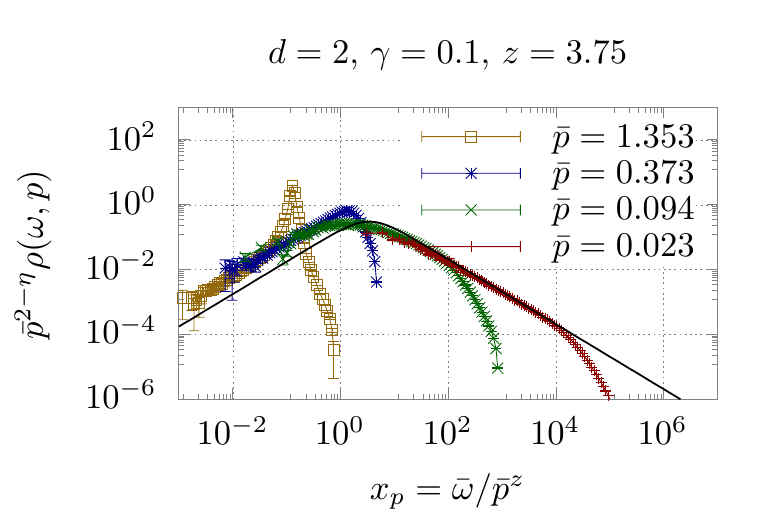}
    \includegraphics{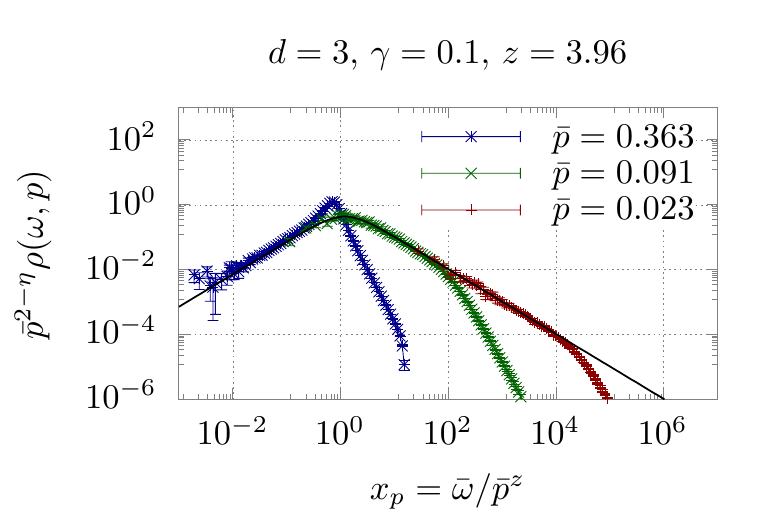}
    \includegraphics{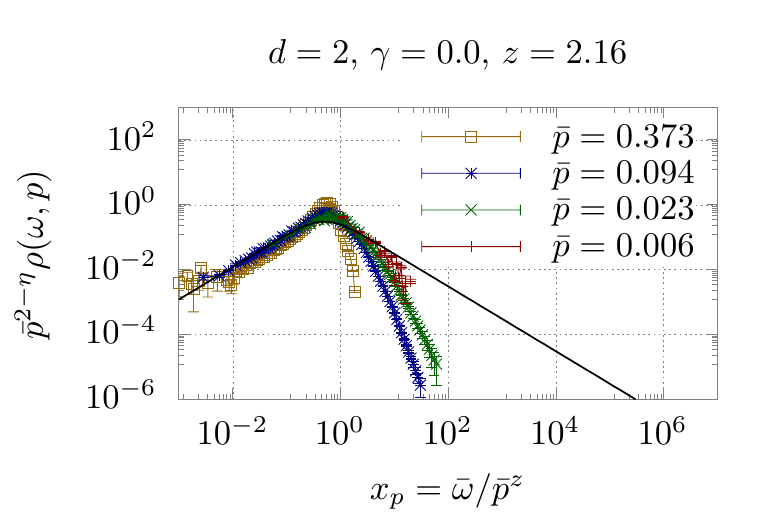}
    \includegraphics{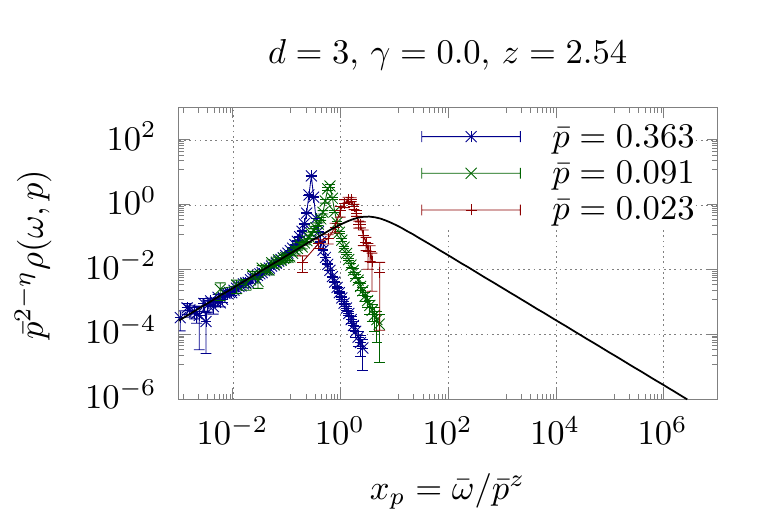}
    \caption{%
        Rescaled fixed-momentum cuts of spectral functions at the critical temperature.
        Regions of overlapping data indicate the universal scaling function $f_p(x_p, 0)$.
        The data sets and parameters are the same as in \cref{fig:fomega_modelbd}.
        Shown as a solid line is the scaling function $f_p(x_p, 0)$ in \cref{eq:fp}.
        This presentation allows to easily identify small- and large-$x_p$ arms of the scaling function $f_p(x_p, 0)$ with the limits of small frequencies $\omega \ll \omega_p$ resp.~large decay rates $\omega_p \ll \omega \ll \Gamma_p$ of the Breit-Wigner functions.
        Deviations observed at large $x_{\omega}$ in \cref{fig:fomega_modelbd} manifest themselves here at small $x_p$.
    }
    \label{fig:fp_modelbd}
\end{figure}

Now that we have established the limiting behavior of the scaling functions, we turn to the analysis of our numerical data from classical-statistical simulations.
By rescaling the critical spectral functions $\rho(\omega,p,\tau=0)$ at different momenta $p$ with the appropriate powers of $\bar\omega $ and $\bar p $ to compensate the explicit scaling factors in \cref{eq:rho_scale_omega,eq:rho_scale_p} and plotting over the scaling variables $x_\omega = \bar p^z/\bar\omega $ and $x_p = \bar\omega/\bar p^z$, we obtain the curves shown in \cref{fig:fp_modelbd,fig:fomega_modelbd}. Regions of overlapping data points then reveal the underlying scaling functions $f_{\omega}(x_{\omega},0)$ and $f_{p}(x_p,0)$.
For  dynamic critical exponent $z$ required to scale both axes in
 \cref{fig:fp_modelbd,fig:fomega_modelbd} we employ $z = 4-\eta$ for the diffusive Model B and $z=4-\eta-z_{\Gamma}$ for the conservative Model BC as discussed above. The respective values used for $z$ are given in the title of each plot.
By comparing the results for different momenta $p$, one observes an excellent scaling collapse of the data obtained with finite coupling to the heat bath $\gamma \in \left\{ 0.1,1.0 \right\}$ (Model B), in both 2+1D and 3+1D.
Conversely, for vanishing heat bath coupling $\gamma=0$ (Model BC), the range of spatial momenta where we observe critical scaling is much narrower, and we can only recognize hints of the onset of a critical scaling behavior for very small momenta in 2+1D.

By closer inspection of \cref{fig:fp_modelbd} and comparison to \cref{eq:fp_bw_D_x0,eq:fp_bw_D_xinf}, one finds that the low- and high-$x_{p}$ tails of the scaling curve $f_{p}$ in correspond to the range of frequencies $\omega \ll \omega_{p}$ and $\omega_{p} \ll \omega \ll \Gamma_{p}$ of the spectral functions in \cref{fig:fp_modelbd}, which exhibit critical scaling behavior.
Conversely, the high-frequency tails $\omega \gg \Gamma_{p}$, where the spectral function behaves approximately as $\lim_{\omega\to\infty} \rho(\omega) = p^2\Gamma_{p}/\omega^3$, do not show dynamic critical behavior, leading to sizeable deviations from the scaling curves at large $x_{p}$ (small $x_{\omega}$) for larger spatial momenta.
We therefore conclude that the scaling window is limited to the range of frequencies $\omega \lesssim \Gamma_{p}$ smaller than the decay width.
Since for the dissipative dynamics of Model B the decay width $\Gamma_{p}$ of infrared modes is determined by the Langevin damping $\gamma$, the scaling window is comparatively large, whereas for the non-disspative dynamics of the Model BC ($\gamma=0$) the decay width $\Gamma_{p}=\Gamma_0 \bar{p}^{z_{\Gamma}}$ decreases rapidly as a function of momentum, resulting in a much narrower scaling window. 

Based on our previous analysis, we expect that the limiting behavior of the scaling functions at small and larger arguments is determined by \cref{eq:fomega_bw_D_xinf,eq:fp_bw_D_x0,eq:fomega_bw_D_x0,eq:fp_bw_D_xinf}.
To include the interpolating region where $x\simeq 1$ we make the following ansatz, for example, first for the scaling function $f_p $, which is inspired by the Breit-Wigner form in \cref{eq:def_BW},
\begin{align}
    f_{p}(x_{p},0) &= \frac{(f_{\xi}^+)^{-2}}{\omega_0^4 \left( \frac{\Gamma_0}{f_t} x_p \right)^{-1} + \frac{\Gamma_0}{f_t} x_p} . \label{eq:fp}
\end{align}
This combines the expected limits \cref{eq:fp_bw_D_x0,eq:fp_bw_D_xinf} in an inverse sum.
If the spectral functions are given by broad Breit-Wigner functions and the decay rates $\Gamma_p \gg 2\omega_p$ are much larger than the central frequencies, then using the expression $\xi_t(p) = \Gamma_p/\omega_p^2$ for the auto-correlation time one can relate the relevant amplitudes via $\omega_0^2 = \Gamma_0/f_t$.
In that case, the scaling function becomes
\begin{align}
    f_p(x_p,0) &= \frac{1}{(f_{\xi}^+ \omega_0)^2} \frac{1}{x_p^{-1} + x_p}.
    \label{eq:bw_scaling}
\end{align}
Comparing to $f_p(x_p, 0) = \rho_0 \Rho(x_p, 1, 0)$, we can thus separate the model-dependent amplitude from the \emph{universal} scaling function and obtain
\begin{align}
    \rho_0 \equiv \frac{1}{(f_{\xi}^+ \omega_0)^2},\;\; \mbox{and} \;\;\;  \Rho(x_p, 1, 0) \equiv \frac{1}{x_p^{-1} + x_p}. \label{eq:unifunc}
\end{align}
Note that universal here means that \eqref{eq:unifunc} describes the functional dependence of the scaling function in all models of the same dynamic universality class, not only those whose spectral functions keep their Breit-Wigner shape exactly.
Model dependencies apart from the constant $\rho_0$ are then hidden in the scaling variable $x_p$, respectively the normalizations of $\bar \omega$ and $\bar p$.

Since especially in 2+1D we are in a region of spatial momenta where $\Gamma_p \gg 2\omega_p$ is not necessarily given, we do not use \eqref{eq:bw_scaling}, but rather the form in \cref{eq:fp} with $f_t$, $\Gamma_0$ and $\omega_0$ as independent parameters, for which we use the results as extracted from fits to the data in \cref{fig:intxits_bd,fig:bwplots_diff_tc}.

To derive the corresponding form for the scaling function $f_{\omega}$ it is best to use \cref{eq:fp_of_fw} which allows to derive $f_\omega  $ from $f_p$. With \cref{eq:bw_scaling} and $x_\omega = x_p^{-1}$ this immediately yields,
\begin{align}
    f_\omega(x_\omega,0) &=  \rho_0 \Rho(1,x_\omega^{1/z},0) =  \frac{\rho_0}{x_\omega^{-1} + x_\omega} \, x_\omega^{-(2-\eta)/z} \, , 
\end{align}
and then combines the limits in \cref{eq:fomega_bw_D_xinf,eq:fomega_bw_D_x0}.
The  curves obtained for $f_{\omega}(x_\omega,0) $ and $f_p(x_p, 0)$ in this way are plotted as the solid lines in \cref{fig:fomega_modelbd,fig:fp_modelbd}.

This rather simple ansatz for the dynamic scaling functions describes the data for the systems with finite Langevin coupling $\gamma>0$ exceptionally well, capturing not only the limits of small and large $x$ nearly perfectly, but also the transition region at the intermediate  $x\sim 1$.
For $\gamma=0$, the small-$x_p$/large-$x_\omega$ behaviour is also nicely described by the dynamic scaling functions based on the Ansatz \eqref{eq:fp}, but one obviously has to compute the critical spectral functions at much smaller spatial momenta to verify their asymptotic large-$x_p$/small-$x_\omega$ behaviour.
In 2+1D at $\gamma=0.1$, we observe a deviation from the scaling function for small $x_p$ resp.~large $x_\omega$.
This is most probably related to the small but significant deviations from the Breit-Wigner shape we already saw in \cref{fig:fit_comparison_BD_tc}, and which also lead to inaccuracies when trying to extract the central frequencies in \cref{fig:bwplots_diff_tc}.
It is also possible that the universal scaling function simply does not approach its limiting behaviour fast enough for $x_p < 1$, and the deviations we observe are just the result of a less simple functional form of the scaling function at the intermediate $x_p\sim 1$.
We observed a similar phenomenon in \cite{schweitzer_spectral_2020} when extracting the scaling function $f_\tau$, encoding the temperature dependence of the spectral function at vanishing spatial momentum.

\section{Conclusion and outlook}\label{sec:conclusion}

We have studied the critical dynamics of relativistic diffusion, by performing classical-statistical simulations of an Israel-Stewart type equation for self-interacting scalar fields in 2+1 and 3+1 space-time dimensions. Close to criticality, we observed a divergent auto-correlation time, which allows us to extract estimates of the dynamic critical exponents $z$. Based on the classification of Halperin and Hohenberg for non-relativistic models, one expects $z=4-\eta$ for models with a dynamically conserved order parameter, irrespective of the presence (Model D) or absence (Model B) of energy conservation. While our simulation results for the dissipative dynamics of the relativistic Model B are in excellent agreement with $z=4-\eta$, the non-dissipative limit of the Israel-Stuart type diffusion equation is realized non-trivially featuring propagating rather than diffusive behavior at tree level. Due to the absence of tree level dissipation, a new infrared power law arises in the momentum dependence of the (thermal) damping rate $\Gamma_{p} \sim p^{z_{\Gamma}}$ of the conservative (Model BC) limit of the relativistic Model B, which leads to a significant decrease of the dynamic critical exponent $z = 4-\eta - z_{\Gamma}$ yielding $z \in [2,2.5]$.

By studying the critical behavior of spectral functions, we have demonstrated that, in the vicinity of the critical point, the spectral function can be described in terms of universal scaling functions~\cite{schweitzer_spectral_2020}, which we determined from our numerical simulations. We observed that even at the critical point, the shape of spectral function stays close to the mean-field Breit-Wigner form, while dispersion relations and thermal damping rates exhibit a power law dependence on the momentum. Based on this result, we obtained additional analytical insights into the universal scaling functions, as we derived the critical scaling function of Breit-Wigner spectral functions under the given constraints for central frequencies and decay widths. Noteably, the same calculation also provided an expression for the dynamic critical exponent $z$ of the relativistic Models B and BC, which is compatible with our results obtained from the divergence of the autocorrelation time.

While our current study focused on the dynamic critical behavior of relativistic diffusion in thermal equilibrium, our framework can easily be extended to investigate non-equilibrium phenomena in the vicinity of a second order phase transition, e.g.~by introducing time-dependent control parameters such as temperature and external fields.
Especially in the context of the search for the QCD critical point, non-equilibrium effects are expected to become highly relevant \cite{berdnikov_slowing_2000,mukherjee_real_2015,mukherjee_universal_2016,mukherjee_universality_2017}. This has been modeled, for example, in \cite{nahrgang_diffusive_2019} where the authors solve a fluid-dynamical diffusion equation with a white noise stochastic current in 1+1D, and find evidence of critical slowing-down and non-equilibrium effects on non-Gaussian cumulants. While these studies aim to implement effective descriptions of the QCD phase transition, the universal aspects of such non-equilibrium phase transitions can also be studied within the microscopic dynamical theories developed in this work. It might also be possible to extend our model to include a coupling to a conserved transverse vector field, replicating the shear modes required for the dynamics of Model H.
This could allow us to numerically investigate the dynamic critical behaviour of QCD matter, which would be of great interest in the search for the location of the chiral end point.

\section*{Acknowledgements}

We thank G.~D.~Moore and O.~Kaczmarek for valuable discussions.
This work was supported by the Deutsche Forschungsgemeinschaft (DFG) through the grant CRC-TR 211 ``Strong-interaction matter under extreme conditions.'' 
D.S.~also received some financial support from the European Union's Horizon 2020 research and innovation programme under grant agreement STRONG -- 2020 - No.~824093. 
The computations in this work were performed on the GPU clusters at JLU Giessen and Bielefeld University, and we thank the Bielefeld HPC.NRW team for their support.

\clearpage

\begin{appendices}
    \section{Breit-Wigner autocorrelation time}\label{app:xit_BW}
Starting with the Breit-Wigner ansatz for the shape of the spectral function \eqref{eq:def_BW}, we find
\begin{align}
    &\int_{0}^{\infty} t \rho(t) \intd t \overset{\rho(t)\text{ odd}}{=} \frac{1}{2} \int_{-\infty}^{\infty} t \rho(t) \intd t = \frac{1}{2} \left. \frac{\intd}{\intd \omega} \rho(\omega) \right|_{\omega=0} = \frac{\Gamma}{2 \omega_p^4},
\end{align}
reminding ourselves that our definition of $\rho(\omega)$ includes an extra factor of $-\iu$ (compare \cref{eq:def_FT_rho}).
For the second integral, we again apply a Fourier transformation to get
\begin{align}
    \int_{0}^{\infty} \rho(t) \intd t &= \int_{-\infty}^{\infty} \Theta(t) \rho(t) \intd t = -\iu \left.\int \intd \omega' \mathcal{F}\left[ \Theta(t) \right](\omega') \rho(\omega - \omega')\right|_{\omega=0} \\
    &= \int \intd \omega' \left( \frac{1}{2\pi \omega'} + \iu\pi\delta(-\omega') \right) \rho(\omega'),
\end{align}
but since the spectral function vanishes at the origin $\rho(\omega = 0) = 0$ due to its symmetry, the $\delta$-term does not contribute.
For the remaining term, we employ the residue theorem.
We note that the function
\begin{equation}
    \rho_{\text{BW}}(\omega, p)/\omega = \frac{\Gamma}{\left( \omega^2 - \omega_p^2 \right)^2 + \Gamma^2\omega^2}
\end{equation}
has four first-order poles in $\omega$, namely
\begin{align}
    \omega_{R}^2 &= - \frac{\Gamma^2 - 2\omega_p^2}{2} \pm \sqrt{\left( \frac{\Gamma^2 - 2\omega_p^2}{2} \right)^2 - \omega_p^4} \\
    &= -\frac{1}{2} \left( A \pm B \right),
\end{align}
where we abbreviate $A \equiv \Gamma^2 - 2\omega_p^2$, $B \equiv \Gamma\sqrt{\Gamma^2 - 4\omega_p^2}$.
One can thus express
\begin{align}
    \frac{\rho_{\text{BW}}(\omega)}{\Gamma \omega} &= \left[ \left( \omega^2 + \frac{1}{2}(A+B) \right)\left( \omega^2 + \frac{1}{2}\left( A-B \right) \right) \right]^{-1} \\
    &= \left[ \left( \omega + \frac{\iu}{\sqrt{2}}(A+B)^{1/2} \right)\left( \omega - \frac{\iu}{\sqrt{2}}\left( A+B \right)^{1/2} \right)\left( \omega + \frac{\iu}{\sqrt{2}}\left( A-B \right)^{1/2} \right)\left( \omega - \frac{\iu}{\sqrt{2}}\left( A-B \right)^{1/2} \right) \right]^{-1} \\
    &\equiv \left[ \left( \omega - \omega_1\right) \left( \omega - \omega_2 \right) \left( \omega - \omega_3 \right) \left( \omega - \omega_4 \right) \right]^{-1},
\end{align}
where of course $\omega_1 = -\omega_2$ and $\omega_3=-\omega_4$.
In order to obtain the locations of these poles on the complex plane, we consider the dependence of $A$ and $B$ on $\Gamma$ and $\omega_p$.
For the critical spectral function in the scaling regime, we find in \cref{sec:crit_sf} that for finite Langevin coupling $\gamma$ one always has $\Gamma \geq \gamma > 2\omega_p$ (case 1) for sufficiently small spatial momentum, since the central frequencies vanish with $p\to 0$.
However, given $\gamma=0$, one finds $\Gamma > 2\omega_p$ (case 2) over a wide range of parameters, even deep into the infrared.

For $A$, $B$ we have in thos cases
\begin{align}
    \text{case 1}:\;\Gamma > 2\omega_p > 0:\quad&\follows A > B > 0, \quad&\follows&\quad \Im(\omega_2)>0,\, \Im(\omega_4) > 0,\\
    \text{case 2}:\;2\omega_p > \Gamma > 0:\quad&\follows A < 0, \Re(B) = 0,\,\Im(B) > 0\quad&\follows&\quad \Im(\omega_2)>0,\, \Im(\omega_3) > 0.
\end{align}

We choose to complete the integration contour by a semi-circle over the positive half plane, where always two of the poles lie.
For the residues one finds
\begin{align}
    \mathrm{Res}(\omega_2) &= \left[ \left( \omega_2-\omega_1 \right)\left( \omega_2 - \omega_3 \right)\left( \omega_{2}-\omega_4 \right) \right]^{-1} \\
    &= \left[ \sqrt{2}\iu\left( A+B \right)^{1/2} \left(\frac{-1}{2}(A+B) + \frac{1}{2}(A-B)  \right) \right]^{-1} \\
    &= \left[ -\sqrt{2}\iu\left( A+B \right)^{1/2} B \right]^{-1} = -\mathrm{Res}(\omega_1), \\
    \mathrm{Res}(\omega_4) &= \left[ \sqrt{2}\iu\left( A-B \right)^{1/2} B \right]^{-1} = - \mathrm{Res}(\omega_3)
\end{align}
In general, case 1 best matches the physical reality in the infrared, and thus have for the integral.
\begin{align}
    \int \intd\omega \frac{\rho_{\text{BW}}(\omega)}{2\pi\omega} &= \iu \Gamma \left( \text{Res}_{\omega_2} + \text{Res}_{\omega_4} \right) \\
    &= \iu \Gamma \left( \frac{\sqrt{2}\iu}{2B} \frac{\sqrt{A-B} - \sqrt{A+B}}{\sqrt{A^2 - B^2}}\right) \label{eq:sqrt_expr}
\end{align}
If the decay width $\Gamma$ is e.g.~bounded from below by the Langevin coupling $\gamma$, we find using $\omega_p \ll \Gamma$
\begin{align}
    \int \intd\omega \frac{\rho_{\text{BW}}(\omega)}{2\pi \omega} &\approx \iu \Gamma\left( \frac{\sqrt{2}\iu}{2B} \frac{- \sqrt{2B}}{2\omega_p^2}\right)
    \approx \frac{1}{2\Gamma} \frac{\Gamma}{\omega_p^2} = \frac{1}{2\omega_p^2}.
\end{align}
where we used that $A^2 - B^2 = 4\omega_p^4$ and, for $\Gamma\gg \omega_p$ one has $A\approx B \approx \Gamma^2$.
This implies for the autocorrelation time the relation
\begin{equation}
    \xi_{t,\text{BW}} = \frac{\int_{0}^{\infty} t \rho(t) \intd t}{\int_{0}^{\infty} \rho(t) \intd t} = \frac{\Gamma}{\omega_p^2}.
    \label{eq:xit_BW}
\end{equation}


    \section{Hydrodynamic Green's functions}\label{sec:israel-stewart}
We compute in the following the propagator of the field evolving under Israel-Stewart hydrodynamics \cite{israel_transient_1979,israel_thermodynamics_1981}.
Starting point is the conservation law and the definition of the current
\begin{align}
    \partial_{\mu}J^{\mu}=0\;, \label{eq:jcons} \quad
    J^{\mu}=\phi u^{\mu} +\nu_{\mu},
\end{align}
such that $\phi=J^{\mu}u_{\mu}$ and $\nu^{\mu}=\Delta^{\mu}_{~\nu}J^{\nu}$ with $\Delta^{\mu\nu}=g^{\mu\nu}-u^{\mu}u^{\nu}$ and metric convention $g^{\mu\nu}=\mathrm{diag}(+,-,-,-)$.  Without loss of generality the evolution equation then takes the form
\begin{equation}
    D_{\tau} \phi + \theta \phi = -\nabla_{\mu} \nu^{\mu}\;,
\end{equation}
where $\nabla^{\mu}=\Delta^{\mu\nu}\partial_{\nu}$ denotes the transverse and $D_{\tau} = u^{\mu}\partial_{\mu}$ the longitudinal derivative, and $\theta = \partial_{\mu}u^{\mu}$ the expansion rate.
In Israel-Stewart hydrodynamics \cite{israel_transient_1979,israel_thermodynamics_1981}, the dissipative field obeys the equation of motion
\begin{equation}
   \Delta^{\mu\nu}  D_{\tau} \nu_{\nu} = -\frac{1}{\tau_R}(\nu^{\mu}-\nu^{\mu}_{\rm NS})\;,
    \label{eq:IS_eom}
\end{equation}
relaxing to the Navier-Stokes limit $\nu^{\mu}_{\rm NS}=D\nabla^{\mu}\phi$ with relaxation time $\tau_R$ and diffusion rate $D$.
Thus, in the limit of \emph{vanishing} relaxation times $\tau_R$ for a static fluid $u^{\mu}={\rm const.}$ and thus $\theta=0$, the evolution equation takes the form of a simple diffusion equation
\begin{equation}
    \label{eq:diff}
    D_{\tau} \phi = D  \Delta \phi\;,
\end{equation}
where $\Delta=-\nabla_{\mu}\nabla^{\mu}$ is the transverse Laplacian.

In the following, we operate under the assumption of $u^{\mu} = (1,0,0,0) = {\rm const.}$~to facilitate notation.
We define the Laplace transform of the field $\phi(\vec x, t)$ as 
\begin{equation}
    \phi({\bf k}, z)=\int_{0}^{\infty}dt e^{\iu zt} \int d^d {\bf x} e^{-i {\bf k}{\bf x }} \phi(\vec x, t)\;,
\end{equation}
and remark that under this transformation, the time derivative transforms as $\dot \phi(t, \vec x) \to -\iu z \phi(z, \vec k) - \phi(t=0, \vec k)$.
Abbreviating the longitudinal components of the dissipative currents as $\nu_{\|}=\nabla_{\mu} \nu^{\mu}$, we find that the constituitive equations \labelcref{eq:jcons,eq:IS_eom} transform as
\begin{align}
    -\iu z \phi({\bf k}, z)+ \nu_{\|}({\bf k}, z) =& \phi({\bf k}, t=0)\;, \\
    -\iu z \tau_R \nu_{\|}({\bf k}, z) =& \tau_{R} \nu_{\|}({\bf k}, t=0) - (\nu_{\|}({\bf k}, z) - D{\bf k}^2 \phi({\bf k}, z)) \;.
\end{align}
Solving for the Laplace transform of the field, we obtain
\begin{equation}
    \phi({\bf k}, z)= \frac{-\tau_R \nu_{\|}({\bf k}, t=0)}{(-\iu z+D{\bf k}^2)(1-\iu z \tau_R)} + \frac{(1-\iu z\tau_R)\phi({\bf k}, t=0)}{(-\iu z(1-\iu z\tau_R)+D{\bf k}^2)}.
\end{equation}
If we further assume that the initial conditions are uncorrelated ($\braket{ \nu_{\|}({\bf k}, t=0) \phi({\bf k}, t=0)} = 0$), we find the retarded propagator
\begin{align}
    G(\vec k, z) &\equiv \int \intd t e^{\iu zt} \int \intd^{d}\vec x e^{-\iu \vec k \vec x} \Theta(t) \Braket{\phi(\vec x, t)\phi({\bf 0}, 0)} \\
    &= \frac{(1-\iu z \tau_R)\chi(\vec k)}{D \vec k^2 - \tau_R z^2 - \iu z}
    \label{eq:IS_propagator}
\end{align}
with the static susceptibility $\chi(\vec k) \equiv \braket{\phi(\vec k, t=0)\phi(-\vec k, t=0)}$.
The two-point function has poles at
\begin{eqnarray}
    z=\frac{-\iu}{2\tau_R} \pm \frac{\iu}{2\tau_R} \sqrt{1-4D {\bf k}^2\tau_R}\;.
\end{eqnarray}
In the limit of small spatial momentum ${\bf k}\to0$, we recover Navier-Stokes dynamics plus an additional non-hydrodynamic mode
\begin{eqnarray}
    z_{\rm hydro}=-\iu D{\bf k}^2\;, \qquad z_{\rm non-hydro}=-\frac{\iu}{\tau_R}\;.
\end{eqnarray}


\end{appendices}

\bibliographystyle{elsarticle-num}
\bibliography{library}

\begin{thebibliography}{10}
\expandafter\ifx\csname url\endcsname\relax
  \def\url#1{\texttt{#1}}\fi
\expandafter\ifx\csname urlprefix\endcsname\relax\def\urlprefix{URL }\fi
\expandafter\ifx\csname href\endcsname\relax
  \def\href#1#2{#2} \def\path#1{#1}\fi

\bibitem{dunlavy_critical_2005}
M.~J. Dunlavy, D.~Venus,
  \href{https://link.aps.org/doi/10.1103/PhysRevB.71.144406}{Critical slowing
  down in the two-dimensional {Ising} model measured using ferromagnetic
  ultrathin films}, Physical Review B 71~(14) (2005) 144406.
\newblock \href {https://doi.org/10.1103/PhysRevB.71.144406}
  {\path{doi:10.1103/PhysRevB.71.144406}}.
\newline\urlprefix\url{https://link.aps.org/doi/10.1103/PhysRevB.71.144406}

\bibitem{honerkamp-smith_experimental_2012}
A.~R. Honerkamp-Smith, B.~B. Machta, S.~L. Keller,
  \href{http://arxiv.org/abs/1104.2613}{Experimental observations of dynamic
  critical phenomena in a lipid membrane}, arXiv:1104.2613 [cond-mat,
  physics:physics, q-bio]ArXiv: 1104.2613 (May 2012).
\newblock \href {https://doi.org/10.1103/PhysRevLett.108.265702}
  {\path{doi:10.1103/PhysRevLett.108.265702}}.
\newline\urlprefix\url{http://arxiv.org/abs/1104.2613}

\bibitem{stephanov_signatures_1998}
M.~Stephanov, K.~Rajagopal, E.~Shuryak,
  \href{http://arxiv.org/abs/hep-ph/9806219}{Signatures of the {Tricritical}
  {Point} in {QCD}}, Physical Review Letters 81~(22) (1998) 4816--4819, arXiv:
  hep-ph/9806219.
\newblock \href {https://doi.org/10.1103/PhysRevLett.81.4816}
  {\path{doi:10.1103/PhysRevLett.81.4816}}.
\newline\urlprefix\url{http://arxiv.org/abs/hep-ph/9806219}

\bibitem{rajagopal_condensed_2001}
K.~Rajagopal, F.~Wilczek, \href{http://arxiv.org/abs/hep-ph/0011333}{The
  {Condensed} {Matter} {Physics} of {QCD}}, arXiv:hep-ph/0011333 (2001)
  2061--2151ArXiv: hep-ph/0011333.
\newblock \href {https://doi.org/10.1142/9789812810458_0043}
  {\path{doi:10.1142/9789812810458_0043}}.
\newline\urlprefix\url{http://arxiv.org/abs/hep-ph/0011333}

\bibitem{odyniec_rhic_2013}
G.~Odyniec, \href{https://pos.sissa.it/185/043}{{RHIC} {Beam} {Energy} {Scan}
  {Program}: {Phase} {I} and {II}}, in: Proceedings of 8th {International}
  {Workshop} on {Critical} {Point} and {Onset} of {Deconfinement} —
  {PoS}({CPOD} 2013), Vol. 185, SISSA Medialab, 2013, p. 043.
\newblock \href {https://doi.org/10.22323/1.185.0043}
  {\path{doi:10.22323/1.185.0043}}.
\newline\urlprefix\url{https://pos.sissa.it/185/043}

\bibitem{bzdak_mapping_2020}
A.~Bzdak, S.~Esumi, V.~Koch, J.~Liao, M.~Stephanov, N.~Xu,
  \href{http://arxiv.org/abs/1906.00936}{Mapping the {Phases} of {Quantum}
  {Chromodynamics} with {Beam} {Energy} {Scan}}, Physics Reports 853 (2020)
  1--87, arXiv: 1906.00936.
\newblock \href {https://doi.org/10.1016/j.physrep.2020.01.005}
  {\path{doi:10.1016/j.physrep.2020.01.005}}.
\newline\urlprefix\url{http://arxiv.org/abs/1906.00936}

\bibitem{star_collaboration_energy_2014}
L.~Adamczyk, et~al. (STAR~Collaboration),
  \href{http://arxiv.org/abs/1309.5681}{Energy {Dependence} of {Moments} of
  {Net}-proton {Multiplicity} {Distributions} at {RHIC}}, Physical Review
  Letters 112~(3) (2014) 032302, arXiv: 1309.5681.
\newblock \href {https://doi.org/10.1103/PhysRevLett.112.032302}
  {\path{doi:10.1103/PhysRevLett.112.032302}}.
\newline\urlprefix\url{http://arxiv.org/abs/1309.5681}

\bibitem{thader_higher_2016}
J.~Thäder, \href{http://arxiv.org/abs/1601.00951}{Higher {Moments} of
  {Net}-{Particle} {Multiplicity} {Distributions}}, Nuclear Physics A 956
  (2016) 320--323, arXiv: 1601.00951.
\newblock \href {https://doi.org/10.1016/j.nuclphysa.2016.02.047}
  {\path{doi:10.1016/j.nuclphysa.2016.02.047}}.
\newline\urlprefix\url{http://arxiv.org/abs/1601.00951}

\bibitem{bluhm_dynamics_2020}
M.~Bluhm, M.~Nahrgang, A.~Kalweit, M.~Arslandok, P.~Braun-Munzinger,
  S.~Floerchinger, E.~S. Fraga, M.~Gazdzicki, C.~Hartnack, C.~Herold,
  R.~Holzmann, I.~Karpenko, M.~Kitazawa, V.~Koch, S.~Leupold, A.~Mazeliauskas,
  B.~Mohanty, A.~Ohlson, D.~Oliinychenko, J.~M. Pawlowski, C.~Plumberg, G.~W.
  Ridgway, T.~Schäfer, I.~Selyuzhenkov, J.~Stachel, M.~Stephanov, D.~Teaney,
  N.~Touroux, V.~Vovchenko, N.~Wink,
  \href{http://arxiv.org/abs/2001.08831}{Dynamics of critical fluctuations:
  {Theory} -- phenomenology -- heavy-ion collisions}, arXiv:2001.08831 [hep-ph,
  physics:nucl-ex, physics:nucl-th]ArXiv: 2001.08831 (Jan. 2020).
\newline\urlprefix\url{http://arxiv.org/abs/2001.08831}

\bibitem{son_dynamic_2004}
D.~T. Son, M.~A. Stephanov, \href{http://arxiv.org/abs/hep-ph/0401052}{Dynamic
  universality class of the {QCD} critical point}, Physical Review D 70~(5)
  (2004) 056001, arXiv: hep-ph/0401052.
\newblock \href {https://doi.org/10.1103/PhysRevD.70.056001}
  {\path{doi:10.1103/PhysRevD.70.056001}}.
\newline\urlprefix\url{http://arxiv.org/abs/hep-ph/0401052}

\bibitem{hohenberg_theory_1977}
P.~C. Hohenberg, B.~I. Halperin,
  \href{https://link.aps.org/doi/10.1103/RevModPhys.49.435}{Theory of dynamic
  critical phenomena}, Reviews of Modern Physics 49~(3) (1977) 435--479.
\newblock \href {https://doi.org/10.1103/RevModPhys.49.435}
  {\path{doi:10.1103/RevModPhys.49.435}}.
\newline\urlprefix\url{https://link.aps.org/doi/10.1103/RevModPhys.49.435}

\bibitem{schweitzer_spectral_2020}
D.~Schweitzer, S.~Schlichting, L.~von Smekal,
  \href{http://arxiv.org/abs/2007.03374}{Spectral functions and dynamic
  critical behavior of relativistic \${Z}\_2\$ theories}, Nuclear Physics B 960
  (2020) 115165, arXiv: 2007.03374.
\newblock \href {https://doi.org/10.1016/j.nuclphysb.2020.115165}
  {\path{doi:10.1016/j.nuclphysb.2020.115165}}.
\newline\urlprefix\url{http://arxiv.org/abs/2007.03374}

\bibitem{aarts_spectral_2001}
G.~Aarts, \href{http://arxiv.org/abs/hep-ph/0108125}{Spectral function at high
  temperature in the classical approximation}, Physics Letters B 518~(3-4)
  (2001) 315--322, arXiv: hep-ph/0108125.
\newblock \href {https://doi.org/10.1016/S0370-2693(01)01081-4}
  {\path{doi:10.1016/S0370-2693(01)01081-4}}.
\newline\urlprefix\url{http://arxiv.org/abs/hep-ph/0108125}

\bibitem{berges_dynamic_2010}
J.~Berges, S.~Schlichting, D.~Sexty,
  \href{http://arxiv.org/abs/0912.3135}{Dynamic critical phenomena from
  spectral functions on the lattice}, Nuclear Physics B 832~(1-2) (2010)
  228--240, arXiv: 0912.3135.
\newblock \href {https://doi.org/10.1016/j.nuclphysb.2010.02.007}
  {\path{doi:10.1016/j.nuclphysb.2010.02.007}}.
\newline\urlprefix\url{http://arxiv.org/abs/0912.3135}

\bibitem{schlichting_spectral_2019}
S.~Schlichting, D.~Smith, L.~von Smekal,
  \href{http://arxiv.org/abs/1908.00912}{Spectral functions and critical
  dynamics of the \${O}(4)\$ model from classical-statistical lattice
  simulations}, arXiv:1908.00912 [cond-mat, physics:hep-lat, physics:hep-ph,
  physics:nucl-th]ArXiv: 1908.00912 (Jul. 2019).
\newline\urlprefix\url{http://arxiv.org/abs/1908.00912}

\bibitem{kubo_statistical-mechanical_1957}
R.~Kubo,
  \href{https://journals.jps.jp/doi/10.1143/JPSJ.12.570}{Statistical-{Mechanical}
  {Theory} of {Irreversible} {Processes}. {I}. {General} {Theory} and {Simple}
  {Applications} to {Magnetic} and {Conduction} {Problems}}, Journal of the
  Physical Society of Japan 12~(6) (1957) 570--586.
\newblock \href {https://doi.org/10.1143/JPSJ.12.570}
  {\path{doi:10.1143/JPSJ.12.570}}.
\newline\urlprefix\url{https://journals.jps.jp/doi/10.1143/JPSJ.12.570}

\bibitem{martin_theory_1959}
P.~C. Martin, J.~Schwinger,
  \href{https://link.aps.org/doi/10.1103/PhysRev.115.1342}{Theory of
  {Many}-{Particle} {Systems}. {I}}, Physical Review 115~(6) (1959) 1342--1373.
\newblock \href {https://doi.org/10.1103/PhysRev.115.1342}
  {\path{doi:10.1103/PhysRev.115.1342}}.
\newline\urlprefix\url{https://link.aps.org/doi/10.1103/PhysRev.115.1342}

\bibitem{folk_critical_2006}
R.~Folk, G.~Moser,
  \href{https://doi.org/10.1088%2F0305-4470%2F39%2F24%2Fr01}{Critical dynamics:
  a field-theoretical approach}, Journal of Physics A: Mathematical and General
  39~(24) (2006) R207--R313, publisher: IOP Publishing.
\newblock \href {https://doi.org/10.1088/0305-4470/39/24/R01}
  {\path{doi:10.1088/0305-4470/39/24/R01}}.
\newline\urlprefix\url{https://doi.org/10.1088%2F0305-4470%2F39%2F24%2Fr01}

\bibitem{rajagopal_static_1993}
K.~Rajagopal, F.~Wilczek, \href{http://arxiv.org/abs/hep-ph/9210253}{Static and
  {Dynamic} {Critical} {Phenomena} at a {Second} {Order} {QCD} {Phase}
  {Transition}}, Nuclear Physics B 399~(2-3) (1993) 395--425, arXiv:
  hep-ph/9210253.
\newblock \href {https://doi.org/10.1016/0550-3213(93)90502-G}
  {\path{doi:10.1016/0550-3213(93)90502-G}}.
\newline\urlprefix\url{http://arxiv.org/abs/hep-ph/9210253}

\bibitem{tauber_critical_2014}
U.~C. Täuber,
  \href{http://scans.hebis.de/HEBCGI/show.pl?33661625_toc.pdf}{Critical
  dynamics : a field theory approach to equilibrium and non-equilibrium scaling
  behavior}, Cambridge University Press, Cambridge {[u.a.]}, 2014.
\newline\urlprefix\url{http://scans.hebis.de/HEBCGI/show.pl?33661625_toc.pdf}

\bibitem{zwanzig_ensemble_1960}
R.~Zwanzig, \href{https://aip.scitation.org/doi/10.1063/1.1731409}{Ensemble
  {Method} in the {Theory} of {Irreversibility}}, The Journal of Chemical
  Physics 33~(5) (1960) 1338--1341, publisher: American Institute of Physics.
\newblock \href {https://doi.org/10.1063/1.1731409}
  {\path{doi:10.1063/1.1731409}}.
\newline\urlprefix\url{https://aip.scitation.org/doi/10.1063/1.1731409}

\bibitem{zwanzig_memory_1961}
R.~Zwanzig, \href{https://link.aps.org/doi/10.1103/PhysRev.124.983}{Memory
  {Effects} in {Irreversible} {Thermodynamics}}, Physical Review 124~(4) (1961)
  983--992, publisher: American Physical Society.
\newblock \href {https://doi.org/10.1103/PhysRev.124.983}
  {\path{doi:10.1103/PhysRev.124.983}}.
\newline\urlprefix\url{https://link.aps.org/doi/10.1103/PhysRev.124.983}

\bibitem{mori_transport_1965}
H.~Mori,
  \href{https://academic.oup.com/ptp/article/33/3/423/1925580}{Transport,
  {Collective} {Motion}, and {Brownian} {Motion}}, Progress of Theoretical
  Physics 33~(3) (1965) 423--455, publisher: Oxford Academic.
\newblock \href {https://doi.org/10.1143/PTP.33.423}
  {\path{doi:10.1143/PTP.33.423}}.
\newline\urlprefix\url{https://academic.oup.com/ptp/article/33/3/423/1925580}

\bibitem{kawasaki_simple_1973}
K.~Kawasaki, \href{https://doi.org/10.1088/0305-4470/6/9/004}{Simple
  derivations of generalized linear and nonlinear {Langevin} equations},
  Journal of Physics A: Mathematical, Nuclear and General 6~(9) (1973)
  1289--1295, publisher: IOP Publishing.
\newblock \href {https://doi.org/10.1088/0305-4470/6/9/004}
  {\path{doi:10.1088/0305-4470/6/9/004}}.
\newline\urlprefix\url{https://doi.org/10.1088/0305-4470/6/9/004}

\bibitem{sen_is_2002}
P.~Sen, S.~M. Bhattacharjee,
  \href{https://doi.org/10.1088/0305-4470/35/11/102}{Is there a true model-{D}
  critical dynamics?}, Journal of Physics A: Mathematical and General 35~(11)
  (2002) L141--L146, publisher: IOP Publishing.
\newblock \href {https://doi.org/10.1088/0305-4470/35/11/102}
  {\path{doi:10.1088/0305-4470/35/11/102}}.
\newline\urlprefix\url{https://doi.org/10.1088/0305-4470/35/11/102}

\bibitem{dzyaloshinskii_poisson_1980}
I.~E. Dzyaloshinskii, G.~E. Volovick,
  \href{https://www.sciencedirect.com/science/article/pii/0003491680901190}{Poisson
  brackets in condensed matter physics}, Annals of Physics 125~(1) (1980)
  67--97.
\newblock \href {https://doi.org/10.1016/0003-4916(80)90119-0}
  {\path{doi:10.1016/0003-4916(80)90119-0}}.
\newline\urlprefix\url{https://www.sciencedirect.com/science/article/pii/0003491680901190}

\bibitem{halperin_renormalization-group_1974}
B.~I. Halperin, P.~C. Hohenberg, S.-k. Ma,
  \href{https://link.aps.org/doi/10.1103/PhysRevB.10.139}{Renormalization-group
  methods for critical dynamics: {I}. {Recursion} relations and effects of
  energy conservation}, Physical Review B 10~(1) (1974) 139--153, publisher:
  American Physical Society.
\newblock \href {https://doi.org/10.1103/PhysRevB.10.139}
  {\path{doi:10.1103/PhysRevB.10.139}}.
\newline\urlprefix\url{https://link.aps.org/doi/10.1103/PhysRevB.10.139}

\bibitem{israel_transient_1979}
W.~Israel, J.~M. Stewart,
  \href{https://www.sciencedirect.com/science/article/pii/0003491679901301}{Transient
  relativistic thermodynamics and kinetic theory}, Annals of Physics 118~(2)
  (1979) 341--372.
\newblock \href {https://doi.org/10.1016/0003-4916(79)90130-1}
  {\path{doi:10.1016/0003-4916(79)90130-1}}.
\newline\urlprefix\url{https://www.sciencedirect.com/science/article/pii/0003491679901301}

\bibitem{israel_thermodynamics_1981}
W.~Israel,
  \href{https://www.sciencedirect.com/science/article/pii/037843718190220X}{Thermodynamics
  of relativistic systems}, Physica A: Statistical Mechanics and its
  Applications 106~(1) (1981) 204--214.
\newblock \href {https://doi.org/10.1016/0378-4371(81)90220-X}
  {\path{doi:10.1016/0378-4371(81)90220-X}}.
\newline\urlprefix\url{https://www.sciencedirect.com/science/article/pii/037843718190220X}

\bibitem{onsager_crystal_1944}
L.~Onsager, \href{https://link.aps.org/doi/10.1103/PhysRev.65.117}{Crystal
  {Statistics}. {I}. {A} {Two}-{Dimensional} {Model} with an {Order}-{Disorder}
  {Transition}}, Physical Review 65~(3-4) (1944) 117--149.
\newblock \href {https://doi.org/10.1103/PhysRev.65.117}
  {\path{doi:10.1103/PhysRev.65.117}}.
\newline\urlprefix\url{https://link.aps.org/doi/10.1103/PhysRev.65.117}

\bibitem{kos_precision_2016}
F.~Kos, D.~Poland, D.~Simmons-Duffin, A.~Vichi,
  \href{http://arxiv.org/abs/1603.04436}{Precision {Islands} in the {Ising} and
  \${O}({N})\$ {Models}}, Journal of High Energy Physics 2016~(8) (2016) 36,
  arXiv: 1603.04436.
\newblock \href {https://doi.org/10.1007/JHEP08(2016)036}
  {\path{doi:10.1007/JHEP08(2016)036}}.
\newline\urlprefix\url{http://arxiv.org/abs/1603.04436}

\bibitem{komargodski_random-bond_2017}
Z.~Komargodski, D.~Simmons-Duffin, \href{http://arxiv.org/abs/1603.04444}{The
  {Random}-{Bond} {Ising} {Model} in 2.01 and 3 {Dimensions}}, Journal of
  Physics A: Mathematical and Theoretical 50~(15) (2017) 154001, arXiv:
  1603.04444.
\newblock \href {https://doi.org/10.1088/1751-8121/aa6087}
  {\path{doi:10.1088/1751-8121/aa6087}}.
\newline\urlprefix\url{http://arxiv.org/abs/1603.04444}

\bibitem{duane_hybrid_1987}
S.~Duane, A.~D. Kennedy, B.~J. Pendleton, D.~Roweth,
  \href{http://www.sciencedirect.com/science/article/pii/037026938791197X}{Hybrid
  {Monte} {Carlo}}, Physics Letters B 195~(2) (1987) 216--222.
\newblock \href {https://doi.org/10.1016/0370-2693(87)91197-X}
  {\path{doi:10.1016/0370-2693(87)91197-X}}.
\newline\urlprefix\url{http://www.sciencedirect.com/science/article/pii/037026938791197X}

\bibitem{dammann_dynamical_1993}
B.~Dammann, J.~D. Reger,
  \href{https://doi.org/10.1209%2F0295-5075%2F21%2F2%2F006}{Dynamical
  {Critical} {Exponent} of the {Two}-{Dimensional} {Ising} {Model}},
  Europhysics Letters (EPL) 21~(2) (1993) 157--162.
\newblock \href {https://doi.org/10.1209/0295-5075/21/2/006}
  {\path{doi:10.1209/0295-5075/21/2/006}}.
\newline\urlprefix\url{https://doi.org/10.1209%2F0295-5075%2F21%2F2%2F006}

\bibitem{matz_dynamic_1994}
R.~Matz, D.~L. Hunter, N.~Jan, \href{https://doi.org/10.1007/BF02188583}{The
  dynamic critical exponent of the three-dimensional {Ising} model}, Journal of
  Statistical Physics 74~(3) (1994) 903--908.
\newblock \href {https://doi.org/10.1007/BF02188583}
  {\path{doi:10.1007/BF02188583}}.
\newline\urlprefix\url{https://doi.org/10.1007/BF02188583}

\bibitem{wang_study_1995}
F.~Wang, N.~Hatano, M.~Suzuki,
  \href{https://doi.org/10.1088%2F0305-4470%2F28%2F16%2F012}{Study on dynamical
  critical exponents of the {Ising} model using the damage spreading method},
  Journal of Physics A: Mathematical and General 28~(16) (1995) 4543--4552.
\newblock \href {https://doi.org/10.1088/0305-4470/28/16/012}
  {\path{doi:10.1088/0305-4470/28/16/012}}.
\newline\urlprefix\url{https://doi.org/10.1088%2F0305-4470%2F28%2F16%2F012}

\bibitem{nightingale_dynamic_1996}
M.~P. Nightingale, H.~W.~J. Blöte,
  \href{http://arxiv.org/abs/cond-mat/9601059}{The {Dynamic} {Exponent} of the
  {Two}-{Dimensional} {Ising} {Model} and {Monte} {Carlo} {Computation} of the
  {Sub}-{Dominant} {Eigenvalue} of the {Stochastic} {Matrix}}, Physical Review
  Letters 76~(24) (1996) 4548--4551, arXiv: cond-mat/9601059.
\newblock \href {https://doi.org/10.1103/PhysRevLett.76.4548}
  {\path{doi:10.1103/PhysRevLett.76.4548}}.
\newline\urlprefix\url{http://arxiv.org/abs/cond-mat/9601059}

\bibitem{yalabik_monte_1982}
M.~C. Yalabik, J.~D. Gunton,
  \href{https://link.aps.org/doi/10.1103/PhysRevB.25.534}{Monte {Carlo}
  renormalization-group studies of kinetic {Ising} models}, Physical Review B
  25~(1) (1982) 534--537, publisher: American Physical Society.
\newblock \href {https://doi.org/10.1103/PhysRevB.25.534}
  {\path{doi:10.1103/PhysRevB.25.534}}.
\newline\urlprefix\url{https://link.aps.org/doi/10.1103/PhysRevB.25.534}

\bibitem{zheng_monte_2000}
B.~Zheng, \href{http://arxiv.org/abs/cond-mat/0103133}{Monte {Carlo}
  {Simulations} of {Critical} {Dynamics} with {Conserved} {Order} {Parameter}},
  Physics Letters A 277~(4-5) (2000) 257--261, arXiv: cond-mat/0103133.
\newblock \href {https://doi.org/10.1016/S0375-9601(00)00658-7}
  {\path{doi:10.1016/S0375-9601(00)00658-7}}.
\newline\urlprefix\url{http://arxiv.org/abs/cond-mat/0103133}

\bibitem{berdnikov_slowing_2000}
B.~Berdnikov, K.~Rajagopal, \href{http://arxiv.org/abs/hep-ph/9912274}{Slowing
  {Out} of {Equilibrium} {Near} the {QCD} {Critical} {Point}}, Physical Review
  D 61~(10) (2000) 105017, arXiv: hep-ph/9912274.
\newblock \href {https://doi.org/10.1103/PhysRevD.61.105017}
  {\path{doi:10.1103/PhysRevD.61.105017}}.
\newline\urlprefix\url{http://arxiv.org/abs/hep-ph/9912274}

\bibitem{mukherjee_real_2015}
S.~Mukherjee, R.~Venugopalan, Y.~Yin,
  \href{http://arxiv.org/abs/1506.00645}{Real time evolution of non-{Gaussian}
  cumulants in the {QCD} critical regime}, Physical Review C 92~(3) (2015)
  034912, arXiv: 1506.00645.
\newblock \href {https://doi.org/10.1103/PhysRevC.92.034912}
  {\path{doi:10.1103/PhysRevC.92.034912}}.
\newline\urlprefix\url{http://arxiv.org/abs/1506.00645}

\bibitem{mukherjee_universal_2016}
S.~Mukherjee, R.~Venugopalan, Y.~Yin,
  \href{http://arxiv.org/abs/1605.09341}{Universal off-equilibrium scaling of
  critical cumulants in the {QCD} phase diagram}, Physical Review Letters
  117~(22) (2016) 222301, arXiv: 1605.09341.
\newblock \href {https://doi.org/10.1103/PhysRevLett.117.222301}
  {\path{doi:10.1103/PhysRevLett.117.222301}}.
\newline\urlprefix\url{http://arxiv.org/abs/1605.09341}

\bibitem{mukherjee_universality_2017}
S.~Mukherjee, R.~Venugopalan, Y.~Yin,
  \href{http://arxiv.org/abs/1704.05427}{Universality regained:
  {Kibble}-{Zurek} dynamics, off-equilibrium scaling and the search for the
  {QCD} critical point}, Nuclear Physics A 967 (2017) 820--823, arXiv:
  1704.05427.
\newblock \href {https://doi.org/10.1016/j.nuclphysa.2017.06.049}
  {\path{doi:10.1016/j.nuclphysa.2017.06.049}}.
\newline\urlprefix\url{http://arxiv.org/abs/1704.05427}

\bibitem{nahrgang_diffusive_2019}
M.~Nahrgang, M.~Bluhm, T.~Schaefer, S.~A. Bass,
  \href{http://arxiv.org/abs/1804.05728}{Diffusive dynamics of critical
  fluctuations near the {QCD} critical point}, Physical Review D 99~(11) (2019)
  116015, arXiv: 1804.05728.
\newblock \href {https://doi.org/10.1103/PhysRevD.99.116015}
  {\path{doi:10.1103/PhysRevD.99.116015}}.
\newline\urlprefix\url{http://arxiv.org/abs/1804.05728}

\end{thebibliography}

\end{document}